\documentclass[a4paper,fleqn,usenatbib]{mnras}

\usepackage{graphicx,multicol}
\usepackage{times}

\usepackage{amssymb}   

\usepackage{newtxtext,newtxmath}

\usepackage[utf8]{inputenc}
\usepackage[T1]{fontenc}
\DeclareRobustCommand{\VAN}[3]{#2}
\let\VANthebibliography\thebibliography
\def\thebibliography{\DeclareRobustCommand{\VAN}[3]{##3}\VANthebibliography}

\usepackage{graphicx}	
\usepackage{amsmath}	
\usepackage{ulem}

\renewcommand{\v}[1]{\ensuremath{\mathbf{#1}}} 

\usepackage{xcolor}
\definecolor{blazeorange}{rgb}{1.0, 0.4, 0.0}
\definecolor{seagreen}{rgb}{0.18, 0.55, 0.34}
\definecolor{rufous}{rgb}{0.66, 0.11, 0.03}
\definecolor{royalfuchsia}{rgb}{0.79, 0.17, 0.57}
\definecolor{scarlet}{rgb}{1.0, 0.13, 0.0}
\definecolor{royalpurple}{rgb}{0.47, 0.32, 0.66}
\definecolor{darkblue}{rgb}{0, 0, 0.66}

\title[internal shocks]{Internal Shocks Hydrodynamics: the Collision of Two Cold Shells in Detail
}

\author[Sk. Minhajur Rahaman]{
Sk. Minhajur Rahaman,$^{1}$\thanks{E-mail: rahaman.minhajur93@gmail.com}
Jonathan Granot,$^{1,2,3}$
Paz Beniamini$^{1,2,3}$
\\
$^{1}$Astrophysics Research Center of the Open University (ARCO), The Open University of Israel, P.O Box 808, Ra’anana 4353701, Israel\\
$^{2}$Department of Natural Sciences, The Open University of Israel, P.O Box 808, Ra’anana 4353701, Israel \\
$^{3}$Department of Physics, The George Washington University, 725 21st Street NW, Washington, DC 20052, USA
}

\date{Accepted XXX. Received YYY; in original form ZZZ}

\begin{document}
\label{firstpage}
\pagerange{\pageref{firstpage}--\pageref{lastpage}}
\maketitle

\begin{abstract}
Emission in many astrophysical transients originates from a shocked fluid. A central engine typically produces an outflow with varying speeds, leading to internal collisions within the outflow at finite distances from the source. Each such collision produces a pair of forward and reverse shocks with the two shocked regions
separated by a contact discontinuity (CD). As a useful approximation, we consider the head-on collision between two cold and uniform shells (a slower leading shell and a faster trailing shell) of finite radial width, and study the dynamics of shock propagation in planar geometry.
We find significant differences between the forward and reverse shocks, in terms of their strength, internal energy production efficiency, 
and the time it takes for the shocks to sweep through the respective shells. We consider the subsequent propagation of rarefaction waves in the shocked regions and explore the cases where these waves can catch up with the shock fronts and thereby limit the internal energy dissipation. We demonstrate the importance of energy transfer from the trailing to leading shell through $pdV$  work across the CD. 
We outline the parameter space regions relevant for models of different transients,e.g., Gamma-ray burst (GRB) internal shock model, fast radio burst (FRB) blastwave model, Giant flare due to magnetars, and superluminous supernovae (SLSN) ejecta. We find that the reverse shock likely dominates the internal energy production for many astrophysical transients. 
\end{abstract}



\section{Introduction}

In many astrophysical scenarios involving different classes of objects, transient electromagnetic emission is thought to arise from internal shocks. In particular, internal shocks have been invoked in blazars \citep[e.g.][]{Rees78,Levinson98,Ghisellini99}, GRBs \citep[e.g.][]{Rees-Meszaros94,Sari-Piran97b,Daigne-Mochkovitch98}, FRBs \citep[e.g.][]{2017ApJ...842...34W,2019MNRAS.485.4091M,2020MNRAS.494.4627M}, superluminous supernova \citep[e.g.][]{2007Natur.450..390W,2014MNRAS.441..289B,2018SSRv..214...59M,2023NatAs...7..779L,2023arXiv230403360K}, magnetar giant flares \citep[e.g.][]{Granot+06,Sculptor-GF-21}, 
etc. In these cases, the central engine generates an outflow whose asymptotic speed varies with time at the ejection site and therefore with the distance from the source. Faster parts of the outflow overtake slower parts leading to collisions that give rise to shocks that are referred to as internal shocks (as they arise within the outflow, in contrast to external shocks that are caused by the outflow's interaction with the external medium). 

It is useful to approximate the outflow as consisting of discrete, uniform shells of finite radial width. In particular, we model here in detail the collision between a pair of uniform, cold shells. Such a collision forms a pair of shock fronts -- a forward shock that accelerates the leading shell and a reverse shock that decelerates the trailing shell, where the two shocked parts of these shells are separated by a contact discontinuity (CD). The reverse/forward shocks 
dissipate the initial kinetic energy of the shells into internal energy,  part of which can be radiated by the particles accelerated in this process and produce the observed emission in different transient astrophysical sources. However, most works that studied the energy dissipation efficiency in internal shocks (e.g. \citealt{Kobayashi+97}, \citealt{Daigne-Mochkovitch98}) used a ballistic model featuring  a completely inelastic (plastic) collision of two infinitely thin shells. 
Such an analysis does not account for the underlying shock physics and hence ignores much of the relevant dynamics.
Few studies \citep[e.g.][]{Peer+17} that do account for the shock physics,
do not study time evolution of the shock fronts for a generic parameter space. 

Therefore, there is a need for 
a comprehensive work that self-consistently studies the hydrodynamics of both shocks 
and the application of the shock dynamics to internal shocks models of various astrophysical objects. This is the aim of the present work. In particular, we study under which conditions the finite widths of the two shells can limit the energy dissipation in each shock, as well as the total internal energy production efficiency.

The paper is structured as follows. \S\,\ref{hydro} introduces our basic model parameters 
and describes the setup for solving the jump conditions across both shocks and the CD, to solve for the system's hydrodynamics. 
\S\,\ref{dissp_limit} describes how the rarefaction waves, which form when a shock finishes crossing a shell, may limit the energy dissipation by the shock fronts. \S\,\ref{sphere_geom} describes the limitation of 
our approximations of cold pre-collision shells and a planar geometry.
In \S\,\ref{appl_cases} we explore the internal shocks hydrodynamic parameter space relevant for  
different astrophysical transients.  Our conclusions are discussed in \S\,\ref{sec:conclusions}.

\section{The setup and jump conditions across the two shocks and the contact discontinuity}\label{hydro}

In this section we describe the setup before and after the collision. We broadly have one global frame -- that lab frame that is the rest frame of the central source (or engine), as well as a number of local frames, namely the rest frame of the fluid in each of the regions in the flow. All quantities measured in the lab frame are unprimed, while quantities measured in the local fluid rest frame are primed. 

\subsection{The description of the ejected shells pre-collision} 

\begin{figure}
    \centering
    \includegraphics[scale=0.7,trim=0.6cm 0.2cm 0.5cm 0.2cm, clip]{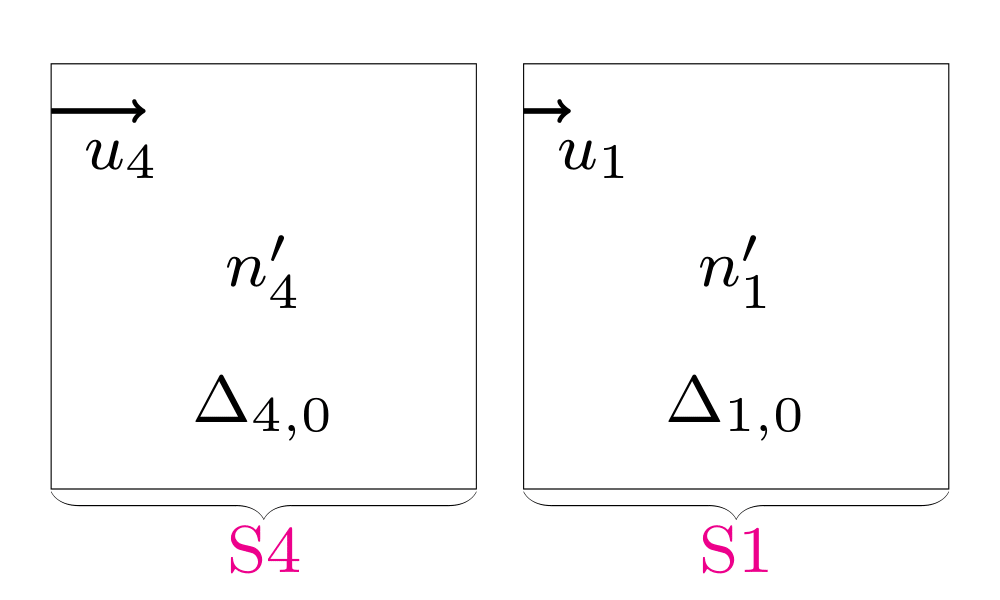} \\
    \includegraphics[scale=0.95,trim=1.8cm 0.5cm 2cm 0.0cm, clip]{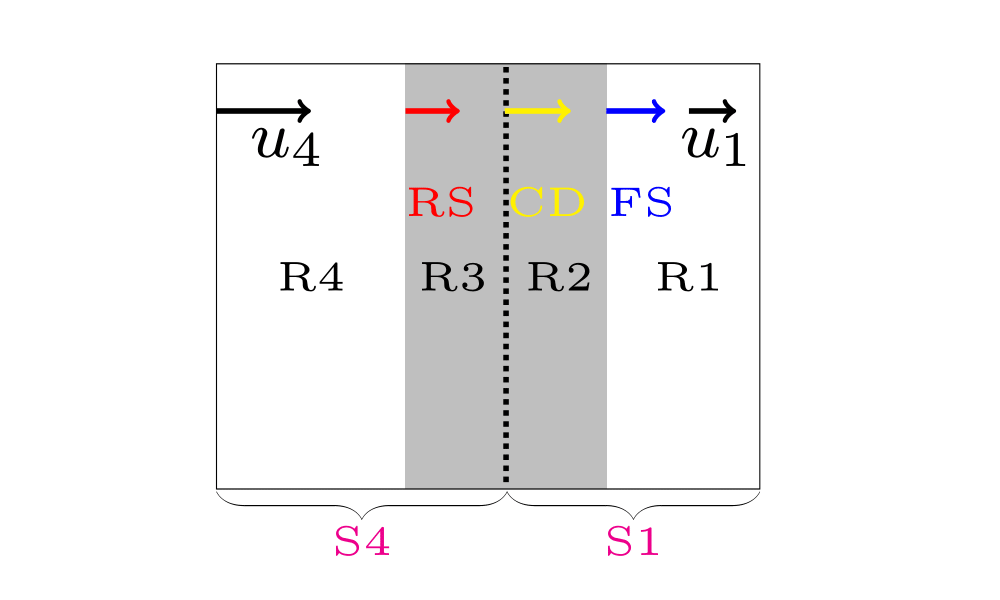}
    \caption{The pre-collision and post-collision setup for the collision of two cold and uniform shells of equal kinetic energy and initial radial width, and initial proper speeds $(u_1,u_4)=(100,200)$ at $t = t_\mathrm{o} + 0.3 t_\mathrm{RS}$.  \textbf{\textit{Top}}: 
    The pre-collision 
    structure -- the leading and trailing shells are denoted by S1 and S4, respectively. The arrow size is proportional to the proper speed of the respective shell.  \textbf{\textit{Bottom}}: The post-collision 
    structure of the two shells, which now divides into 4 regions (R1 to R4), where each shell develops a shocked region (the \textit{shaded regions}), and the two shocked regions (R2 and R3) are separated by a CD. The structure is shown at a time when the reverse shock front has swept through 40\% of shell S4. The radial width of the four regions is to scale. The arrow sizes scale as the proper speeds of the shells, the CD, and the two shock fronts.  } 
    \label{fig1}
\end{figure}

\begin{table}
    \centering
    \caption{List of seven basic parameters for two cold-shell collision. Here the subscript $i=(1,4)$ stands for the leading and the trailing shell respectively.}
    \begin{tabular}{c|c} \hline 
        Symbol & Definition  \\ \hline
        $E_\mathrm{k,i,0}$ & Available kinetic energy in shell $i$ just before collision \\
         $\Delta_\mathrm{i,0}$ & The radial width of shell $i$ just before collision \\ 
        $u_\mathrm{i}$ & The proper speed of shell $i$ \\
        $t_\mathrm{off}$  &  Time between ejection of shells S1 and S4  \\ \hline 
    \end{tabular}
    \label{basic_param}
\end{table}

\begin{table}
    \centering
    \caption{List of  derived parameters to be used throughout the text. Note that $a_\mathrm{u} >1$ is required to ensure the shells S1 and S4 collide.}
    \begin{tabular}{ccc} 
    \hline 
     Symbol & Definition &  Expression 
     \\ \hline 
     $R_\mathrm{o}$   & Collision radius & $\frac{\beta_1 \beta_4 c t_\mathrm{off}}{(\beta_4 - \beta_1)}$ \\ \\ 
      $\chi$ & Radial width ratio of S1 to S4  & $\frac{\Delta_{1,0}}{\Delta_{4,0}}$ \\ \\ 
    $a_\mathrm{u}$ & Proper speed ratio of S4 to S1  & $\frac{u_{4}}{u_{1}}$ > 1 \\  \\ 
    $f$ & Proper density ratio of S4 to S1 & $\frac{n'_{4}}{n'_{1}} = \chi 
    \frac{E_\mathrm{k,4,0}}{E_\mathrm{k,1,0}} 
    \frac{\Gamma_1 (\Gamma_1 - 1)}{\Gamma_4 (\Gamma_4 - 1)} $  \\ \hline 
    \end{tabular}
     \label{der_param}
\end{table}

In this subsection we describe the setup before the shells collide.
Our initial setup is illustrated in the \textit{top panel} of Figure~\ref{fig1}. 
The central engine produces a cold leading shell (labeled S1) and a cold trailing shell (labeled S4) of  initial kinetic energies $(E_\mathrm{k,1,0},E_\mathrm{k,4,0})$ with radial widths $(\Delta_\mathrm{1,0},\Delta_\mathrm{4,0})$ and proper speeds ($u_1,u_4$). Assuming both shells are initially cold, the \textit{available} energy is entirely due to the initial kinetic energy of the outflow and is given as
\begin{equation}
  E_\mathrm{k,1,0} = \left(\Gamma_1 - 1 \right) M_{1,0} c^2 , \quad \quad  E_\mathrm{k,4,0} = \left(\Gamma_4 - 1 \right) M_{4,0} c^2  , \label{Ek}
\end{equation}
where $(M_{1,0},M_{4,0})$ are the \textit{rest} masses of the shells. The rest mass energy has been subtracted from the initial total energy of the shells as it is unavailable for internal energy dissipation at the shocks that form in the  collision. 

As shown in Table \ref{basic_param}  our setup has seven basic parameters  viz., the time $t_\mathrm{off}$ between the ejection of the two shells , and,  the proper speeds $(u_{1},u_{4})$, the initial radial widths $(\Delta_{1,0},\Delta_{4,0})$ and the initial kinetic energies $(E_\mathrm{k,1,0},E_\mathrm{k,4,0})$ of the shells. The number of free parameters can be reduced depending on the frame of reference and assuming particular conditions viz., equal mass and equal energy shells in the ultra-relativistic and newtonian limits. As shown in Table \ref{der_param} there are four derived parameters required to  describe shock hydrodynamics post-collision in the lab frame viz., the collision radii $R_\mathrm{o}$, the ratio of the initial radial width of shell S1 to S4 $\chi$, the proper speed contrast $a_\mathrm{u}$ and the proper density contrast $f$. Since we assume planar geometry, $t_\mathrm{off}$ only decides the collision radii $R_\mathrm{o}$ but does not decide the shock hydrodynamics (see \S \ref{sphere_geom} for discussion on the effects due to spherical geometry). Morever, if the shock hydrodynamics were to be studied not in the lab frame but in the rest frame of shell S1, only two quantities would suffice for the description of shocked fluid viz., the proper density ratio $f$ and the relative proper speed $u_{41}$ of shell S1 and S4 (see \S \ref{restS1_soln}). In order to estimate the ratio of the time taken (in the lab frame) by the FS/RS to sweep to the front/rear edge of the respective shell one needs the ratio $\chi$ the radial widths of the respective shells. Moreover, if the source power $L$ of the central engine is constant during ejection of both shells at ultra-relativistic speeds $\beta \rightarrow 1$, the ratio $\chi = \frac{t_\mathrm{on1}}{t_\mathrm{on4}}$. In this instance, only three free parameters are required to describe shock hydrodynamics.  To illustrate this point,  we consider the collision of two equal energy shells of equal radial width as our prototypical case for all of our illustrations.

In the next subsection we describe the hydrodynamics of shock propagation post collision.

\subsection{Hydrodynamics of the reverse and the forward shock fronts}

In this subsection we describe the hydrodynamics of shock propagation after the shells collide. Post ejection of the shells the trailing shell S4 collides with the leading shell S1 at the lab frame $t_\mathrm{o}$ and at a distance $R_\mathrm{o}$ from the central engine. As seen in the \textit{bottom panel} of Fig.~\ref{fig1}, for $u_4>u_1$ the shells S1 and S4 collide and the collision launches a pair of reverse (hereafter RS) and forward shock (hereafter FS) fronts. The two shocked regions are separated by a contact discontinuity (hereafter CD). The FS sweeps through shell S1 while the RS sweeps through shell S4. Post collision the two shells develop four regions (R1,\,R2,\,R3,\,R4). Region R1 (R4) is the portion of S1 (S4) that is not yet shocked by the FS (RS). Region R2 (R3) is the portion of S1 (S4) shocked by the FS (RS). Before collision the internal energy in both shells is zero, and this still holds for regions R1 and R4. Post collision, as both the forward and the reverse shock fronts dissipates energy in regions R2 and R3 respectively, there is non-zero internal energy in both of these regions. As a result, there is a non-zero pressure in both of these regions which leads to $pdV$ work across CD (see discussion preceding equation~(\ref{WpdV})). In summary, post collision four regions exist: two unshocked regions (R1,R4) and two shocked regions that develop as a result of the collision (R2,R3).

To study shock hydrodynamics we assume a \textit{planar} geometry wherein the number density in regions $(R1,R2,R3,R4)$ does not change with time (the volume of each fluid element in these regions remains constant, both in the comoving frame and in the lab frame). The quantities determined by shock hydrodynamics are summarized in Table \ref{shock_quant}. Subsequently, all physical quantities are homogeneous in all 4 regions at all times. In particular, the propagation velocities of the shock fronts remain constant. As a result, all changes in all 4 regions scale linearly with time (see Table \ref{time_evol}). The limitation of this approach will be discussed in section \ref{sphere_geom}.

\begin{table}
    \centering
     \caption{Symbols and definitions for quantities required to describe post-collision hydrodynamics. The comoving quantities in each region are primed and the regions are referred to by a subscript (Here $j= (1,2,3,4)$ refers to regions 1,2, 3, and 4 respectively). For cold shells the internal energy density in regions 1 and 4 are zero  ($e'_\mathrm{int,1},e'_\mathrm{int,4} $) = 0.   }
    \begin{tabular}{c|c} \hline 
        Symbol & Definition \\ \hline  
        $n'_\mathrm{j}$ & Proper particle number density in region $j$ \\
          $e'_\mathrm{int,j}$ & The comoving internal energy density in regions $j$ \\  
        $\Gamma_\mathrm{ij}$ & The relative LF of regions R$i$ and R$j$ \\ 
          $u $ & The proper speed of the shocked fluid in regions R2 and R3 \\ 
           $u_\mathrm{i}$ & The proper speed of the shock front $i=(FS,RS)$ \\ 
           $t_\mathrm{i}$ & The shell crossing time by   shock front $i=(FS,RS)$ \\ 
          $\Gamma_\mathrm{ij} - 1$ & Internal energy per unit 
        rest energy in region R$j\;(j=2,3)$ \\
         $E_\mathrm{j, int}$ & Internal energy in R$j$ $(j\!=\!2,3)$ at shock crossing ($t_{\rm FS},\,t_{\rm RS}$) 
        \\
        $E_\mathrm{j, k}$ & Kinetic energy in R$j$ $(j\!=\!2,3)$ at shock crossing ($t_{\rm FS},\,t_{\rm RS}$)   \\
        $E_\mathrm{j, int} (t)$ & Internal energy in R$j$ $(j\!=\!2,3)$ at time $t$ \\ 
        $E_\mathrm{j, k} (t)$ & Kinetic energy in R$j$ $(j\!=\!1,2,3,4)$ at time $t$ \\\hline  
    \end{tabular}
    \label{shock_quant}
\end{table}

Our objective is to estimate the proper speed $u$ of the shocked fluid given the proper densities $(n'_1,n'_4)$ and the lab frame proper speeds ($u_1,u_4$) of the shells (S1,S4) just before collision.  The hydrodynamical shock jump  conditions for the collision of two cold shell collisions can be summarized \citep[e.g.,][]{BM76} as 
\begin{subequations}
\begin{align}
 & \frac{e'_\mathrm{2,int}}{n'_{2} m_\mathrm{p} c^2 } = \left(\Gamma_{21} - 1 \right),  \label{BM1}  \\ 
     & \frac{n'_2}{n'_1} = 4 \Gamma_{21} , \label{BM2} \\
     & \frac{e'_\mathrm{3,int}}{n'_{3} m_\mathrm{p} c^2 } = \left(\Gamma_{34} - 1 \right) , \label{BM3}  \\ 
     & \frac{n'_3}{n'_4} = 4 \Gamma_{34}, \label{BM4}
\end{align}
\end{subequations}
(see Appendix A for the full derivation), where $m_\mathrm{p}$ is the proton mass and the other physical quantities appearing in the equations are summarized in Table~\ref{shock_quant}. The relative LFs are given as 
\begin{equation}
\Gamma_{21} = \Gamma_2 \Gamma_1 (1 - \beta_1 \beta_2) \\ 
, \quad \quad \Gamma_{34} = \Gamma_3 \Gamma_4 (1 - \beta_3 \beta_4) 
\end{equation}

Equations (\ref{BM1}) and (\ref{BM3}) relate the internal energy per baryon to the shock strength $(\Gamma_{21},\Gamma_{34}) - 1$. In other words, the efficiency of energy dissipation associated with forward/reverse shock front increases if the proper speed of the shocked fluid $(u_2,u_3)$ is significantly different from $(u_1,u_4)$. Thus, the internal energy per baryon is small for \textit{Newtonian} shocks ,$(\Gamma_{21},\Gamma_{34}) - 1 \ll 1$, and is significant for \textit{relativistic shocks} $(\Gamma_{21}, \Gamma_{34}) \gg 1$. Equations (\ref{BM2}) and (\ref{BM4}) show that the proper densities of particles in shocked regions are higher than those of the unshocked regions by a shock \textit{compression ratio}. 

The velocities and the pressure across the CD are equal
\begin{subequations}
\begin{align}
  & u_2 = u_{3} = u , \label{bc1} \\ 
& p_{2} = p_{3} ,   \label{bc2}
\end{align}
\end{subequations}

Using equations (\ref{bc1})-(\ref{bc2}) in equations (\ref{BM1})-(\ref{BM4}) gives 
\begin{equation}\label{fund_eqn}
     \left( \Gamma^2_{21} - 1 \right) = f (\Gamma^2_{34} - 1) \ \Leftrightarrow\ u^2_{21} = f u^2_{34}\ , 
\end{equation}
corresponding to equal ram pressures across the CD in its rest frame. It can be seen that for $f<1$ the reverse shock strength ($u_{34}$ or $\Gamma_{34}$) is higher than the forward shock strength ($u_{21}$ or $\Gamma_{21}$) and vice versa. In particular, the shock strengths are equal for $f=1$. Equation~(\ref{fund_eqn}) has the symmetry that under transformation $f \rightarrow 1/f$ the ratio undergoes the  
transformation $\frac{u_{21}}{u_{34}} \rightarrow \frac{u_{34}}{u_{21}}$, which simply corresponds to switching the labels of the two shocked regions (R2 and R3) and the two unshocked regions (R1 and R4), as in the CD's rest frame it makes no difference which shell is leading and which shell is trailing in the lab frame. 

It can also be instructive to analyse the shock hydrodynamics in the CD frame. In Appendix I, we analyse the $f=1$ scenario in the CD frame and compare our results with those by \cite{2004ApJ...611.1021K}, who performed a numerical study in CD frame for a collision of ultra-relativistic shells. The principal difficulty in a CD frame approach is associated with estimating the thermal energy dissipated in the lab frame using quantities in the CD frame. Specifically, in the CD frame there is no $p dV$ work across the CD from region R3 to R2, and as a result the thermal efficiency is underestimated when calculated using quantities in the CD frame. In \S 3 we circumvent this difficulty by estimating the thermal efficiencies in the lab frame (for an expanded discussion see the last paragraph in Appendix I).

In \ref{restS1_soln} we will solve for the proper speed of the shocked fluid in the rest frame of shell S1 (where one can explicitly see that the results depend only on the density and LF ratio between the shells) and then in \ref{lab_soln} Lorentz transform the solution from rest frame of S1 to the lab frame (which adds an additional parameter, the absolute proper speed of S1 but which is useful for considering observed properties resulting from internal shocks).

\subsubsection{Solution in the rest frame of shell S1}\label{restS1_soln}

Equation (\ref{fund_eqn}) can be solved in the rest frame of region R1 to obtain the proper speed of the shocked fluid relative to frame 1 (see Appendix B for a full derivation) 
\begin{equation}\label{soln}
    u_{21} = u_{31} =  u_{41} \sqrt{\frac{  2 f^{3/2} \Gamma_{41} - f(1+f)  }{ 2 f( u^2_{41} + \Gamma^2_{41} ) - (1 + f^2)}}\ . 
\end{equation} 
The solution in equation~(\ref{soln}) is the general solution in the rest frame of region R1. It depends only on two parameters, namely the relative initial proper speed $u_{41}$ and  proper density contrast $f$ of S4 and S1. 

\begin{figure}
    \centering
     \includegraphics[scale= 0.55]{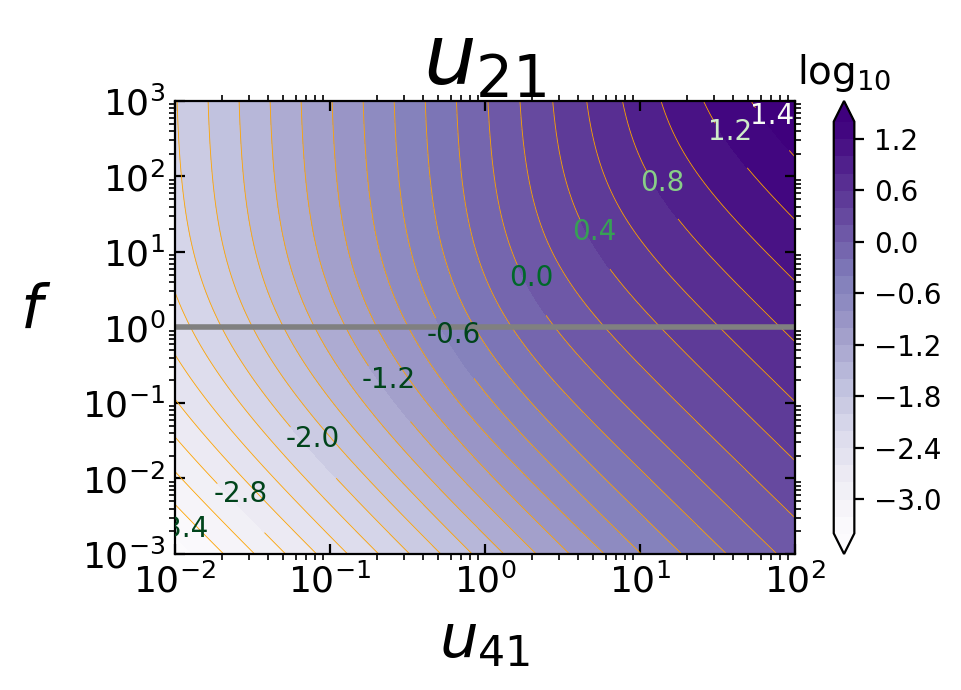}\\
      \includegraphics[scale= 0.55]{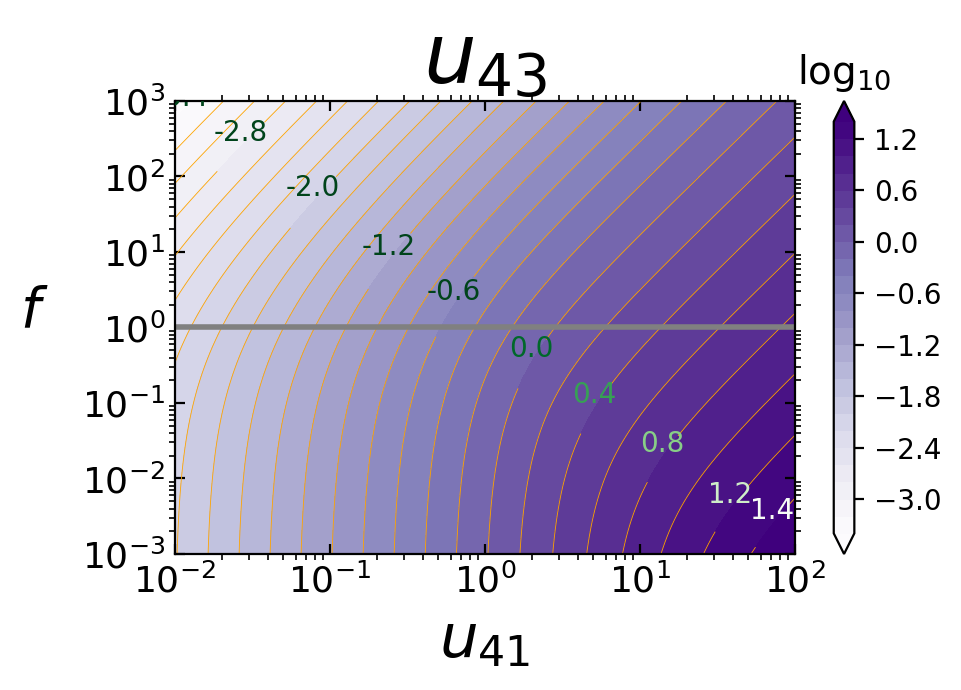}\\
      \includegraphics[scale=0.55]{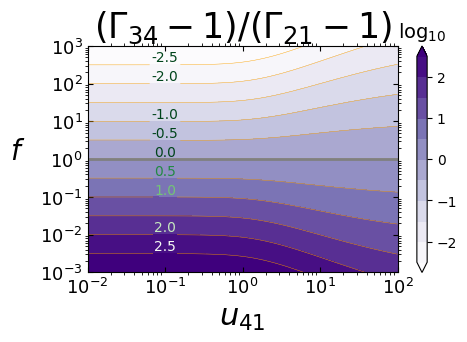} \\
      \vspace{-0.2cm}
    \caption{ The general solution for the proper speed of the shocked fluid.  \textbf{\textit{Top:}} shows a logarithmic contour plot of the relative proper velocity of regions 2 and 1, $u_{21}$, as a function of the relative proper speed $u_{41}$ and proper density ratio $f=n'_4/n'_1$ of the unshocked parts of the two shells (S4 and S1). \textbf{\textit{Middle:}}  
    the relative proper velocity of regions 3 and 4, $u_{43}$, as a function of 
    $u_{41}$ and $f$. \textbf{\textit{Bottom:}} 
    the shock strength ratio $(\Gamma_{34}-1)/(\Gamma_{21} - 1)$ as a function of $u_{41}$ and $f$. The mirror symmetry of the ratio of the shock strength reflects the symmetry inherent in equation (\ref{fund_eqn}) under the transformation $f \rightarrow 1/f$.  }
    \label{gen_soln}
\end{figure}

The upper and middle panels of Fig.~\ref{gen_soln} show the general solution of $u_{21}$ and $u_{43}$, respectively, as a function of relative proper speed $u_{41}$ and proper density contrast $f$. They correspond to each other upon reflection about the $f=1$ line due to the symmetry mentioned above.
The lower panel shows the ratio of the strengths of the reverse ($\Gamma_{34} - 1$) and forward ($\Gamma_{21}-1$) shocks. It can be seen that the ratio is the \textit{mirror} reflected about the $f=1$ line, reflecting the symmetry of equation~(\ref{fund_eqn}). For $f=1$, the reverse and the forward shock strengths are equal and given by
\begin{equation}
    \Gamma_{21} - 1 = \Gamma_{34} - 1 = \sqrt{\frac{1 + \Gamma_{41}}{2}} - 1 \quad\quad \text{(for $f = 1$)}\ . 
\end{equation}
Besides, it can be seen that for $f<1$ (e.g. as is the case in equal energy or mass collisions), the reverse shock is stronger than the forward shock strength. Additionally, it can be seen that for $u_{41}\ll1$, the shock strength ratio goes as $f^{-1}$ and is independent of $u_{41}$. This can be understood as follows: for $u_{41} \ll 1$, both shock fronts are \textit{Newtonian}. Thus, one can use the approximation $(\Gamma_{21},\Gamma_{34}) \sim 1$ in equation (\ref{fund_eqn}) to get,
\begin{equation}
    \frac{(\Gamma_{34} + 1) (\Gamma_{34} - 1) }{(\Gamma_{21} + 1) (\Gamma_{21} - 1)} = f^{-1} \Rightarrow \frac{\Gamma_{34} - 1}{\Gamma_{21}-1} \approx f^{-1}. 
\end{equation}

\subsubsection{Solution in the lab frame}\label{lab_soln}

In order to calculate the proper speed $u$ of the shocked fluid in the lab frame, we need one more parameter -- the proper speed $u_1$ of S1 in the lab frame. The proper speed $u$ of the shocked fluid in the lab frame  can be obtained by the Lorentz transformation of equation~(\ref{soln}) from the rest frame of shell 1 to the lab frame as 
\begin{equation}\label{eq:u-lab}
    u = \Gamma_{21} \Gamma_1 (\beta_1 + \beta_{21})\ . 
\end{equation}

\begin{figure}
\begin{tabular}{c}
      \includegraphics[scale=0.55]{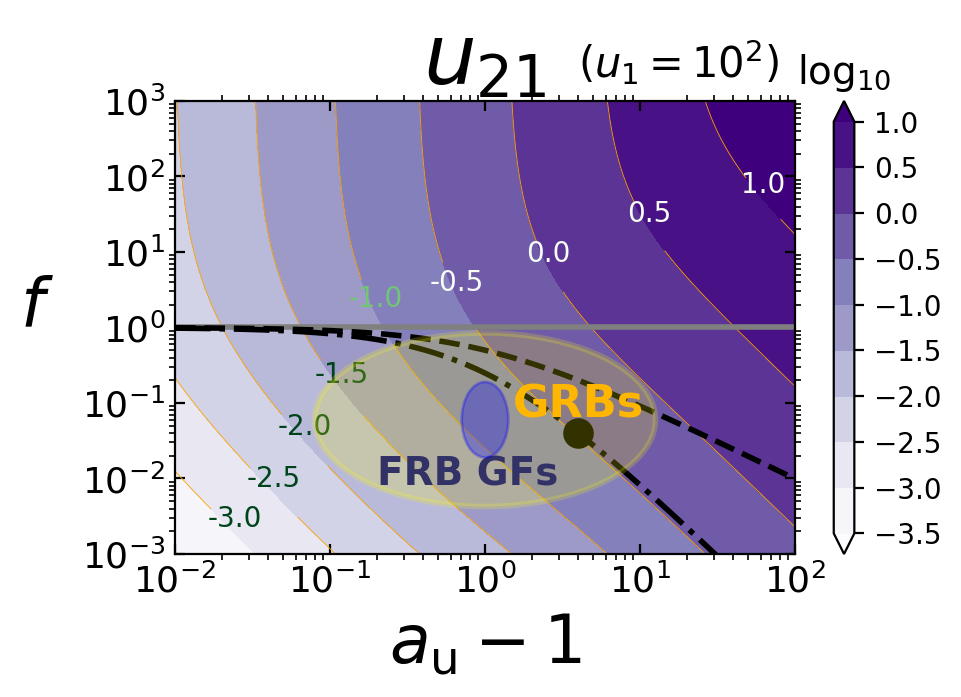} \\ \vspace{-0.15cm}
      \includegraphics[scale=0.55]{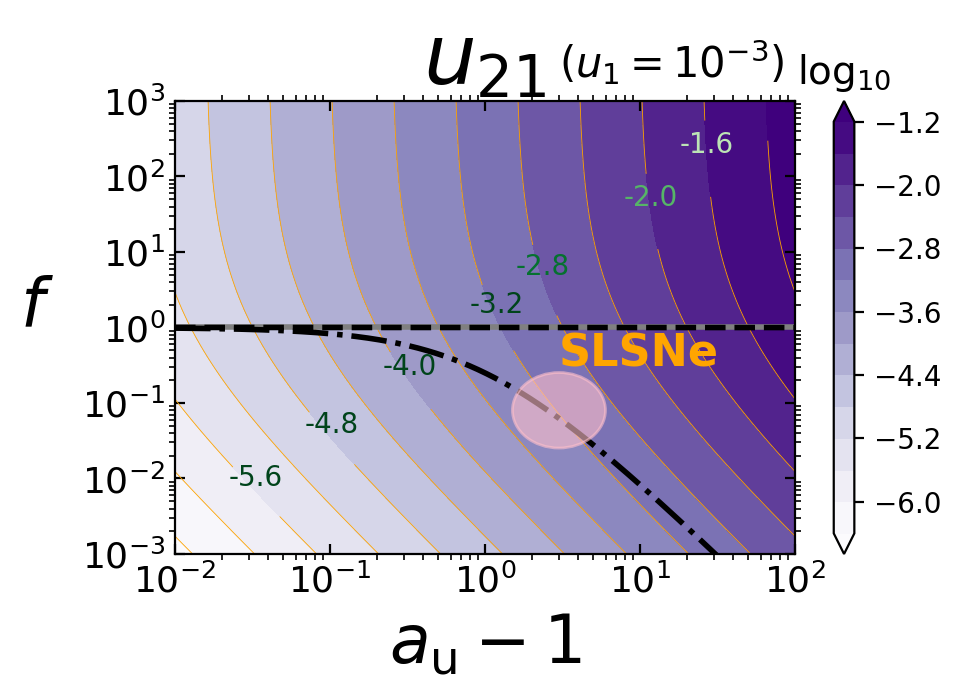} \\ \vspace{-0.15cm}
      \includegraphics[scale=0.55]{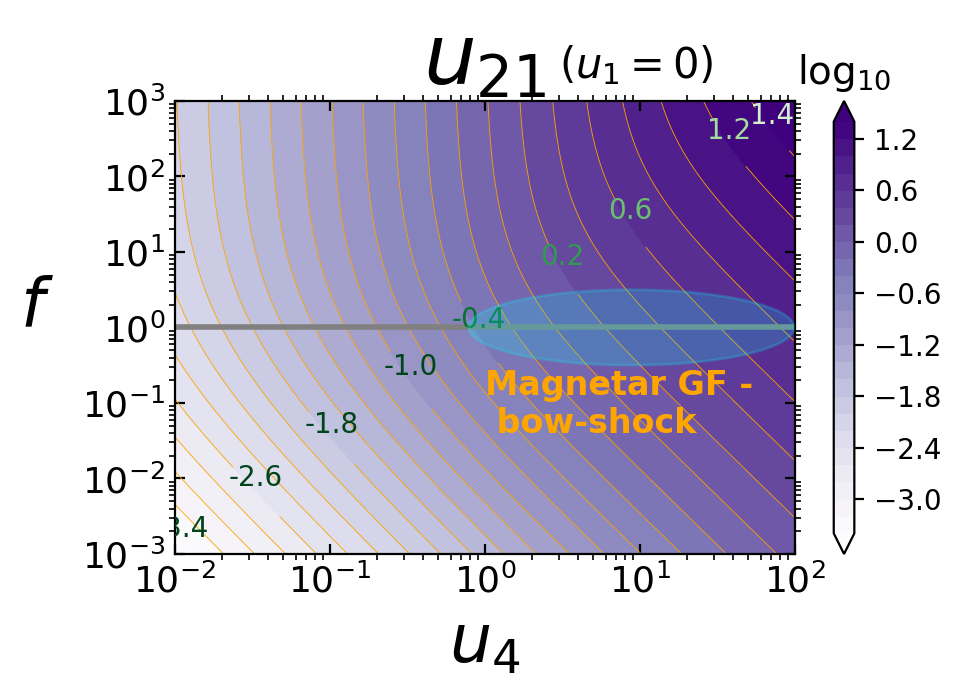}
\end{tabular}
    \caption{ Parameter space for astrophysical transients. In the top and middle panels the equal proper density ($f\!=\!1$), the equal mass ($M_{4,0} \!=\! M_{1,0}$), and equal kinetic energy ($E_\mathrm{k,4,0} \!=\! E_\mathrm{k,1,0}$) are represented by a grey horizontal line, dashed black line and black dot-dashed line respectively. \textbf{\textit{Top:}} shows the proper speed $u$ of the shocked fluid, for ultra-relativistic shells with a fixed $u_1\!=\!10^2$ and $\chi=1$, as a function of the proper speed contrast $a_\mathrm{u} -1$ and the proper density contrast $f$. The yellow and blue ellipses indicate the phase space for Gamma-ray bursts (GRBs) and blast wave models of Fast radio bursts (FRBs) (see subsections \ref{app_1} and \ref{app_2} ). \textbf{\textit{Middle:}} shows  the proper speed $u$ of the shocked fluid, for Newtonian shells with a fixed $u_1\!=\!10^{-3}$ and $\chi=1$, as a function $a_\mathrm{u} -1$ and $f$. The pink ellipse represents the phase space of superluminous supernovae (SLSNe) (see subsection \ref{app_3}).  \textbf{\textit{Bottom:}} shows a collision of shell S4 with shell S1 at rest in the lab frame as a function of the proper speed $u_{4}$, $a_\mathrm{u} -1$ and  $f$. The blue ellipse represents the phase space of the magnetar giant flare interaction with a bow-shock shell (see subsection \ref{app_4}).
    }
    \label{part_soln}
\end{figure}

Thus, while the general solution in the rest frame of region R1 depends only on $(u_{41},f)$, the lab frame solution (which we refer to as the particular solution) depends on $(u_{41},f,u_1)$. Figure \ref{part_soln} shows particular solutions for a few illustrative cases. The shaded region in each panel shows the relevant parameter space for a few models of astrophysical transients that feature
internal shocks. A detailed discussion of various internal shocks models for astrophysical transients is presented in \S\;\ref{appl_cases}. From this point onwards all our analysis will be  carried out in the lab frame.

The speed of the forward/reverse shock fronts are given by  (see Appendix C for the full  derivation)
\begin{subequations}
\begin{align}
&\  \beta_\mathrm{FS} = \frac{\left( \frac{\Gamma_1 n'_1}{\Gamma_2 n'_2} \right) \beta_1 - \beta_2 }{\left( \frac{\Gamma_1 n'_1}{\Gamma_2 n'_2} \right) - 1} = \frac{ \frac{1}{4 \Gamma_{21}} \left( \frac{u_1 }{\Gamma } \right) - \beta }{ \frac{1}{4 \Gamma_{21}} \left( \frac{\Gamma_1}{\Gamma} \right) - 1}\ , \\
&\  \beta_\mathrm{RS} = \frac{\beta_4 - \beta_3 \left(\frac{\Gamma_3 n'_3}{\Gamma_4 n'_4} \right) }{1 - \left(\frac{\Gamma_3 n'_3}{\Gamma_4 n'_4} \right) } = \frac{\beta_4 - 4 \Gamma_{34}  \left(\frac{u }{\Gamma_4 }  \right) }{1 - 4 \Gamma_{34} \left(\frac{\Gamma }{\Gamma_4 } \right) }\ .   
\end{align}
\end{subequations}
The time  it takes the FS to reach the front edge of shell S1 ($t_\mathrm{FS}$) and the RS to reach the rear edge of shell S4 ($t_\mathrm{RS}$) are given by (see Appendix C , also see \citealt{1995ApJ...455L.143S}) 
\begin{subequations}
\begin{align}
   &\ t_\mathrm{FS} = \frac{\Delta_{1,0}}{c(\beta_\mathrm{FS} - \beta_1)}  = \frac{\Delta_{1,0} }{c(\beta - \beta_1)} \left[ 1 - \left( \frac{\Gamma_1}{\Gamma} \right) \left( \frac{1}{4 \Gamma_{21}} \right) \right] , \label{t_FS} \\
    &\ t_\mathrm{RS} = \frac{\Delta_{4,0}}{c(\beta_4 - \beta_\mathrm{RS})} =  \frac{\Delta_{4,0} }{c(\beta_4 - \beta)} \left[ 1 - \left( \frac{\Gamma_4}{\Gamma} \right) \left( \frac{1}{4 \Gamma_{34}} \right) \right], \label{t_RS}
\end{align}
\end{subequations}

The internal energy that is produced at the FS (RS), as it disipates the kinetic energy of the relative bulk motion of regions R1 and R2 (R4 and R3), resides in the shocked region R2 (R3) and over the shock crossing time $t_\mathrm{FS}$ ($t_\mathrm{RS}$) accumulates to
(see Appendix D for the full derivation)
\begin{subequations}
\begin{align}
 &\ E_\mathrm{2,int}  = \Gamma M_{1,0} \; c^2  \left[ 1 + \beta^2 \left( \frac{\Gamma_{21} + 1}{3 \Gamma_{21}} \right) \right] (\Gamma_{21} - 1) , \label{E2int}\\
&\ E_\mathrm{3,int} =  \Gamma M_{4,0} \; c^2 \left[ 1 + \beta^2 \left(\frac{\Gamma_{34} + 1}{3 \Gamma_{34}} \right) \right] (\Gamma_{34} - 1) ,\label{E3int} 
\end{align}
\end{subequations}

The maximum bulk kinetic energy in region R2 (R3) at the shock crossing time $t_\mathrm{FS}$ ($t_\mathrm{RS}$) is given by (see Appendix D for a full derivation) 
\begin{subequations}
\begin{align}
 &\ E_\mathrm{2,k} = (\Gamma - 1) M_{1,0} \; c^2, \label{E2k} \\
 &\ E_\mathrm{3,k} = (\Gamma - 1)  M_{4,0}  \; c^2 , \label{E3k}
\end{align}
\end{subequations}

The final radial width of region R2 (R3) at the shock crossing time $t_\mathrm{FS}$ ($t_\mathrm{RS}$) is given by (see Appendix E for full derivation)
\begin{subequations}
\begin{align}
&\ \frac{\Delta_\mathrm{2f}}{\Delta_{1,0}} = \frac{1}{4 \Gamma_{21}} \; \left( \frac{\Gamma_1}{\Gamma} \right) \;, \label{Delta2f}\\ 
&\  \frac{\Delta_\mathrm{3f}}{\Delta_{4,0}} = \frac{1}{4 \Gamma_{34}}  \left( \frac{\Gamma_4}{\Gamma}\right)  \label{Delta3f},
\end{align}
\end{subequations}

The FS and the RS  produce internal energy in regions R2 and R3, respectively, resulting in non-zero pressures across the CD. As a result, region R3 performs a positive $pdV$ work on region R2 across the CD. From equation (4b) (and from energy conservation)
an equal amount of negative $pdV$ amount of work is done by region R2 on Region R3. This $pdV$ work leads to a transfer of energy from S4 to S1. In this setup, as viewed in the lab frame, the CD essentially acts as a piston which allows the pdV work done across it. The $pdV$ work done by region R3 on region R2 by the RS shell crossing time $t_\mathrm{RS}$ is given by (see Appendix F for the full derivation)
\begin{equation}
    \frac{W_\mathrm{pdV,RS}}{E_\mathrm{k,4,0}}  = \frac{4}{3} \frac{(\Gamma^2_{34} - 1)}{\Gamma_4 (\Gamma_4 - 1)} \frac{\beta}{(\beta_4 - \beta)} \left[ 1 - \frac{1}{4 \Gamma_{34}} \left( \frac{\Gamma_4}{\Gamma}\right) \right]\ .  \label{WpdV} 
\end{equation}
The details of how the $pdV$ work is re-distributed into the kinetic and the internal energy in region R2 are explored below.

Table~\ref{time_evol} shows the time evolution of different quantities (in the lab frame). 
To illustrate the basic ideas we consider the collision of two shells of equal energy and radial width, moving with proper speeds  $(u_1,u_4)$ in the lab frame.  
While there is a transfer of energy from shell S4 to S1, there is no mass transfer between them as no mass flows across the CD (equation~(\ref{bc1})).

To summarize, the collision produces two shock fronts (FS and RS), where the corresponding shocked parts of the shells (regions R2 and R3) are separated by a CD.
The unshocked parts of leading and trailing shells are labeled 1 and 4, respectively. The shock fronts dissipate the available kinetic energy into internal energy and heat up the gas. For cold shells, the pressure (and internal energy) in regions R1 and R4 is zero, while the pressures in shocked regions R2 and R3 are non-zero. As is shown later, the non-zero equal pressure across the CD has very important consequences. We find that the $p dV$ work done across the CD acts as an important mechanism of energy transfer from region R3 to R2. Note that all quantities involved vary \textit{linearly} with time. This is a consequence of assuming a planar geometry. In \S\;\ref{appl_cases} we will discuss the limitation of our approach.

\begin{table*}
    \centering
     \caption{Time evolution in lab frame of the various physical quantities of regions $j = (1,2,3,4 )$. Here the quantities $(E_\mathrm{2,int}, E_\mathrm{3,int})$ are defined in equations (\ref{E2int}) - (\ref{E3int}), $(E_\mathrm{k,2},E_\mathrm{k,3})$ are defined in equations (\ref{E2k})-(\ref{E3k}) and the quantities $(\Delta_\mathrm{2f}, \Delta_\mathrm{3f})$ are defined in equations (\ref{Delta2f})-(\ref{Delta3f}). The quantity $W_\mathrm{pdV,RS}$ is defined in equation (\ref{WpdV}). In all these expressions we put the datum of zero at the time of collision $t_\mathrm{o} = 0$. }
    \begin{tabular}{c|c|c|c|c|c} \hline 
       Region  & $M_\mathrm{j} c^2$ &$E_\mathrm{int,j} (t)$  & $E_\mathrm{k,j} (t)$ & $\Delta_\mathrm{j} (t)$ & $p dV$ \\ \hline
           R1   & $ M_{1,0} c^2 \left[ 1 - \frac{t}{t_\mathrm{FS}} \right] $ &  0  &  $E_\mathrm{k,1,0} \left[ 1 - \frac{t}{t_\mathrm{FS}} \right]$   &    $\Delta_{1,0} \left[ 1 - \frac{t}{t_\mathrm{FS}}\right]$  & 0 \\ \\ 
           R2  & $  M_{1,0} c^2 \left(\frac{t}{t_\mathrm{FS}} \right) $  &  $ E_\mathrm{int,2} \left(\frac{t}{t_\mathrm{FS}} \right) $   &   $E_\mathrm{k,2} \left( \frac{t}{t_\mathrm{FS}} \right) $   &  $\left(\frac{t}{t_\mathrm{FS}} \right) \Delta_\mathrm{2f}$   & $-\!\,W_\mathrm{pdV,RS} \left( \frac{t}{t_\mathrm{RS}} \right) $ \\  \\ 
           R3   & $M_{4,0} c^2 \left(\frac{t}{t_\mathrm{RS}} \right) $ & $ E_\mathrm{int,3} \left(\frac{t}{t_\mathrm{RS}} \right) $   &  $E_\mathrm{k,3} \left( \frac{t}{t_\mathrm{RS}} \right) $   &  $\left(\frac{t}{t_\mathrm{RS}} \right) \Delta_\mathrm{3f}$  & $+\!\,W_\mathrm{pdV,RS} \left( \frac{t}{t_\mathrm{RS}} \right) $   \\ \\ 
           R4 & $M_{4,0} c^2 \left[ 1 - \frac{t}{t_\mathrm{RS}} \right]$ &  0   &   $E_\mathrm{k,4,0} \left[ 1 - \frac{t}{t_\mathrm{RS}} \right]$   &   $\Delta_{4,0} \left[ 1 - \frac{t}{t_\mathrm{RS}}\right]$  & 0  \\ \\ \hline 
    \end{tabular}
    \label{time_evol}
\end{table*}

\subsection{Shell S$1$ is at rest in the lab frame}

When region R1 is at rest with respect to the central engine frame, the lab frame and the rest frame of region 1 are coincident and the proper speed of the shocked fluid is given by 
\begin{equation}
    u =      u_{4} \sqrt{\frac{  2 f^{3/2} \Gamma_{4} - f(1+f)  }{ 2 f( u^2_{4} + \Gamma^2_{4} ) - (1 + f^2)}}  \quad\quad \text{(for $u_1 = 0$)\ .}    \label{ext_shock}
\end{equation}
Equation~(\ref{ext_shock}) corresponds to the solution presented in \citet{1995ApJ...455L.143S} for an external shock scenario for semi-infinite shell S1 $(\chi \rightarrow \infty)$ and for $(u_{4},f) \gg 1$.

This scenario is an illustrative example of the possibility that the FS can dissipate internal energy higher than the initially available kinetic energy in shell S1. Here the leading shell is at rest. Thus, the initial available kinetic energy in shell S1 is zero, $E_\mathrm{k,1,0} = 0$, and the entirety of the energy dissipated by the forward shock front in region R2 comes from the initially available kinetic energy in shell 4. This raises the important question what leads to this energy transfer from the trailing shell to the leading shell? The only possible source of energy transfer is the $pdV$ work done by region R3 on R2 across the CD. The $pdV$ work done goes towards increasing both the kinetic energy and the internal energy of region R2. Thus, the forward shock dissipates more energy than the initial available kinetic energy in the leading shell S1 and the internal energy dissipation occurs at the expense of energy transfer from S4 to S1 via $pdV$ work across the CD (in particular from region R3 to R2). 
 
\subsection{Both shells are moving in the lab frame}

In this case, the proper speed of the shocked fluid $u$ is a function of three parameters $(u_4,u_{1},f)$, which is given by substituting equation~(\ref{soln}) into equation~(\ref{eq:u-lab}). Here, we make use of the proper speed contrast $a_\mathrm{u}$ where $u_4 = a_\mathrm{u} u_1$, such that the proper speed of the shocked fluid is a function of the three parameters $(a_\mathrm{u},u_{1},f)$. In the next two subsections we present some key results for collision of shells moving at ultra-relativistic and Newtonian speeds, respectively.  

\subsubsection{Both shells move with ultra-relativistic speeds}\label{ultra_shells}

For collision between ultra-relativistic shells ($u_4>u_1\gg1$), the proper velocity of the shocked fluid is given by
\begin{equation}
    u \approx \Gamma \,\,\approx\,\,
    \sqrt{\frac{\sqrt{f} a^2_\mathrm{u} + a_\mathrm{u}}{a_\mathrm{u} + \sqrt{f}}} \Gamma_1\ , 
\end{equation}
such that the shock strengths are given by
\begin{subequations}
\begin{align}
 &\ \Gamma_{21} \;\approx\; \frac{1}{2} \frac{2 a_\mathrm{u} + \sqrt{f} (1 + a^2_\mathrm{u}) }{\sqrt{(a_\mathrm{u} + \sqrt{f}) (\sqrt{f} a^2_\mathrm{u} + a_\mathrm{u})} }\ , 
 \\ 
&\ \Gamma_{34} \;\approx\; \frac{1}{2} \frac{a^2_\mathrm{u} + 2 \sqrt{f} a_\mathrm{u} + 1 }{\sqrt{(a_\mathrm{u} + \sqrt{f}) (\sqrt{f} a^2_\mathrm{u} + a_\mathrm{u})} }\ ,
\end{align}
\end{subequations}
while $\Gamma_{41}\approx \frac{1}{2}(a_\mathrm{u}+a_\mathrm{u}^{-1})$ and $u_{41}\approx \frac{1}{2}(a_\mathrm{u}-a_\mathrm{u}^{-1})$.

Let us consider the expression for the ratio of the initial kinetic energies of the two colliding shells
\begin{equation}\label{ratio_kin}
    \frac{E_\mathrm{k,4,0}}{E_\mathrm{k,1,0}} = \frac{f}{\chi} \; \frac{\Gamma_{4} (\Gamma_4 - 1)}{\Gamma_1 (\Gamma_1 - 1)} \approx \frac{a^2_\mathrm{u} f}{\chi}\ . \hspace{1cm} 
\end{equation}

Next we summarize certain key results at high proper speed contrast $a_\mathrm{u} \gg 1$. The proper density contrast $f$ for a collision between two equal energy or equal mass ultra-relativistic shells in the high proper speed contrast limit ($a_u\gg1$) given by
\begin{equation}
f  =
     \begin{cases} &\ \chi \frac{\Gamma_1 (\Gamma_1 -1
    )}{\Gamma_4 (\Gamma_4 - 1)} \approx \frac{\chi}{a^2_\mathrm{u}},   \hspace{1cm} \text{For $E_\mathrm{k,4,0} = E_\mathrm{k,1,0}$} \\
    &\  \chi \frac{\Gamma_1}{\Gamma_4} \approx \frac{\chi}{a_\mathrm{u}},  \hspace{1.8cm} \text{For $M_{4,0} = M_{1,0}$} \\
    &\  1    \hspace{3.1cm}  \text{For $n'_{4} = n'_{1}$}  
\end{cases}\label{f_ultra}
\end{equation}

The proper speed of the shocked fluid is ($a_\mathrm{u} \gg1$)
\begin{equation}
    u \approx  
    \begin{cases}
        &\ \sqrt{2} u_{1} \hspace{2.5cm} \text{For $E_\mathrm{k,4,0} = E_\mathrm{k,1,0}$} \\ 
        &\ a^{1/4}_\mathrm{u} u_{1} \hspace{2.3cm} \text{For $M_{4,0} = M_{1,0}$} \\ 
        &\ a^{1/2}_\mathrm{u} u_{1}  \hspace{2.3cm}  \text{For $n'_{4} = n'_{1}$}  \\ 
    \end{cases}
\end{equation}

 The FS shock strength is given by ($a_\mathrm{u} \gg1$)
\begin{equation}
   \Gamma_{21} - 1 \approx  
      \begin{cases}
          &\ \frac{3}{2 \sqrt 2} - 1 \approx 0.0607 \hspace{0.2cm} \text{For $E_\mathrm{k,4,0} = E_\mathrm{k,1,0}$} \\
          &\ \frac{a^{1/4}_\mathrm{u}}{2} - 1 \hspace{1.5cm} \text{For $M_{4,0} = M_{1,0}$}\\
          &\ \frac{a^{1/2}_\mathrm{u}}{2} -1   \hspace{1.5cm}  \text{For $n'_{4} = n'_{1}$}  
      \end{cases}
\end{equation}

The RS shock strength is given by ($a_\mathrm{u} \gg1$)
\begin{equation}
   \Gamma_{34} - 1 \approx 
      \begin{cases}
          &\ \frac{a_\mathrm{u}}{2 \sqrt{2}}-1 \gg 1 \hspace{0.9cm} \text{For $E_\mathrm{k,4,0} = E_\mathrm{k,1,0}$} \\
          &\  \frac{a^{3/4}_\mathrm{u}}{2} - 1 \hspace{1.5cm} \text{For $M_{4,0} = M_{1,0}$} \\
          &\ \frac{a^{1/2}_\mathrm{u}}{2} -1  \hspace{1.5cm}  \text{For $n'_{4} = n'_{1}$}  
      \end{cases}
\end{equation}

The FS crossing timescale is given by ($a_\mathrm{u} \gg1$)
\begin{equation}
   \frac{t_\mathrm{FS}}{\Delta_{1,0}/c} \approx
   \begin{cases}
       &\ \frac{5}{3} \Gamma^2 \hspace{1.9cm} \text{For $E_\mathrm{k,4,0} = E_\mathrm{k,1,0}$}  \\  
       &\ 2 \Gamma^2_1 \hspace{2cm} \text{For $M_{4,0} = M_{1,0}$}  \\
       &\ 2 \Gamma^2_1 \hspace{2cm}  \text{For $n'_{4} = n'_{1}$}  
   \end{cases}
\end{equation}

The RS crossing timescale is given by ($a_\mathrm{u}\gg1$)
\begin{equation}
    \frac{t_\mathrm{RS}}{\Delta_{4,0}/c} \approx
    \begin{cases}
        &\ \Gamma^2 \hspace{2.3cm} \text{For $E_\mathrm{k,4,0} = E_\mathrm{k,1,0}$}  \\
        &\ a^{1/2}_\mathrm{u} \Gamma^2_1 \hspace{1.8cm} \text{For $M_{4,0} = M_{1,0}$}  \\ 
        &\ a_\mathrm{u} \Gamma^2_1  \hspace{2.0cm}\text{For $n'_{4} = n'_{1}$}  
    \end{cases} \label{cross_RS}
\end{equation}

The final radial width of region R2 post FS passage is given by ($a_\mathrm{u} \gg1$)
\begin{equation}
    \frac{\Delta_\mathrm{2f}}{\Delta_{1,0}} \approx
    \begin{cases}
        &\ \frac{1}{6}  \hspace{2.3cm} \text{For $E_\mathrm{k,4,0} = E_\mathrm{k,1,0}$}      \\ 
        &\  \frac{1}{2 a^{1/2}_\mathrm{u}}\hspace{1.8cm} \text{For $M_{4,0} = M_{1,0}$}  \\ 
        &\ \frac{1}{2 a_\mathrm{u}} \hspace{2.0cm}\text{For $n'_{4} = n'_{1}$}  
    \end{cases} \label{cross_FS}
\end{equation}

The final radial width of region R2 post RS passage is fixed for relativistic reverse shock  ($a_\mathrm{u} \gg1$)
\begin{equation}
    \frac{\Delta_\mathrm{3f}}{\Delta_{4,0}} \approx \frac{1}{2}
\end{equation}

\begin{figure*}
    \begin{tabular}{c|c}
      \includegraphics[scale=0.55]
      {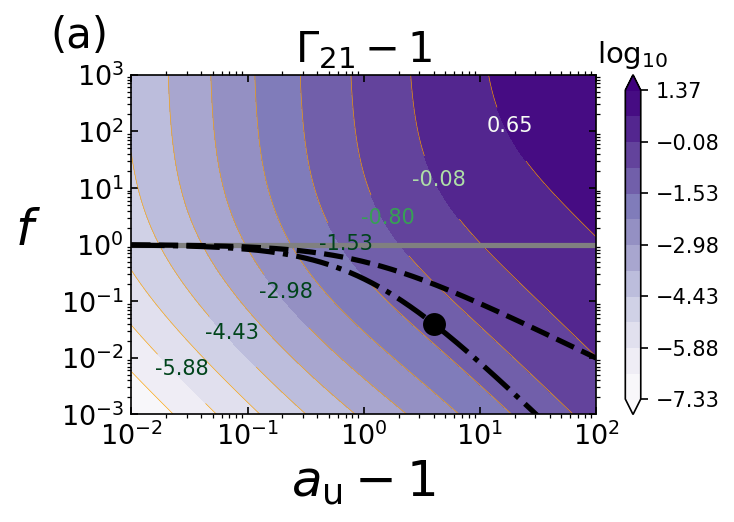}   &  \includegraphics[scale=0.55]{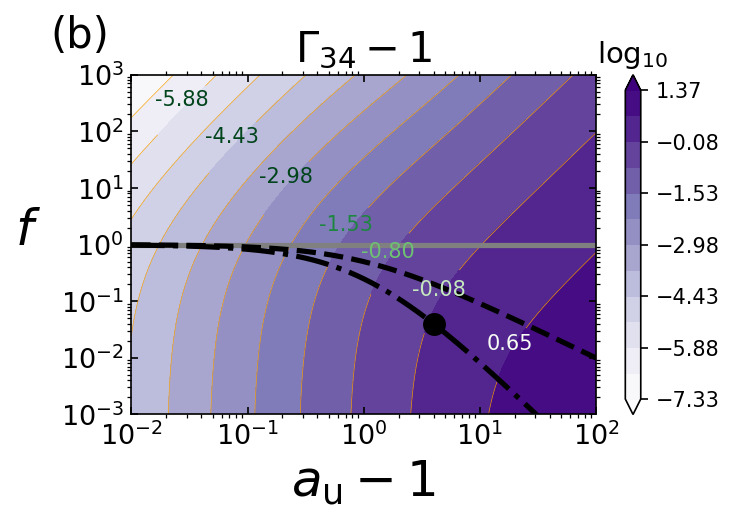}\\
      \includegraphics[scale=0.55]{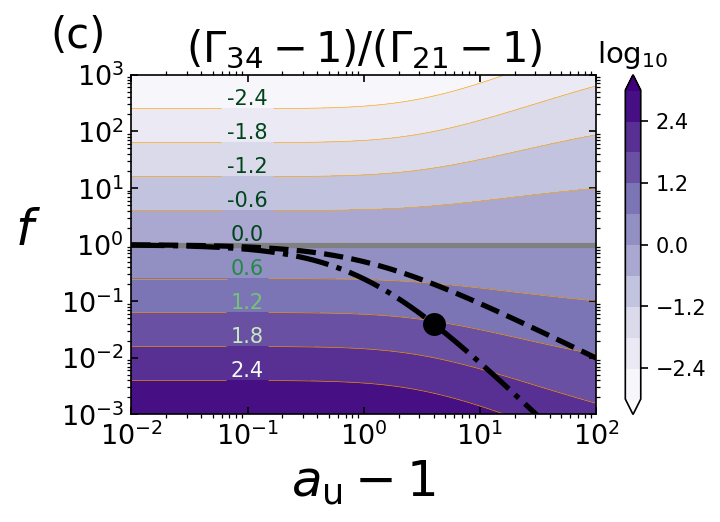}   &  \includegraphics[scale=0.55]{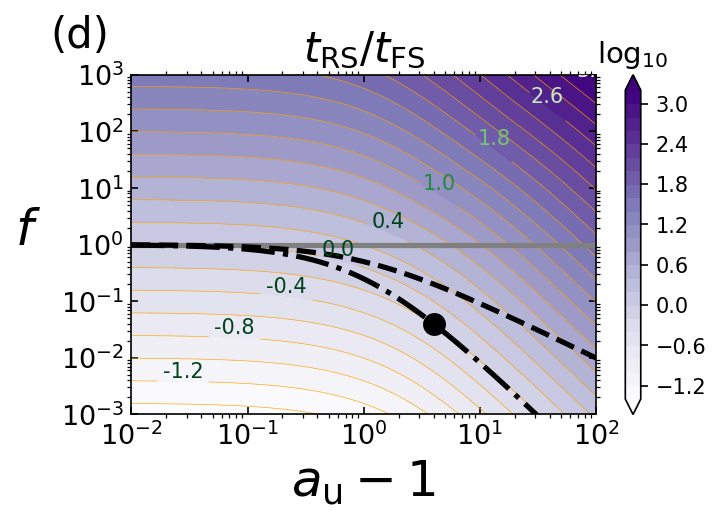}\\ 
      \includegraphics[scale=0.55]{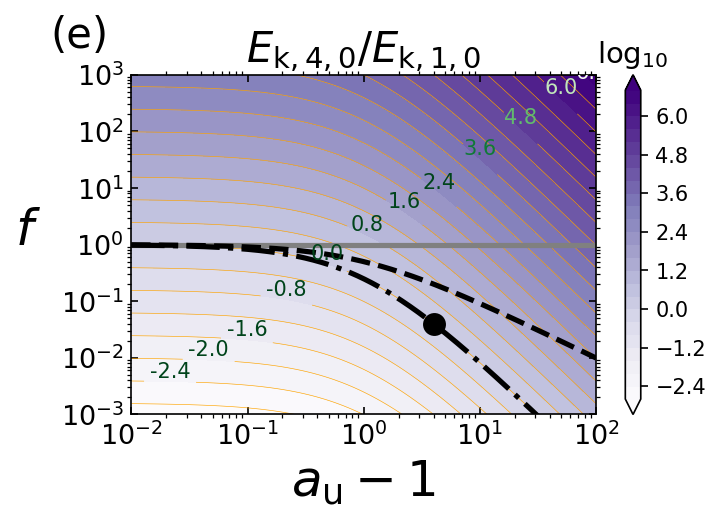}   &  \includegraphics[scale=0.55]{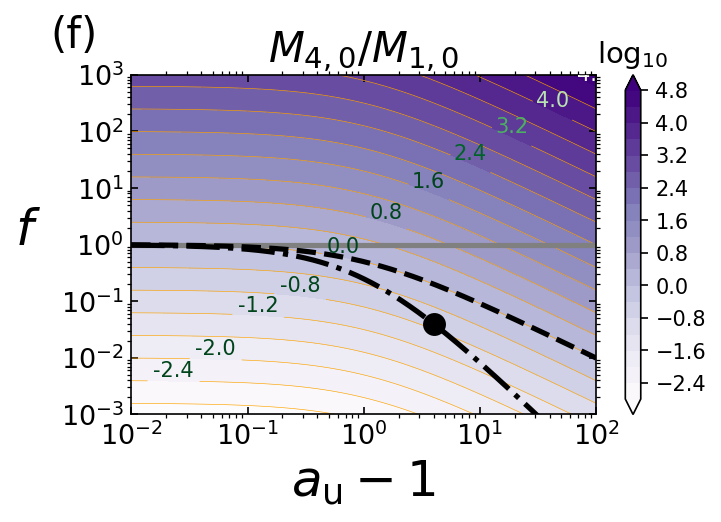}\\
    \end{tabular}
    \caption{Hydrodynamic parameter space  for the collision of two cold \textbf{ultra-relativistic shells} of equal initial radial width ($\chi = 1$) at a fixed proper speed $u_1 = 10^2$ for shell S1. In all panels the equal proper density ($f=1$), the equal mass ($M_{4,0} = M_{1,0}$), and equal kinetic energy ($E_\mathrm{k,4,0} = E_\mathrm{k,1,0}$) are represented by a grey horizontal line, dashed black line and  dot-dashed black line, respectively. In all panels, the black-filled circle on the dot-dashed line represents the collision of two equal kinetic energy shells with proper speeds $(u_1,u_4) = (100,500)$, which is used in all illustrations in Fig.~\ref{EqEn_time}.  \textbf{\textit{Top}}: Panels (a) and (b) show  the forward shock strength $\Gamma_{21} - 1$  and the reverse shock strength $\Gamma_{34} - 1$  as a function of the proper speed contrast $a_\mathrm{u}$ and the proper density contrast $f$. \textbf{\textit{Middle}}: Panels (c) and (d) show the ratio of the shock strength $(\Gamma_{34} - 1)/(\Gamma_{21} -1)$ and ratio of the shock crossing timescale $t_\mathrm{RS}/t_\mathrm{FS}$ and  as a function of $a_\mathrm{u}$ and $f$. \textbf{\textit{Bottom}}: Panels (e) and (f) show  the ratio of the initial kinetic energy $E_\mathrm{k,4,0}/E_\mathrm{k,1,0}$  and the ratio of the masses $M_{4,0}/M_{1,0}$ as a function of $a_\mathrm{u}$ and $f$. For a detailed explanation see the text. }\label{ultra_param}
\end{figure*}

Fig. ~\ref{ultra_param} shows the hydrodynamical shock parameter space for  the collision of two ultra-relativistic shells of equal initial radial widths ($\chi = 1$). In all panels equal energy, equal mass, and equal proper density shells are shown by the black dot-dashed line, black dashed line, and a grey line, respectively. In the low proper speed contrast limit ($a_\mathrm{u} - 1\ll 1$), the $f=1$ collision is the asymptotic limit for the equal energy, and equal mass shell collision. This is due to the fact that at low proper speed contrast the ratio of the Lorentz factor of both shells tends to unity. This can be seen directly from equation~(\ref{ratio_kin}).  In fact, the scaling in equation~(\ref{f_ultra}) is a reasonable approximation even for $a_\mathrm{u}-1\ll 1$. 

Next, let us consider the trend as we move from the equal energy collision towards $f=1$ at the high proper speed contrast $a_\mathrm{u} \gg 1$ limit. Equation~(\ref{f_ultra}) for $\chi=1$ shows the proper density contrast  
$f \approx a^{-2}_\mathrm{u} \ll 1$ 
for equal energy and $f \approx a^{-1}_\mathrm{u} \ll 1$ for equal mass. The consequence is reflected in panel (a) of Fig. ~\ref{ultra_param}. It shows that the FS strength for the equal energy collision approaches a constant, almost Newtonian value of $\Gamma_{12}-1=2^{-3/2}3-1\approx0.0607$ for $a_\mathrm{u} \gg1$, while for the equal mass case it gradually increases with $a_u$ (asymptotically as $\Gamma_{12}-1\approx\frac{1}{2}a_u^{1/4}$ for extremely high $a_\mathrm{u}$ values), and is typically mildly relativistic. Panel (b) shows that the RS for both is typically relativistic, but the strength of the RS is stronger for equal energy collisions than equal mass collisions. Asymptotically, for $a_u\gg1$, 
we have $\Gamma_{34}-1\approx2^{-3/2} a_\mathrm{u}$ for the equal energy case and $\Gamma_{34}-1\approx\frac{1}{2} a_\mathrm{u}^{3/4}$ for the equal mass case. We note that panels (a) and (b) are exact mirror images of each other, symmetric to reflection about the $f=1$ line ($f\!\to\!1/f$). This arises for the following reason. Since $u_1$ is fixed, the value of $a_u=u_4/u_1$ determines that of $u_{41}=\Gamma_4\Gamma_1(\beta_4-\beta_1)$, i.e. the relative proper speed between the two shells. Now, the strength of the two shocks depend only on $u_{41}$ and on the proper density ratio of the two shells, $f=n'_4/n'_1$. This problem is symmetric to relabeling of the shells ($1\leftrightarrow4$, RS\,$\leftrightarrow$\,FS and $f\leftrightarrow1/f$), such that for the same value of $a_u$ (and therefore $u_{41}$) $\Gamma_{34}-1$ for a given proper density contrast $f$ must equal $\Gamma_{21}-1$ for a proper density contrast $1/f$, and that is the origin of this mirror symmetry.

This induces mirror anti-symmetry in Panel (c), where the shock strength ratio, $\frac{\Gamma_{34}-1}{\Gamma_{21}-1}$, switches to its inverse value (i.e. its log switches sign) upon reflection about the $f=1$ line ($f\!\to\!1/f$).
Panel (c) also shows that this shock strength ratio is higher for equal energy collision ($\approx a_\mathrm{u}/(3-2^{3/2})$ for $a_\mathrm{u}\gg1$) compared to equal mass collision ($\approx a_\mathrm{u}^{1/2}$ for $a_\mathrm{u}\gg1$). In \S \ref{ultra_high} we present  a detailed breakdown of the shock hydrodynamics associated with the three scenarios.

Panel (d) shows that for equal energy collisions the RS front reaches the rear edge of shell S4 somewhat before the FS front can reach the front edge of shell S1 ($t_{\rm RS}<t_{\rm FS}$). However, this trend is reversed for equal mass collision, while
for $f=1$ and $a_u\gg1$ we have $t_{\rm RS}\gg t_{\rm FS}$. The ratio of the crossing times varies
by orders of magnitude, particularly between the top right corner, $(f,a_\mathrm{u})\gg 1$, and the bottom left corner, $(f,a_\mathrm{u} -1)\ll 1$. The consequence of different shock crossing times for the two shells
will be explored in \S\;\ref{dissp_limit}. Lastly, panels (e) and (f) show that as we move towards the $f=1$ line from the equal energy collision,  both the initial kinetic energy and the mass is dominated by  the trailing shell S4.

\begin{figure}
\begin{tabular}{l}
 \includegraphics[scale=0.50]{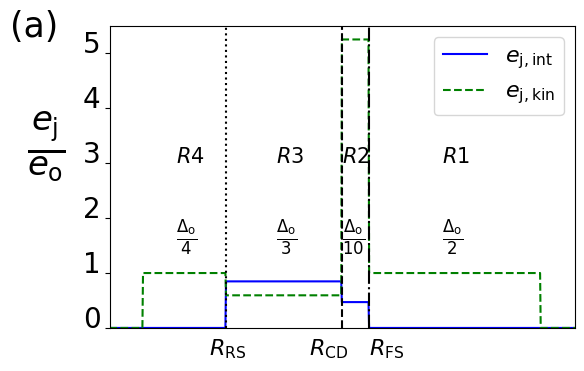}\\
\includegraphics[scale=0.50]{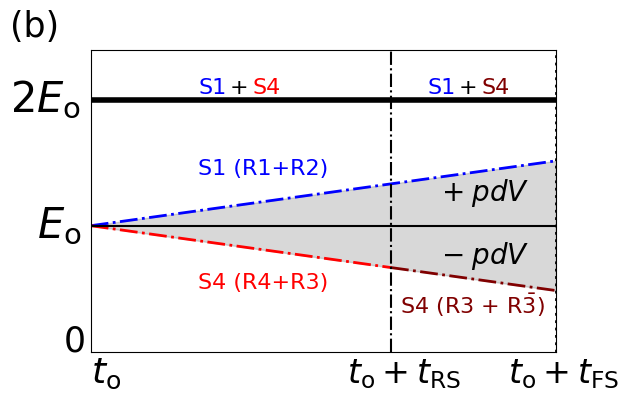}  \\
\includegraphics[scale=0.50]{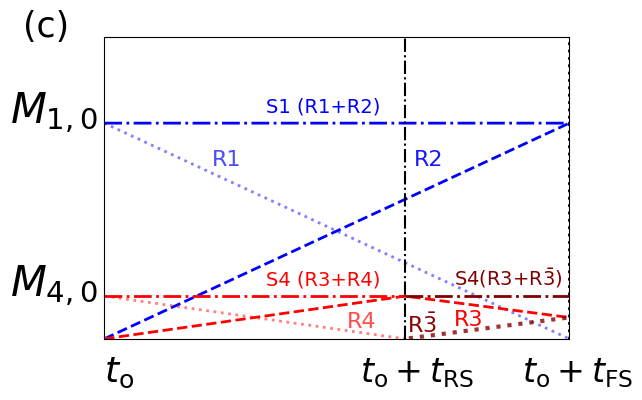} \\
\includegraphics[scale=0.50]{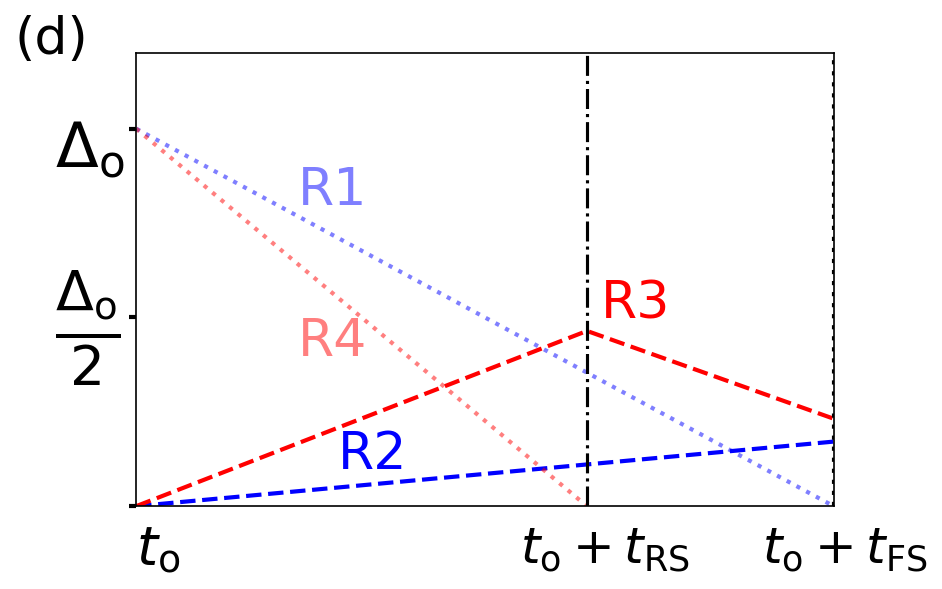}  \\
\end{tabular}
\caption{
The  distribution of the (internal\,+\,kinetic) energy, rest mass and radial width, as measured in the lab frame, for different regions post-collision of two cold equal kinetic energy shells with equal initial radial width with proper speeds $(u_1,u_4) = (100,500)$. \mathbfit{(a)} a snapshot of the lab frame energy density at time $t = t_\mathrm{o} + \frac{3}{4} t_\mathrm{RS}$. \mathbfit{(b)} temporal evolution
of the total energy in different regions. 
\mathbfit{(c)} temporal evolution
of the rest mass in different regions. 
\mathbfit{(d)} temporal evolution of 
the radial width of regions R1, R2, R3, R4.} \label{EqEn_time}
\end{figure}

Figure~\ref{EqEn_time} shows the breakdown of the physical quantities as a function of time elapsed post-collision for the collision of two equal energy shells with proper speeds $(u_1,u_4) = (100,500)$, which is shown by the black-filled circle on the black dot-dashed line in Fig.~\ref{ultra_param}. Panel (a) of Fig.~\ref{EqEn_time} shows that the lab frame internal energy density in the reverse shocked region is higher than that in the forward shock region,
while the kinetic energy density in the forward shocked region is much higher than the kinetic energy density in the reverse shocked region, both of these arise since the RS is significantly stronger than the FS, and the two shocked regions have the same velocity and pressure. Panel (b)  shows that the total energy (kinetic and internal) of the two shells is conserved at all times and is equal to its initial pre-collision value.
However, while the total energy of both shells remains constant, their individual energies change with time -- the energy in the trailing shell S4 decreases while the energy in the leading shell S1 increases. 
This illustrates the energy transfer via $pdV$ work across the CD from region R3 of shell S4 to region R2 of shell S1. Panel (c) of Fig.~\ref{EqEn_time} shows the rest mass in each individual shell remains constant, as there is no bulk flow of particles across the CD (e.g. equation~(\ref{bc1})). Lastly, panel (d) shows that although the FS is weaker than the RS, the lab frame compression ratio is larger for the forward shocked region R2 than the reverse shocked region R3.  All physical quantities change linearly with time (also see Table ~\ref{time_evol}), which is a consequence of assuming a planar geometry.

Lastly, we 
summarize the following important results for  \textit{relativistic RS} ($\Gamma_{34} \gg 1$) for one complete sweep of shell S4, i.e. at $t_\mathrm{o}+t_{\rm RS}$ when the RS reaches the rear edge of shell S4 (see Appendix G): 
(i)  at $t_\mathrm{o}+t_{\rm RS}$ the lab frame radial width of region R3 is half of that of the initial radial width of shell S4 ($\Delta_\mathrm{3f} \approx \frac{1}{2}\Delta_{4,0}$). (ii) as relativistic RS implies $u \ll u_4$,  at $t_\mathrm{o}+t_{\rm RS}$
the bulk energy of region R3 becomes $E_\mathrm{k,4,0}(\Gamma-1)/(\Gamma_4-1)\approx E_\mathrm{k,4,0}\,u/u_4\ll E_\mathrm{k,4,0}$ or $\sim E_\mathrm{k,4,0}/a_\mathrm{u}$ for $a_\mathrm{u} \gg 1$, i.e. it becomes negligible. (iii) 
 at $t_\mathrm{o}+t_{\rm RS}$ the maximum energy that is dissipated at the RS is $\frac{2}{3}E_\mathrm{k,4,0}$, independent of the FS strength. The deficit energy of $\frac{1}{3} E_\mathrm{k,4,0}$ is channeled by the $p dV$ work done by the CD to the combination of (kinetic+internal) energies of the region R2.  If the FS is \textit{relativistic}, the $p dV$ work is mostly channeled into internal energy increase and if it is \textit{Newtonian} the $p dV$ work done is mostly channeled into increasing the bulk kinetic energy. (iv) for $a_\mathrm{u} \gg 1$, we have $\frac{E_\mathrm{k,4,0}}{E_\mathrm{k,1,0}} \approx a^2_\mathrm{u} f $. Thus, for $f > a_\mathrm{u}^{-2}$, the combined available initial kinetic energy of both shells is dominated by the kinetic energy of shell 4. In particular, for $f=1$, almost all the available kinetic energy is in shell 4. 

To summarize, for a collision of equal energy and equal mass ultra-relativistic shells, the reverse shock is relativistic. However, for equal initial radial width of both shells, if the shells have equal energy the reverse shock finishes crossing the trailing shell S4 before the forward shock can finish crossing the leading shell S1, while the trend is reversed for a collision of equal mass shells.

In the next subsection, we consider the collision of two Newtonian shells and then compare it to the results obtained in this subsection.

\subsubsection{Both shells are moving with Newtonian velocities}\label{newt_shells}

\begin{figure*}
    \begin{tabular}{c|c}
      \includegraphics[scale=0.55]
      {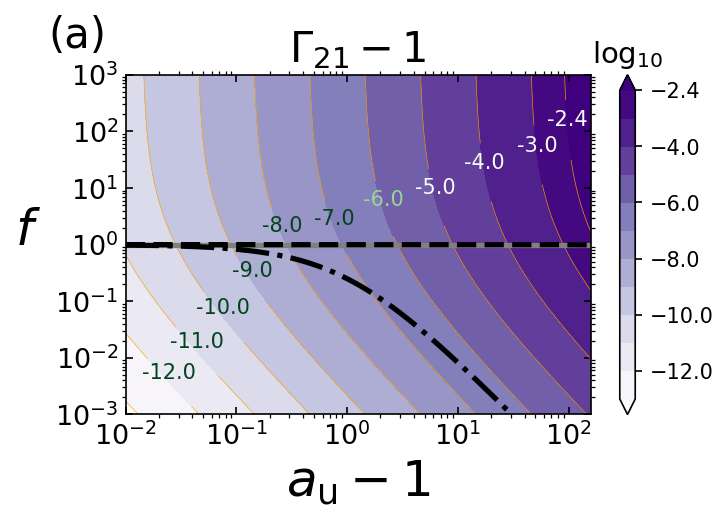}   &  \includegraphics[scale=0.55]{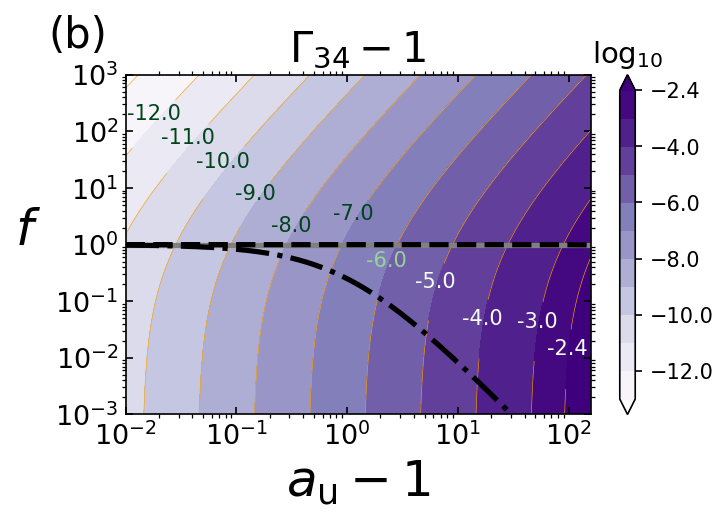}\\
      \includegraphics[scale=0.55]{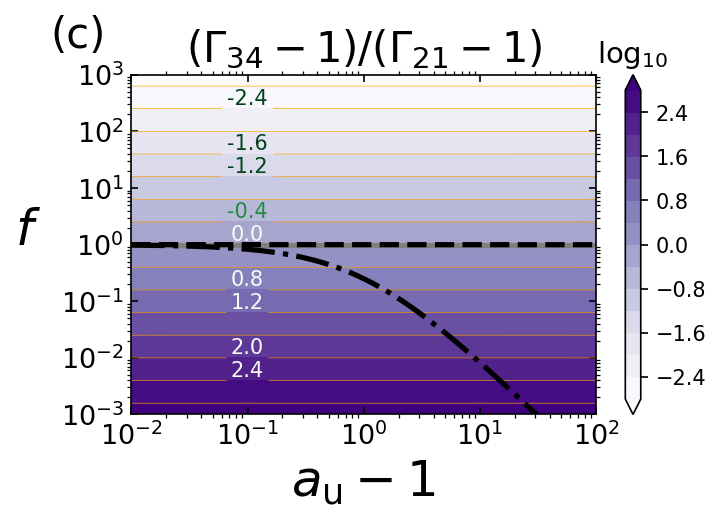}   &  \includegraphics[scale=0.55]{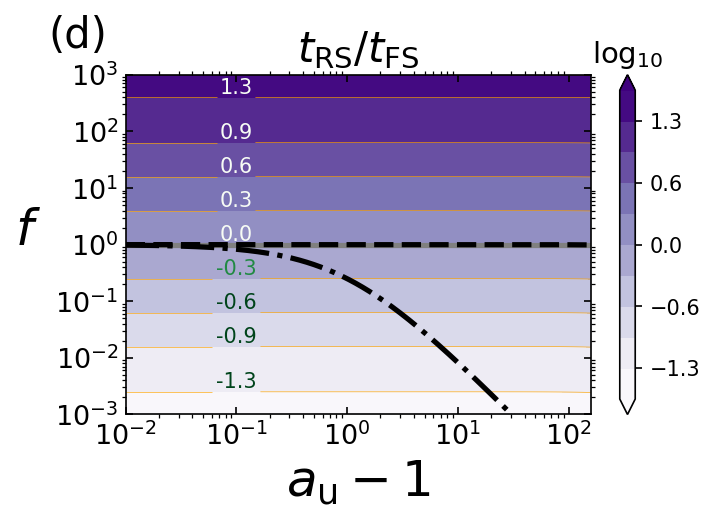}\\ 
      \includegraphics[scale=0.55]{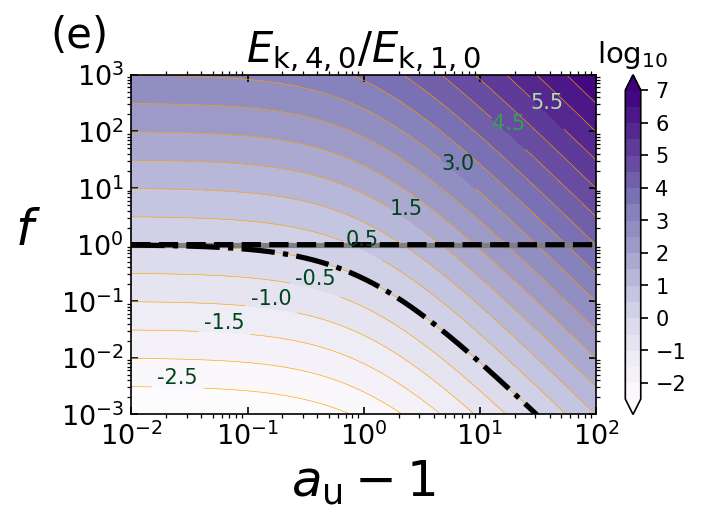}   &  \includegraphics[scale=0.55]{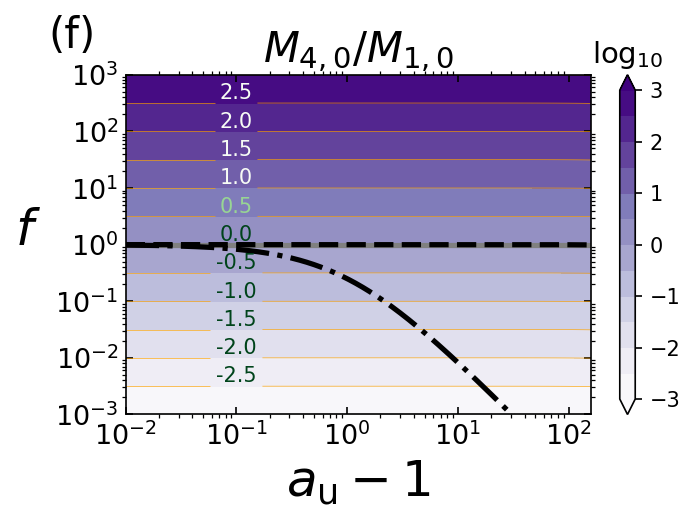} \\
    \end{tabular}
    \caption{The figure corresponds to collision of two \textbf{Newtonian shells} of equal initial radial width ($\chi = 1$) for a fixed proper speed $u_1 = 10^{-3}$ for shell S1. The panel description remains the same as Fig. \ref{ultra_param}.  }
    \label{newt_param}
\end{figure*}

For collision between shells moving with Newtonian velocities, i.e. $u_1<u_4\ll1$, the proper speed of the  shocked fluid is given by
\begin{equation}
    u \approx \beta =   \beta_1  \frac{ (1+\sqrt{f}\,a_\mathrm{u})  }{(1 + \sqrt{f})}\ , 
\end{equation}
such that 
\begin{subequations}
\begin{align}\label{forward_newt}
\beta_{21} &=  \beta_1 \frac{(a_\mathrm{u} - 1) \sqrt{f}}{ (1+ \sqrt{f}) }\ , 
\\ \label{rev_newt}
\beta_{43} & = - \beta_{34} = \beta_1 \frac{(a_\mathrm{u} - 1)}{ ( 1 + \sqrt{f} )}\ ,    
\end{align}
\end{subequations}
and the shock strengths are given by 
\begin{equation}\label{shock_strength_newt}
    \Gamma_{21} - 1 \approx \frac{1}{2} \beta^2_{21} \ll 1\ ,      \quad\quad\quad  \Gamma_{34} - 1  \approx \frac{1}{2}\beta^2_{34} \ll 1\ . 
\end{equation}
This shows that both shocks are Newtonian and using equations~(\ref{forward_newt})-(\ref{rev_newt}) we infer the ratio of the shock strengths, 
\begin{equation}
    \frac{\Gamma_{34}-1}{\Gamma_{21}-1} \approx \frac{1}{f}\ . 
\end{equation}

In order to gain physical insight we consider the density contrast $f$ for collision between two equal mass and equal energy shells moving at Newtonian speeds ($u_1<u_4\ll1$),
\begin{equation}
         f \;\approx\; \begin{cases}  \chi & \text{for $M_{1,0} = M_{4,0}$\ ,  } 
         \vspace{0.2cm}\\ 
            \frac{\chi}{a^2_\mathrm{u}} & \text{for $E_\mathrm{k,4,0} = E_\mathrm{k,1,0}$\ . } \\
    \end{cases} \label{fnewt}
\end{equation}

Thus, we can use the approximation $(\Gamma_{21},\Gamma_{34}) \approx 1$ in equations (\ref{Delta2f})-(\ref{Delta3f}), to obtain the shock crossing timescales $(t_\mathrm{FS},t_\mathrm{RS})$,
\begin{equation}
    t_\mathrm{FS} \approx \frac{3}{4} \frac{\Delta_\mathrm{1,0}}{v_1} \; \frac{(1 + \sqrt{f})}{(a_\mathrm{u} + 1) \sqrt{f}}\ , \hspace{0.8cm}  t_\mathrm{RS} \approx \frac{3}{4} \frac{\Delta_{4,0}}{v_1}  \frac{(1 + \sqrt{f})}{ (a_\mathrm{u} + 1)}\ ,
\end{equation}
where $v_1$ is the pre-collision speed of shell S1, leading to a ratio of shock crossing times (for $u_1<u_4\ll1$),
\begin{equation}
    \begin{split}
        \frac{t_\mathrm{RS}}{t_\mathrm{FS}} &\  \approx  \frac{\sqrt{f}}{\chi} \approx \begin{cases} 
        &\ 1/\sqrt{\chi} \hspace{1cm }\text{for $M_\mathrm{1,0} = M_\mathrm{4,0}$}\ , \\ 
        & \ 1/\sqrt{\chi}\,a_u
        \hspace{0.6cm} \text{for $E_\mathrm{k,1,0} = E_\mathrm{k,4,0}$}\ , \end{cases}
    \end{split} \label{time_newt}
\end{equation}
 where we have used equation~(\ref{fnewt}) to eliminate the dependence on $f$ in the second and the third line. As both shocks are Newtonian, the final radial width after shock passage can be obtained by substituting $(\Gamma_{21},\Gamma_{34}) \approx 1$ in equations~(\ref{Delta2f})-(\ref{Delta3f}), 
\begin{equation}
    \Delta_\mathrm{2f}  \approx \frac{1}{4} \Delta_\mathrm{1,0}\ ,    \hspace{1cm}  \Delta_\mathrm{3f} \approx \frac{1}{4} \Delta_\mathrm{4,0}\ . 
\end{equation}
Thus, both shells have the same lab frame shock compression ratio, which is the familiar Newtonian strong shock compression ratio of 4 (as the lab frame densities approach the comoving ones in the Newtonian limit).

Fig. ~\ref{newt_param} shows the hydrodynamical parameter space for a collision of Newtonian shells  with  equal  initial radial width ($\chi  = 1$). In all panels the equal mass collision coincides with the $f=1$ line at both low and high proper speed contrast. This is because for Newtonian velocities, the Lorentz factor is always very close to unity, such that 
the lab frame number density equals the comoving number density. Thus, shells of equal mass and radial width have not only equal lab frame density but also equal proper density ($f=1$).
Panels (a) and (b) show that both shocks are Newtonian (as seen, e.g., from equation~(\ref{shock_strength_newt})). Moreover, panels (a), (b) and (c) show the same mirror symmetry properties about the $f=1$ line ($f\!\to\!1/f$) as the corresponding panels in  Fig.~\ref{ultra_param}. Panel (c) shows that while both shocks are equally strong for equal mass collision (at both high and low proper speed contrast), for equal energy collision the reverse shock is stronger at high proper speed contrast, and the ratio of the shock strengths depends inversely on the proper density contrast $f$. Panel (d) shows that the shell crossing times are equal for the equal mass collision (see equation~(\ref{time_newt})). For the equal energy collision the RS finishes crossing before the FS ($t_{\rm RS}<t_{\rm FS}$). Panel (e) shows that for equal mass collision the total initial kinetic energy is dominated by the kinetic energy in shell S4. Panel (f) shows that for equal kinetic energy collision the mass in shell S4 is much less than that in shell S1.

Before concluding this subsection, we want to emphasize the difference between collision of shells moving with Newtonian and ultra-relativistic speeds. As a particular illustrative example, we consider the collision of two equal mass shells and equal initial radial widths ($\chi=1$). For Newtonian shells $\chi=1$ implies  $f = 1$ at both low and high proper speed contrast limit, since for Newtonian velocities the lab densities are equal to the comoving densities. It is to be noted that for ultra-relativistic speeds, the $f=1$ is attained only in low proper speed contrast limit.

To summarize, for the collision of two shells moving with Newtonian velocities, both shock strengths are naturally Newtonian. However, for an equal energy collision the reverse shock is stronger than the forward shock and therefore reaches the rear edge of shell S4 before the the forward shock can reach the front edge of shell S1. The same is true for the collision of ultra-relativistic shells considered in the previous subsection.     

Panels (d) of Fig.~\ref{ultra_param} and Fig.~\ref{newt_param} show that the ratio of shock crossing times, $t_{\rm RS}/t_{\rm FS}$, varies significantly over a wide parameter space. 
This begs the question as to what happens when one of the shock fronts reaches the edge of its respective shell before the other can. As we will see, this is an important consideration for the total energy dissipated at both shocks. Equations~(\ref{E2int})-(\ref{E3int}) provide the internal energy dissipated assuming both shocks 
manage to reach the edge of their respective shells.  In the next section, we pursue this question of whether each shock can complete crossing its shell or whether some other process hinders it.

\section{Limits on kinetic energy dissipation due to rarefaction waves}\label{dissp_limit}

In the next subsections, we motivate the need for the inclusion of rarefaction waves in our analysis and explore limits on the energy dissipation by the shock fronts. We provide in-depth analysis for equal proper density, equal kinetic energy, and equal mass collisions.  

\subsection{The need for a rarefaction wave}
\label{sec:rarefactionmotivation}
In the previous section, we saw that in general $t_\mathrm{RS} \neq t_\mathrm{FS}$. In order to derive physical insight, we consider an ``external'' shock scenario where shell S1 is at rest while its radial width is semi-infinite such that the reverse crossing timescale $t_\mathrm{RS}$ is finite while the forward crossing time $t_\mathrm{FS}$ is infinite. Now, consider the situation when the reverse shock reaches the edge of shell S4. If no additional process kicks in beyond this instant, the CD continues to perform $p dV$ work indefinitely and as a consequence, the forward shock front will also continue to dissipate energy indefinitely. But clearly, this is unphysical as the $p dV$ work done by CD comes at the expense of $E_\mathrm{kin,4,0}$ which is finite.  So what happens physically is that once the RS reaches the edge of shell S4, it produces a high pressure at its matter-vacuum interface and a  rarefaction (hereafter rf) wave is launched toward the CD. The head of the rarefaction wave moves at the local sound speed relative to the fluid into which it propagates. Once the head of the rf wave reaches the CD, it leads to a drop in pressure, and hence the $p dV$ work done also decreases until the head of the rarefaction wave catches up with the forward shock front. At this point, the FS quickly weakens and its subsequent energy dissipation is severely suppressed. A rf wave is an inevitable consequence of the finite width and energy of the shell(s). During the propagation of the rf wave from the edge to the CD, the $p dV$ work continues to be done at the CD, but since it is done at the expense of the  energy in the region R3, the latter decreases (by the rf wave). Thus, a fraction of the \textit{internal energy dissipated by the reverse shock is reprocessed} into the (bulk+internal) energy of region R2. Table ~\ref{rf_defns} summarizes the quantities required for analysis of rf wave propagation. 

\begin{figure}
    \centering
    \includegraphics[scale=0.93,trim=2cm 0.55cm 1.9cm 0.8cm, clip]{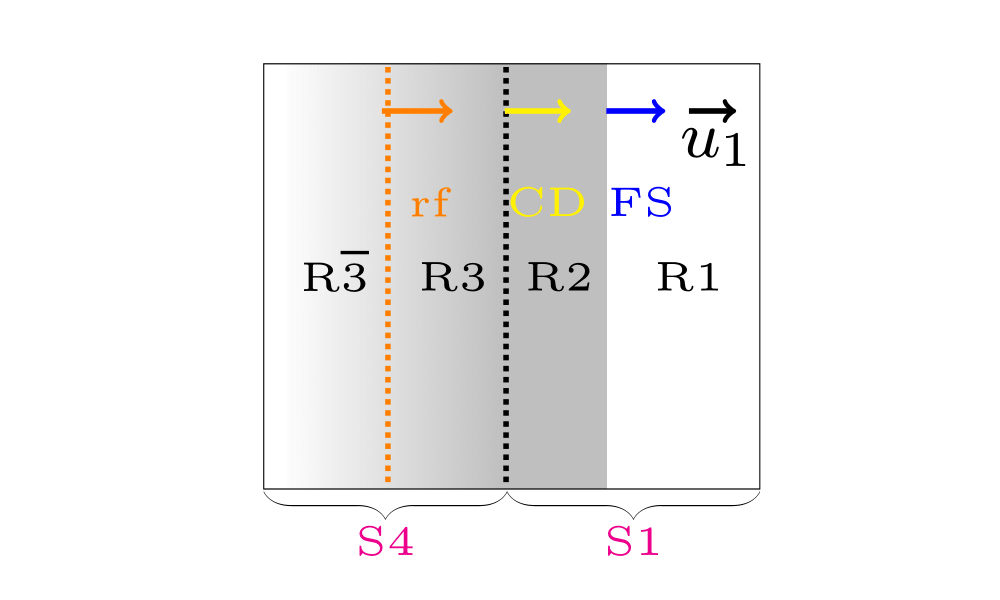}
    \caption{The launch of a rarefaction wave chasing a shock front. This particular illustration corresponds to the collision of two equal energy shells with equal initial radial width $\chi= 1$ and $(u_1,u_4) = (100,200)$ at $t = t_\mathrm{o} + 1.2 t_\mathrm{RS}$ ( since $t_\mathrm{RS} < t_\mathrm{FS}$ ). After the reverse shock reaches the rear edge of shell S4, a  rarefaction wave with proper speed $u_\mathrm{3rf+}$ is launched that chases after the forward  shock front. The arrows show in scale the proper speed of the rarefaction wave, the CD and the forward shock front.}
    \label{rf_wave}
\end{figure}

Fig. ~\ref{rf_wave} shows a particular case for collision of two equal energy shells of equal initial radial width ($\chi=1$). As shown in \S\,\ref{ultra_shells} for equal energy collision the RS reaches the rear edge of shell S4 before the FS reaches the front edge of shell S1. After the RS reaches the edge of S4, a rf wave is launched towards the FS. The case is reversed for equal mass collision where the FS reaches the front edge of S1 before the RS can reach the rear edge of S4 (see expanded discussion in \S\,\ref{ultra_high}.
Panel (d) in Figs.~\ref{ultra_param} and ~\ref{newt_param} show that in a wide parameter space the shell crossing timescales are significantly different. This points to the possibility that the rf wave can catch up with the shock front with the longer crossing timescale and halt the internal dissipation. 
Below we explore the parameter space where the rf wave can cross the CD and catch up with the shock front with the longer crossing timescale, before the latter reaches the edge of the corresponding shell leading to a halting of the energy dissipation by that shock.

\begin{table*} 
    \centering
     \caption{Symbols and their definitions to be used for the analysis of the limitation of the internal energy dissipation by either of the shock fronts due to rarefaction waves. The symbols $+$ and $-$ in quantities refer to a rarefaction wave propagating towards the forward and the reverse fronts respectively. The primed superscript refers to a comoving frame of the relevant fluid numbered by a subscript. The subscript $j=(2,3)$ stands for regions R2 and R3 shocked by forward and reverse shock respectively.}
    \begin{tabular}{c|c} \hline 
        Symbol & Definition \\ \hline 
        $\beta'_\mathrm{sj}$ & Sound speed in the comoving frame of region $j$ \\ 
           $\beta_\mathrm{rfj+}$ & Speed of rarefaction ($+$) waves in region $j$ \\ 
        $\beta_\mathrm{rfj-}$ & Speed of rarefaction ($-$) waves in region $j$ \\  \hline 
         $t_\mathrm{3rf+}$ & The time taken by the rf wave ($+$) to reach CD from the back edge of shell S4  \\
        $t_\mathrm{2rf+}$ & The time taken by the rarefaction wave ($+$) to reach forward shock front starting from CD \\
        $t_\mathrm{3rf-}$ & The time taken by the rf wave ($-$) to reach RS starting from CD \\
        $t_\mathrm{2rf-}$ & The time taken by the rarefaction wave ($-$) to reach CD from front edge of shell S1 \\ \hline 
        $W_\mathrm{pdV}$ & The p dV work done by the CD against region 3 and on region 2 \\ 
        $E_\mathrm{j,int}$ & The total internal energy dissipated in region $j$ \\
         $E_\mathrm{j, int, max}$ & The maximum energy that can be dissipated in region $j$ \\
          $M_\mathrm{j}$ & Mass in region $j$ \\
          $\alpha_3$ & Defined as $M_\mathrm{3} / M_\mathrm{4}$  \\ 
        $\alpha_2$ & Defined as $M_\mathrm{2} / M_\mathrm{3}$  \\ 
      \hline 
    \end{tabular}
    \label{rf_defns}
\end{table*}

\begin{table*}
    \centering
    \caption{List of the various scenarios of the rf waves chasing either the FS or the RS. The propagation of  ($\pm$) rf waves is not tracked beyond the time at which the forward/reverse reaches the edge of the corresponding shell. The five critical initial radial widths which satisfy the lines in time (L1-L5) are summarized in Table \ref{5crit_lines}.  }
    \begin{tabular}{c|c|c|c|} \hline
        Cases & Description & $\alpha_2$  & $\alpha_3$  \\ \hline

        I & $t_\mathrm{RS} + t_\mathrm{3rf+} + t_\mathrm{2rf+}  < t_\mathrm{FS}$ & $\frac{t_\mathrm{RS} + t_\mathrm{3rf+} \; + \;   t_\mathrm{2rf+}}{t_\mathrm{FS}} = \frac{\chi_\mathrm{c1}}{\chi}$ & 1\\  
     \hline  
      \hspace{-0.3cm} L1: $t_\mathrm{RS} + t_\mathrm{3rf+} + t_\mathrm{2rf+}  = t_\mathrm{FS}$  & \hspace{-3cm}(+)rf wave catches up with FS  at the front edge of S1  \\  \hline 

     II & $t_\mathrm{RS} + t_\mathrm{3rf+} + t_\mathrm{2rf+}  >  t_\mathrm{FS}$ & 1 & 1  \\  \hline 
       
    \hspace{-1.0cm}L2: $t_\mathrm{RS} + t_\mathrm{3rf+} = t_\mathrm{FS}$ & \hspace{-1cm} (+)rf wave reaches CD and FS reaches the front edge of S1 simultaneously     \\ \hline

      III & $t_\mathrm{RS} + t_\mathrm{3rf+}  >  t_\mathrm{FS}$ & 1 & 1 \\  
       \hline \hline 
       
      \hspace{-1.8cm}  L3: $t_\mathrm{RS} = t_\mathrm{FS}$ & \hspace{0.1cm} FS reaches the front edge of S1 and RS reaches the rear edge of S4 simultaneously\\ \hline \hline 

     IV & $t_\mathrm{FS} + t_\mathrm{2rf-}   >  t_\mathrm{RS}$ & 1 & 1  \\  \hline 

     \hspace{-1.0cm}L4: $t_\mathrm{FS} + t_\mathrm{2rf-} = t_\mathrm{RS}$ & \hspace{-1.1cm}(-)rf wave reaches CD and RS reaches the rear edge of S4 simultaneously \\ \hline

     V & $t_\mathrm{FS} + t_\mathrm{2rf-} + t_\mathrm{3rf-}  >  t_\mathrm{RS}$ & 1 & 1 \\  \hline 

     \hspace{-0.2cm}L5: $t_\mathrm{FS} + t_\mathrm{2rf-} + t_\mathrm{3rf-} = t_\mathrm{RS}$  &  \hspace{-3cm}(-)rf wave catches up with RS at the rear edge of S4 \\ \hline
     VI & $t_\mathrm{FS} + t_\mathrm{2rf-} + t_\mathrm{3rf-}  < t_\mathrm{RS}$ & 1 & $\frac{t_\mathrm{FS} + t_\mathrm{2rf-} \; + \;   t_\mathrm{3rf-}}{t_\mathrm{RS}} = \frac{\chi}{\chi_\mathrm{c5}} < 1$ \\  \hline  
    \end{tabular}
    \label{6cases_table}
\end{table*}

\begin{table*} 
    \centering
    \caption{Expression for the five critical initial radial width ratio that divides the $a_\mathrm{u} -f$ parameter space into six cases  } 
    \begin{tabular}{c|c} \hline 
       Critical lines  & Expressions \\ \hline 
      $\chi_\mathrm{c1}$  &  $ (\beta_\mathrm{FS} - \beta_1) \left[ 1 + \frac{(\beta_\mathrm{FS} - \beta)}{(\beta_\mathrm{2rf+} - \beta_\mathrm{FS})}\right] \left[ \frac{1}{(\beta_4 - \beta_\mathrm{RS})}  + \frac{1}{4 \Gamma_{34}} \left( \frac{\Gamma_4}{\Gamma}\right) \frac{1}{(\beta_\mathrm{3rf+} - \beta )}   \right]$   \\ \\ 
      $\chi_\mathrm{c2}$  & $ (\beta_\mathrm{FS} - \beta_1) \left[ \frac{1}{(\beta_4 - \beta_\mathrm{RS})}  + \frac{1}{(\beta_\mathrm{3rf+} - \beta )}  \frac{1}{4 \Gamma_{34}} \left( \frac{\Gamma_4}{\Gamma}\right) \right] $   \\ \\ 
      $\chi_\mathrm{c3}$  & $\frac{(\beta_\mathrm{FS} - \beta_1)}{(\beta_4 - \beta_\mathrm{RS})}$   \\  \\ 
      $\chi^{-1}_\mathrm{c4}$  & $ (\beta_4 - \beta_\mathrm{RS}) \left[ \frac{1}{(\beta_\mathrm{FS} - \beta_1)} + \frac{1}{4 \Gamma_{21}} \left( \frac{\Gamma_1}{\Gamma} \right) \left( \frac{1}{(\beta - \beta_\mathrm{2rf-})}\right) \right] $  \\ \\ 
     $\chi^{-1}_\mathrm{c5}$   &  $(\beta_4 - \beta_\mathrm{RS}) \left[ 1 + \left( \frac{\beta - \beta_\mathrm{RS}}{\beta_\mathrm{RS} - \beta_\mathrm{3rf-}} \right) \right] \left[ \frac{1}{(\beta_\mathrm{FS} - \beta_1)} + \frac{1}{4 \Gamma_{21}} \left( \frac{\Gamma_1}{\Gamma} \right) \left( \frac{1}{(\beta - \beta_\mathrm{2rf-})}\right) \right]    $  \\ \\ \hline 
    \end{tabular}
    \label{5crit_lines}
\end{table*}

In Table \ref{6cases_table} we summarize 5 critical lines (L1-L5) in time. As shown in the Appendix H, the lines L1-L5 in time can be inverted to define five critical ratios of the initial radial width  of shell S1 to shell S4 as $\chi_\mathrm{cX}$ where $X = (1,2,3,4,5)$ are 
summarized in Table \ref{5crit_lines}. The five critical ratios $\chi_\mathrm{cX}$ can be used to define six different cases:
\begin{itemize}
    \item \textbf{Case I} ($\chi> \chi_\mathrm{c1}$): shell S1 is partially shocked; the forward ($+$) rf wave catches up with the FS front before reaching the front edge of shell S1, and the shocked fraction of S1 is given by
    \begin{equation}
        \alpha_{2} = \frac{\chi_\mathrm{c1}}{\chi} < 1\ , \hspace{4cm} \text{(for $\chi> \chi_\mathrm{c1}$)\ .} 
    \end{equation}

    \item \textbf{Case II} ($\chi_\mathrm{c1} < \chi < \chi_\mathrm{c2}$): the FS front reaches the edge of S1 after the forward ($+$) rf wave reached the CD but before it reaches the front edge of S1 (i.e. when its head is propagating into region R2).

    \item \textbf{Case III} ($\chi_\mathrm{c2} < \chi< \chi_\mathrm{c3}$): the FS front reaches the edge of S1 before the forward ($+$) rf wave reaches the CD (i.e. when its head is propagating into region R3).

    \item \textbf{Case IV} ($\chi_\mathrm{c4} < \chi < \chi_\mathrm{c3}$): the RS front reaches the rear edge of shell S4 before the backward ($-$) rf wave reaches the CD (i.e. when its head is propagating into region R2). 

    \item \textbf{Case V} ($\chi_\mathrm{c5} < \chi < \chi_\mathrm{c4}$): the RS front reaches the rear edge of shell S4 after the backward ($-$) rf wave reaches the CD but before it reaches the back edge of S4 (i.e. when its head is going into R3).

    \item \textbf{Case VI} ($\chi < \chi_\mathrm{c5}$): the shell S4 is partially shocked; the backward ($-$) rf wave catches up with the RS front before it reaches the rear edge of shell S4, and the shocked fraction of S4 is given by 
    \begin{equation}
        \alpha_\mathrm{3} = \frac{\chi}{\chi_\mathrm{c5}} < 1\ , \hspace{4cm} \text{(for $\chi < \chi_{c5}$)\ .} 
    \end{equation}
\end{itemize}

The dissipation efficiency into internal (or thermal -- subscript `th') energy, of the FS and the RS, can be expressed as 
\begin{subequations}\label{eq:e_th}
\begin{align}
\epsilon_\mathrm{th2} &=  \frac{\alpha_2 \; E_\mathrm{int,2}}{ E_\mathrm{k,1,0} + E_\mathrm{k,4,0}} = \alpha_2 \epsilon_\mathrm{th2,max}\ ,
\\
\epsilon_\mathrm{th3} &= \frac{\alpha_3 \; E_\mathrm{int,3}}{ E_\mathrm{k,1,0} + E_\mathrm{k,4,0}} = \alpha_3 \epsilon_\mathrm{th3,max}\ ,
\end{align}
\end{subequations}
where the weighting factors $(\alpha_2,\alpha_3)$ characterize the fraction of the shells (S1,S4) shocked by the forward/reverse shock front respectively. As discussed before, the shells (S1, S4) are completely shocked ($\alpha_2=1,\alpha_3=1$) by the (forward, reverse) shock fronts except for case I where S1 is partially shocked ($\alpha_\mathrm{2} < 1$), and case VI  where S4 is partially shocked ($\alpha_\mathrm{3} < 1$). Thus, the energy dissipated by both shock fronts taken together is
\begin{equation}
    \epsilon_\mathrm{th,tot} = \epsilon_\mathrm{th2} + \epsilon_\mathrm{th3} = \alpha_2 \epsilon_\mathrm{th2,max} + \alpha_3 \epsilon_\mathrm{th3,max}\ . \label{total_effi}
\end{equation}

Note that equation~(\ref{total_effi}) is an addition of $\epsilon_\mathrm{th2}$ and $\epsilon_\mathrm{th3}$, which are estimated at two different times. Therefore, the internal energy dissipated by the RS can be reprocessed by the $pdV$ transfer of work across CD from shell S4 to shell S1, where a part of it can be used by the FS front to dissipate internal energy in shell S1. As a result, the combined thermal efficiency as defined in equation~(\ref{total_effi}) can also exceed unity (see discussion in subsection \ref{upp_lim}). However, this does not violate  energy conservation as the internal energy dissipated by the two shocks are evaluated at different times. The usefulness of this definition is that if some fraction of the thermal energy can be converted to radiation, this efficiency will be a proxy for the radiated energy which is a measurable quantity.

Next, in order to gain physical insights we consider three scenarios of internal shocks and see how they map to the six cases, I-VI. The three scenarios correspond to the collision between (i) two equal energy shells ($E_\mathrm{k,1,0} = E_\mathrm{k,4,0}$), (ii) two equal mass shells ($M_{1,0} = M_{4,0}$), and (iii) two equal proper density shells ($n'_{1} = n'_{4} \Leftrightarrow f=1$). For these scenarios, the ratio of the initial radial widths of the shells is taken to be unity, $\chi = 1$.

\begin{figure}
    \begin{tabular}{l}
     \includegraphics[scale=0.54]{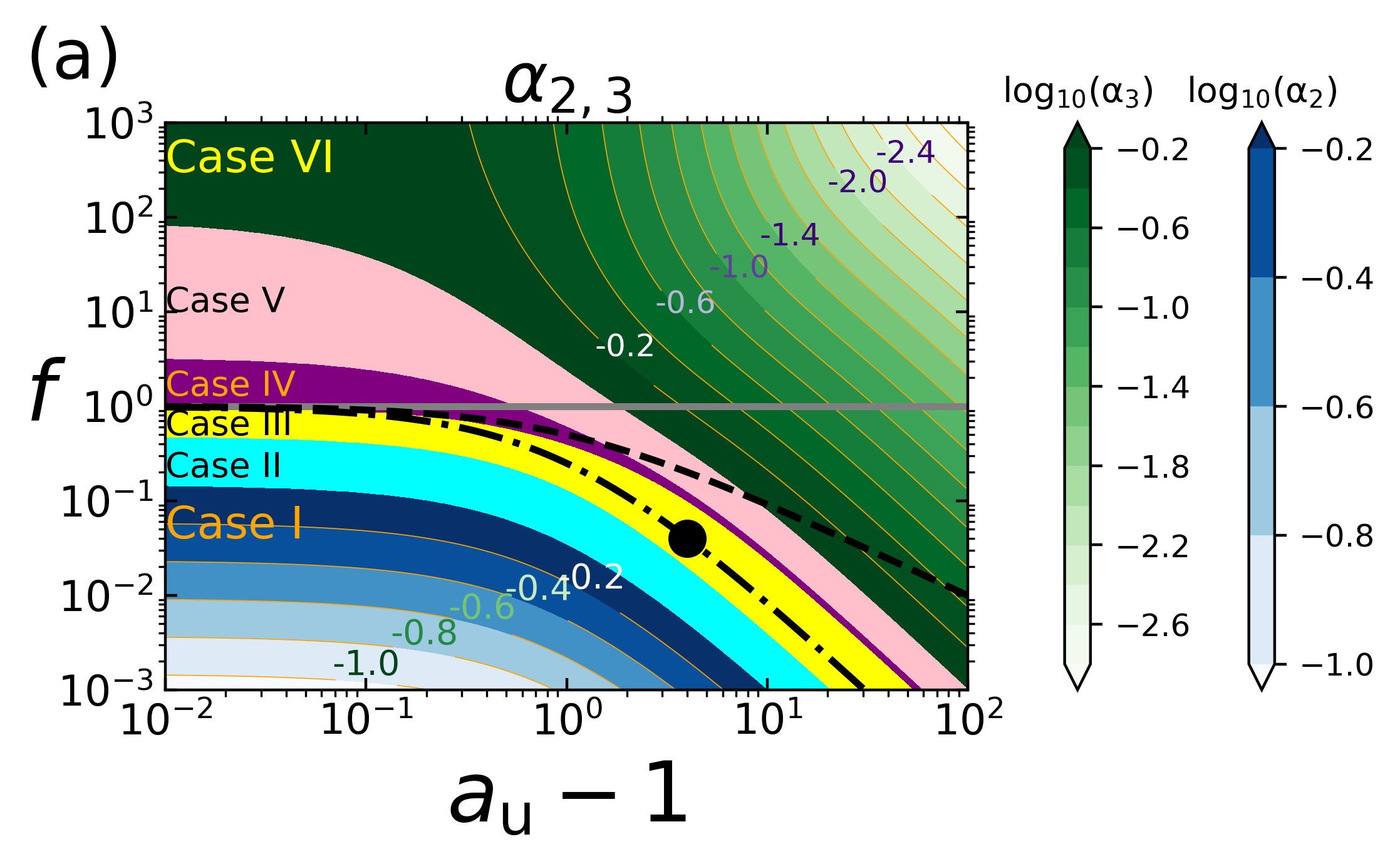} \\ 
     \vspace{-0.2cm}
     \includegraphics[scale=0.54]{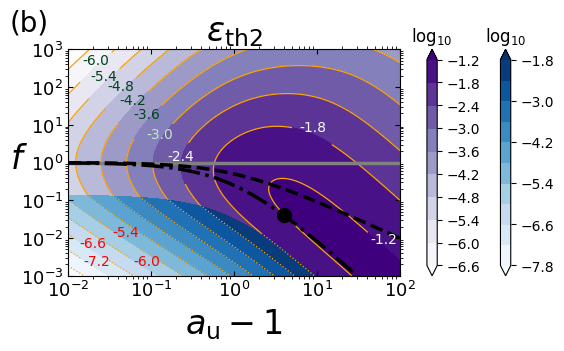} \\ 
     \includegraphics[scale=0.54]{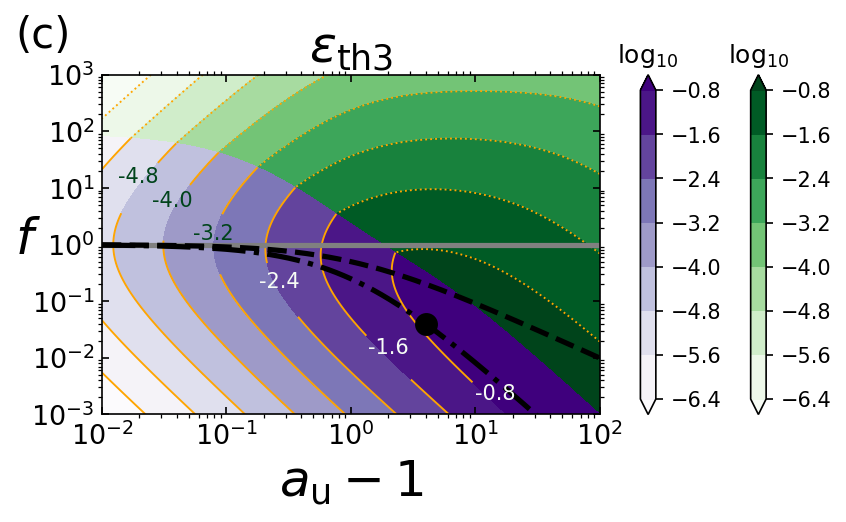} \\
     \includegraphics[width=0.453\textwidth]{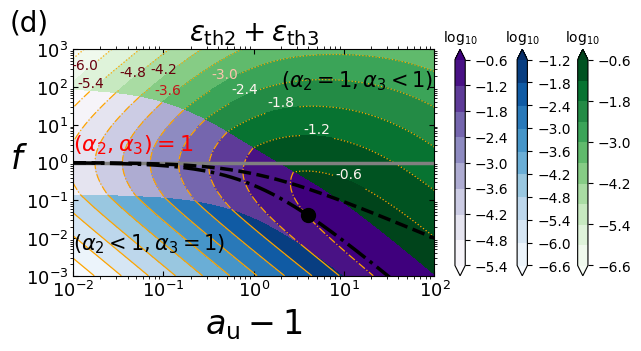}
     \end{tabular} \vspace{-0.3cm}
    \caption{This figure depicts the collision of two \textbf{ultra-relativistic} shells of equal initial radial widths ($\chi=1$) for a fixed proper speed of $u_1 = 100$. Panel (a) shows the six cases corresponding to Table \ref{5crit_lines}. Logarithmic contours for $(\alpha_2,\alpha_3)<1$ as a function  of the proper speed contrast $a_\mathrm{u}-1$ and the density contrast  $f$ are shown for cases (I) and (VI), respectively. Note that $(\alpha_2,\alpha_3)=1$ for all other cases (II)-(V). Panels (b) and (c) show the fraction of the initial total kinetic energy dissipated into internal energy by the FS ($\epsilon_\mathrm{th2}$) and by the RS ($\epsilon_\mathrm{th3}$), respectively. Panel (d) shows the fraction of the combined initial kinetic energy dissipated by both shock fronts, $\epsilon_\mathrm{th,tot}=\epsilon_\mathrm{th2}+\epsilon_\mathrm{th3}$. All contours use a logarithmic scale. The thick lines are as described in Fig.~\ref{ultra_param}. Like all previous figures the black filled circle in all panels corresponds to the collision of equal energy shells with proper speeds $(u_1,u_4)=(100,500)$. Figure \ref{EqEn_breakdown} corresponds to this specific point in the phase space.}
    \label{ultra_critical}
\end{figure}

Fig.~\ref{ultra_critical} shows the parameter space of $(\alpha_2, \alpha_3)$ (panel (a)) as well as $\epsilon_\mathrm{th2}$,  $\epsilon_\mathrm{th3}$ and $\epsilon_\mathrm{th,tot}=\epsilon_\mathrm{th2}+\epsilon_\mathrm{th3}$ (panels (b), (c) and (d)), for a collision of ultra-relativistic shells of equal initial radial width. The 5 critical lines (L1-L5) divide the proper density and proper speed contrast parameter space into six cases. It can be seen that equal energy collisions correspond to case III throughout, while equal mass collisions corresponds to case III at low proper speed contrast, but transition to case IV and V at moderate values of proper speed contrast and finally enter the case VI regime at very high values of proper speed contrast. The behaviour is similar for $f=1$ collisions, except that they enter case VI already at more moderate values of proper speed contrast. 

\subsection{Collision between two ultra-relativistic shells at high proper speed contrast} \label{ultra_high}

In the following subsections we describe the physics of shock propagation  for several cases of interest.

\subsubsection{Two equal kinetic energy and equal radial width shells }

From subsection \ref{ultra_shells} for collision of two ultra-relativistic ($u_4>u_1\gg1$) equal energy shells ($E_\mathrm{k,1,0} = E_\mathrm{k,4,0} = E_\mathrm{o}$) of equal radial width $\Delta_\mathrm{1,0} = \Delta_\mathrm{4,0} = \Delta_\mathrm{0}$ ($\chi=1$), the proper density contrast $f$ is given by $f \approx \frac{1}{a^2_\mathrm{u}} \ll 1$. Thus, the RS is much stronger than the FS. The proper speed of the shocked fluid reaches the asymptotic value $u \approx \sqrt{2}\,u_{1}$. The strength of the RS is given by $\Gamma_{34} - 1 \approx \frac{a_\mathrm{u}}{2 \sqrt{2}} \gg 1$ while the FS has shock strength $\Gamma_{21} - 1 \approx 0.0607 \ll 1$. Thus, the RS is ultra-relativistic while the FS is Newtonian and independent of $a_\mathrm{u}$. Besides, the RS and the FS crossing timescales are given by $t_\mathrm{RS} \approx \frac{\Gamma^2\Delta_\mathrm{4,0}}{c}$ and $t_\mathrm{FS} \approx \frac{5}{3} \frac{\Gamma^2 \Delta_{1,0}}{c}$ respectively.
Thus, since $\Delta_{1,0} = \Delta_{4,0}$ the RS reaches the rear edge of shell S4 before the FS reaches the front edge of shell S1. After the RS reaches the rear edge of shell S4, region R4 no longer exists. The final radial width of the region R3 is $\Delta_\mathrm{3f} \approx \frac{1}{2}\Delta_\mathrm{4,0}$ (see Appendix  G). 
. 

After the RS reaches the rear edge of shell S4, a forward propagating $(+)$ rf wave is launched. Since the strength of the RS is ultra-relativistic, the co-moving sound speed in region R3 reaches the asymptotic value $\beta'_\mathrm{s3} \rightarrow 1/\sqrt{3}$. The speed of the head of the rf wave in the lab frame is $\beta_\mathrm{3rf+} = (1 + \sqrt{3} \beta)/(\sqrt{3} + \beta)$. The time taken by the $(+)$ rf wave to reach the CD is given by
\begin{equation}
    t_\mathrm{3rf+} = \frac{\Delta_\mathrm{3f}}{c(\beta_\mathrm{3rf+} - \beta)} \approx \left( \frac{1 + \sqrt{3}}{2} \right)  \frac{\Gamma^2 \Delta_{4,0}} {c} \approx  1.37 \; t_\mathrm{RS} 
\end{equation}

Since, $t_\mathrm{3rf+} + t_\mathrm{RS} =  2.37 t_\mathrm{RS} > t_\mathrm{FS}$, the forward shock front reaches the front edge of shell S1 before the forward propagating ($+$) rf wave reaches the CD.  This corresponds to case III  (see \S \ref{sec:rarefactionmotivation}). Thus, the weighting factors are $(\alpha_2,\alpha_3) = 1$. The internal energy dissipated by the FS and the RS are given by
\begin{equation}
\begin{split}
&\ E_\mathrm{int,3} \approx \frac{2}{3} E_\mathrm{k,4,0} \approx 0.67 \; E_\mathrm{K,0} \\
&\ E_\mathrm{int,2} \approx \left(\frac{14}{9}-\sqrt{2}\right) E_\mathrm{k,1,0} \approx 0.14 E_\mathrm{k,0}   
\end{split}  
\end{equation}
The RS dissipates internal energy $\sim4.7$  times more efficiently than the FS. The thermal efficiencies of the FS and the RS front are given by 
\begin{equation}
    \epsilon_\mathrm{th,2} \approx 0.071\ ,\quad\quad\quad 
    \epsilon_\mathrm{th,3} \approx 0.33\ . 
\end{equation}

Next, we can look at the kinetic energies of the shells after one complete sweep by the RS and the FS:
\begin{equation}
    \begin{split}
       &\ E_\mathrm{k,3} \approx \left( \frac{\Gamma}{\Gamma_\mathrm{4}} \right)  E_\mathrm{k,4,0} \approx \frac{\sqrt{2}}{a_\mathrm{u}}  E_\mathrm{k,0} \ll E_\mathrm{0}\ , 
       \\
       &\ E_\mathrm{k,2} \approx \left( \frac{\Gamma}{\Gamma_1}\right) E_\mathrm{k,1,0} \approx \sqrt{2} E_\mathrm{k,0} > E_\mathrm{0}\ .
    \end{split}
\end{equation}
Thus, after the RS sweeps through shell S4, the kinetic energy of region R3 is negligible compared to the initially available kinetic energy $(E_\mathrm{k,4,0} = E_\mathrm{k,0})$ in S4. However, the kinetic energy of region R2 after one complete sweep of shell S1 by the FS is $\sim 1.41$ times higher than the initially available kinetic energy $(E_\mathrm{k,1,0} = E_\mathrm{0})$.

Next, we can estimate the total energies in regions R2 and R3 after one complete sweep by the FS and the RS, respectively, as
\begin{equation}
    \begin{split}
        &\ E_\mathrm{tot,3} = E_\mathrm{k,3} + E_\mathrm{int,3} \approx 0.67 E_\mathrm{0}\ , 
        \\
        &\ E_\mathrm{tot,2} = E_\mathrm{k,2} + E_\mathrm{int,2} \approx 1.55 E_\mathrm{0}\ . 
    \end{split}
\end{equation}
Thus, the passage of the FS increases the net energy of shell S1 by $\sim 0.55 E_\mathrm{0}$, which ultimately comes from region R3 to region R2 via  $p dV$ work across the CD. It can be estimated explicitly as follows. The $pdV$ work done during $t_\mathrm{RS}$ for a relativistic RS is $W_\mathrm{pdV,RS} \approx \frac{1}{3} E_\mathrm{k,4,0}$. Due to the planar geometry, the work done in $t_\mathrm{FS}$ scales linearly with time. Using equation~(\ref{eq:e_th}) the $p dV$ work done till the forward shock front reaches the front edge of shell S1 can be estimated as
\begin{equation}
    W_\mathrm{pdV,FS} = W_\mathrm{pdV,RS} \left( \frac{t_\mathrm{FS}}{t_\mathrm{RS}}\right) \approx \frac{5}{9} E_\mathrm{k,4,0} \approx 0.55 E_\mathrm{0}\ .
\end{equation}

Besides, the final radial widths ($\Delta_\mathrm{3f},\Delta_\mathrm{2f}$) of the regions (R3, R2) after a full sweep of shells (S4, S1) by the (RS, FS) are $ \Delta_\mathrm{3f} \approx  \frac{\Delta_{0}}{2}$ and $\Delta_\mathrm{2f} \approx \frac{\Delta_{0}}{6}$ (see \S \ref{ultra_shells}). Thus, the lab frame compression ratio for the FS is higher than for the RS by a factor of three. 

Notice that the sum total energies of the shells after a complete sweep by  both shock fronts ($\sim 2.22 E_{0}$) is more than the initially available kinetic energy of both shells ($2 E_\mathrm{0}$). However, this does not violate energy conservation as the energies of the two shells are evaluated at different times, and part of the energy of region R3 at $t_{\rm o}+t_{\rm RS}$ is transferred to region R2 by $t_{\rm o}+t_{\rm FS}$ 
through the $pdV$ work across the CD.

\begin{figure}
    \centering
\begin{tabular}{c}
     \includegraphics[scale=0.50]{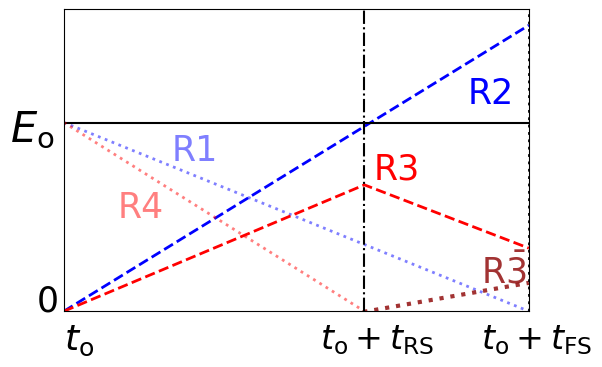} \\
     \includegraphics[scale=0.50]{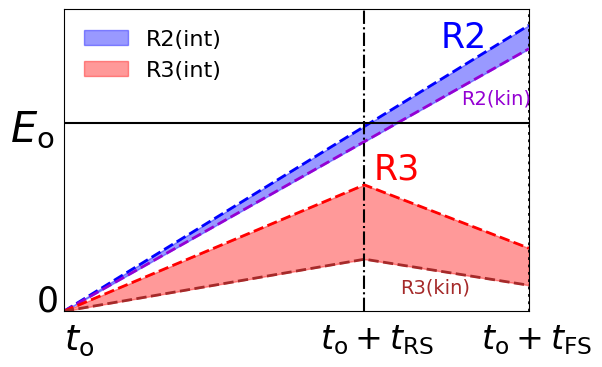} \\
\end{tabular}    
    \caption{The breakdown of total (kinetic + internal ) energy between regions (R1,R2,R3,R4) as a function of time for collision of two equal energy ultra-relativistic shells with equal initial radial width. \textbf{\textit{Top:}} shows the breakdown of total energy in different regions as a function of time. \textbf{\textit{Bottom:}} shows the breakdown of energy in regions R2 and R3 as a function of time. }
    \label{EqEn_breakdown}
\end{figure}
Fig. ~\ref{EqEn_breakdown} shows that the internal energy in region R3 (in \textit{shaded red}) remains larger than that in region R2 (in \textit{shaded blue}) at any instant, even at $t_{\rm o}+t_{\rm FS}$ when the FS has completely swept through shell S1.  

To summarize, the collision of two ultra-relativistic shells of equal radial width launches a relativistic RS and a Newtonian FS. The RS reaches the rear edge of shell S4 before the FS reaches the front edge of shell S1. Thereafter, a forward ($+$) propagating rf wave is launched towards CD, but before it can reach the CD the FS reaches the front edge of shell S1. From the launch of the FS till it finishes crossing  shell S1 around 55\% of the initially available kinetic energy in shell S4 is transferred to shell S1 via $pdV$ work from region R3 to region R2. The FS reprocesses the $pdV$ work done into both accelerating and increasing the internal energy of the material in region R2. Since the FS is Newtonian, the $pdV$ work done is used in primarily increasing the kinetic energy of region R2. The thermal efficiency of the FS and the RS is $\sim$7\% and $\sim$33\%, respectively, corresponding to a total thermal efficiency of $\sim$40\%. Thus, the RS dissipates internal energy almost five times more efficiently than the FS. This is despite the $pdV$ transfer of work from shell S4 to S1. The reason being the RS is ultra-relativistic and very strong compared to the FS.  

\subsubsection{Two equal mass and radial width ultra-relativistic shells}

From subsection \ref{ultra_shells} for the collision of equal mass shells ($M_{1,0} = M_{4,0} $), equal radial widths ($\chi = 1$) and large proper speed contrast ($a_u\gg1$), the proper density contrast is given by
$f \approx \frac{1}{a_\mathrm{u}}$ and the proper speed of the shocked fluid is given by $u \approx a^{1/4}_\mathrm{u} u_\mathrm{1}$ such that the FS and the RS strengths are given by $ \Gamma_{21} \approx \frac{a^{1/4}_\mathrm{u}}{2}$ and $\Gamma_{34} \approx  \frac{a^{3/4}_\mathrm{u}}{2}$, 
which shows that the RS is relativistic while the FS can be mildly relativistic.

The forward and the reverse crossing timescales are given by 
\begin{equation}
    t_\mathrm{FS} \approx 2 \Gamma^2_1 \frac{\Delta_{1,0}}{c}\ ,\quad\quad 
    t_\mathrm{RS} \approx \sqrt{a_\mathrm{u}} \frac{\Gamma^2_{1} \Delta_{1,0}}{c} = \frac{\sqrt{a}_\mathrm{u}}{2} t_\mathrm{FS}\ , 
\end{equation}
which shows the FS reaches the edge of shell S1 before the RS can reach the edge of shell S4.  Since $t_\mathrm{RS} \propto a^{1/2}_\mathrm{u}$, it is not surprising the RS is halted at higher values of proper speed contrast as it provides sufficient time for the $(-)$ rf wave to catch-up with it.

\subsubsection{Collision of two equal proper density ultra-relativistic shells}

From subsection \ref{ultra_shells} for $f = 1$ the proper speed of the shocked fluid is given by $ \approx \sqrt{a}_\mathrm{u} u_1$ and the shock strengths of both shock fronts are equal. For ultra-relativistic shells ($u_4>u_1\gg1$) with very high proper speed contrast ($a_u\gg1$), or altogether $u_4\gg u_1\gg1$,
both shocks are ultra-relativistic as well,
\begin{equation}
    \Gamma_{21} = \Gamma_{34} \approx \frac{\sqrt{a_\mathrm{u}}}{2} \;\gg1\ .
\end{equation}

The ratio of the initially available kinetic energies in both shells is
\begin{equation}
    \frac{E_\mathrm{k,1,0}}{E_\mathrm{k,4,0}} \approx \frac{1}{a^2_\mathrm{u}}\ ,
\end{equation}
showing that almost all the initial kinetic energy resides in shell S4. 

The reverse crossing timescales are given by
\begin{equation}
      t_\mathrm{RS} \approx a_\mathrm{u} \Gamma^2_{1} \frac{\Delta_{4,0}}{c} = \frac{1}{2} a_\mathrm{u} \chi  t_\mathrm{FS}\ ,
\end{equation}
which shows that for $\Delta_{1,0} = \Delta_{4,0}$ ($\chi = 1$), we have $t_\mathrm{RS}=\frac{1}{2}a_\mathrm{u}t_\mathrm{FS}$. Thus, for equal initial radial widths, the FS reaches the front edge of shell S1 much earlier than the RS can reach the rear edge of shell S4. The final radial width of the region R2 at $t_{\rm o}+t_{\rm FS}$ is
\begin{equation}
    \Delta_\mathrm{2f} \approx \frac{\Delta_{1,0}}{2 a_\mathrm{u}} =  \frac{\Delta_{0}}{2 a_\mathrm{u}} .
\end{equation}
This shows that for $a_\mathrm{u} \gg 1$, the radial width of S1 is drastically reduced by the passage of the FS. Since the FS is ultra-relativistic the comoving sound speed in region R2 reaches the value $\beta'_\mathrm{s2} \rightarrow 1/\sqrt{3}$. The speed of the backward $(-)$ propagating rf wave is given by $\beta_\mathrm{2rf-} = (\beta - \beta'_{s2})/(1 - \beta \beta'_\mathrm{s2}) \rightarrow (\sqrt{3} \beta - 1)/(\sqrt{3} - \beta)$. The time it takes the backward propagating rf wave to reach the CD is  
\begin{equation}
    t_\mathrm{2rf-} = \frac{\Delta_\mathrm{2f}/c}{\beta - \beta_\mathrm{2rf-}} \approx \frac{(\sqrt{3} - 1)}{2} 
    \frac{\Gamma^2_1 \Delta_{1,0}}{c}  = \frac{\sqrt{3} - 1}{4} t_\mathrm{FS} \approx \; 0.183 t_\mathrm{FS}\ .
\end{equation}
Thus, the ($-$) rf wave reaches the CD in $\sim$18\%  of the FS crossing timescale. This is due to the drastically compressed radial width of shell S1 post-FS passage. Since the strengths of both shocks are equal, so is the sound speed at  regions R2 and R3,  
($\beta_\mathrm{3rf-} = \beta_\mathrm{2rf-}$). At the instant the ($-$) rf wave reaches CD, the separation between the CD and the RS is given by $\Delta_{3} = \Delta_\mathrm{3f}(t_\mathrm{FS}+t_\mathrm{2rf-})/t_{\rm RS}$ where $\Delta_\mathrm{3f}\approx\frac{1}{2}\Delta_{4,0}$ is the (hypothetical) width of region R3 upon complete crossing of S4 by the RS (which is prevented by the ($-$) rf).  The time taken by the $(-)$ rf propagating into region R3 to catch up with the RS is 
\begin{equation}
   \begin{split}
        t_\mathrm{3rf-} &\ = \frac{\Delta_3/c }{\beta_\mathrm{RS} - \beta_\mathrm{3rf-}} \approx \frac{\Delta_{4,0}/a_\mathrm{u} c}{(\beta_\mathrm{RS} - \beta_\mathrm{3rf-})} \; \frac{(3+\sqrt{3})}{4}
        \\ &\ \approx 
        \frac{(\sqrt{3}+1)}{2}\frac{\Gamma^2_1 \Delta_{4,0}}{c}\approx \frac{(\sqrt{3}+1)}{4}t_\mathrm{FS}\approx 0.683\,t_\mathrm{FS}\ .
   \end{split}
\end{equation}
Thus, the $(-)$ rf wave propagating into region R3 catches up with the RS in around 68\% of $t_\mathrm{FS}$. The fraction of mass in shell S4 swept by the RS before it is halted is given by
\begin{equation}
    \alpha_{3} = \frac{t_\mathrm{FS} + t_\mathrm{2rf-}+ t_\mathrm{3rf-}}{t_\mathrm{RS}} \approx \frac{2+\sqrt{3}}{a_\mathrm{u} } \approx \frac{3.73}{a_\mathrm{u}}   \ll 1 \quad \text{(for $a_\mathrm{u} \gg 1$)\ ,} 
\end{equation}
which shows the RS is halted by the backward propagating rf wave very close to the CD. The shocked fraction $\alpha_3$ must be an invariant in all frames of reference (as shown in Eq. I20-21 of Appendix I where the analysis has been performed in the CD frame.) 

The internal energy generated at the FS and RS, with weighting factors $\alpha_2 = 1$ and $\alpha_3 = 3.73/a_\mathrm{u} $, respectively, are
given by
\begin{equation}
    \begin{split}
        &\ E_\mathrm{int,2} \approx \frac{2}{3} {a_\mathrm{u}}{E_\mathrm{k,1,0}} = \frac{2}{3} \frac{E_\mathrm{k,4,0}}{a_\mathrm{u}}\ ,\quad  \frac{E_\mathrm{int,3}}{E_\mathrm{k,4,0}} \approx \frac{2(2+\sqrt{3}) }{3a_\mathrm{u} }\approx\frac{2.48}{a_\mathrm{u} }   .
    \end{split}
\end{equation}
Thus, the thermal efficiency for the relativistic FS and RS ($a_\mathrm{u}\gg 1$) for a collision of two equal proper density and radial width shells is given by
\begin{equation}
    \epsilon_\mathrm{2,th} \approx \frac{2}{3 a_\mathrm{u}} \ll 1 \hspace{1cm}; \hspace{1cm} \epsilon_\mathrm{3,th} \approx \frac{ 2.48}{a_\mathrm{u}} \ll 1 
\end{equation}

To summarize, for $f=1$ collision while both shock fronts are relativistic, the thermal efficiency for both shock fronts is much less than unity. The RS persists till timescales $\sim$ 1.87 times that of the FS crossing timescale.  

\subsection{Collision between two Newtonian shells} \label{ultra_newt} 

\begin{figure}
    \begin{tabular}{l}
     \includegraphics[scale=0.55]{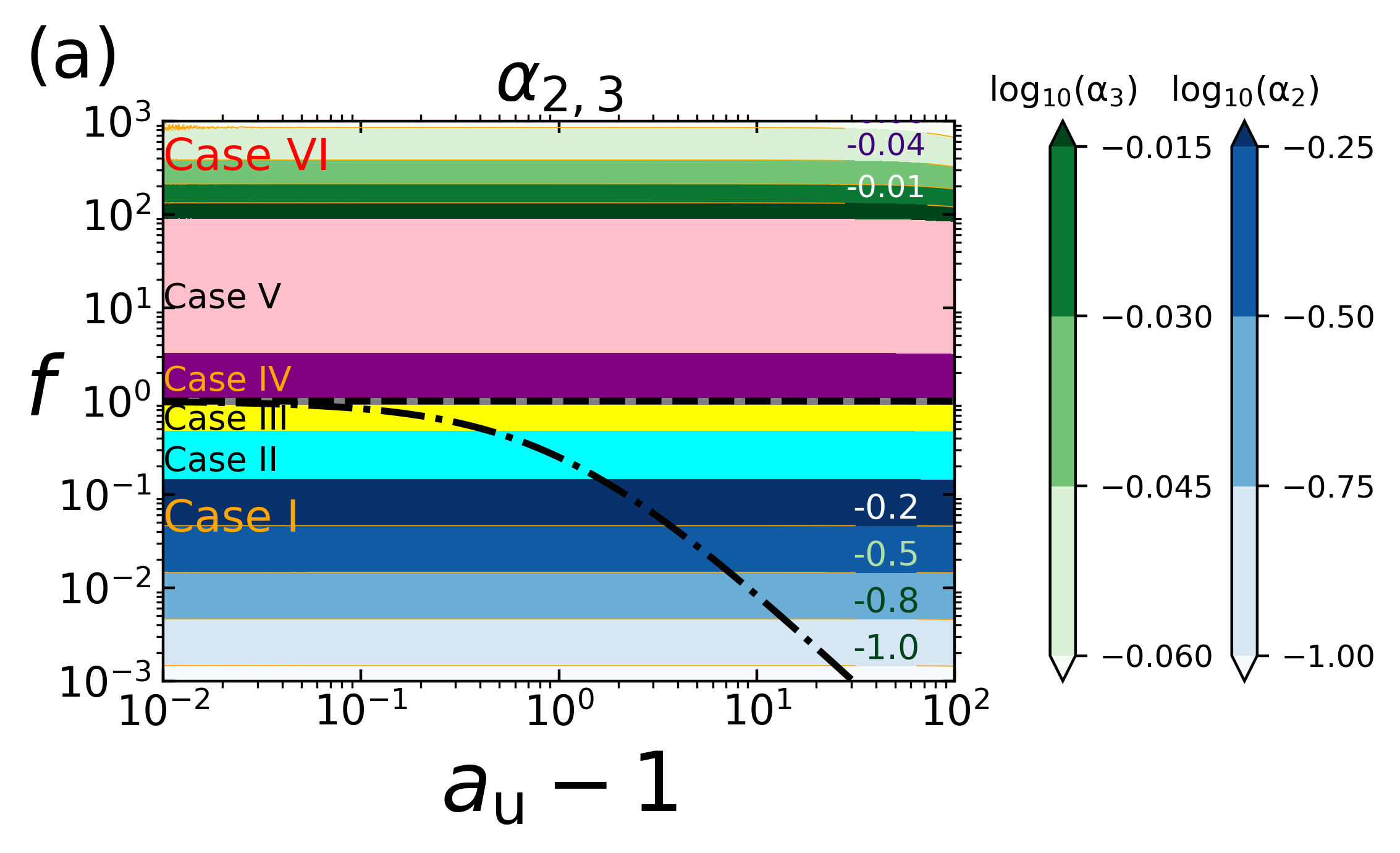} \\
     \vspace{-0.3cm}
     \includegraphics[scale=0.55]{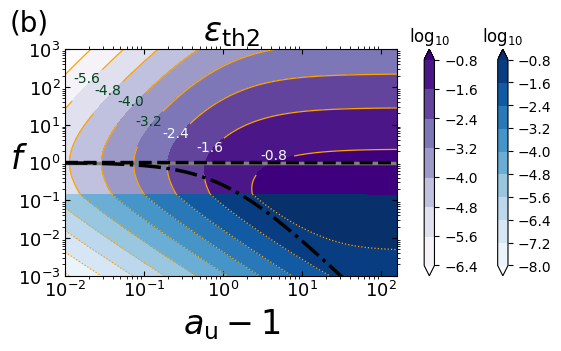} \\ 
     \vspace{-0.3cm}
     \includegraphics[scale=0.55]{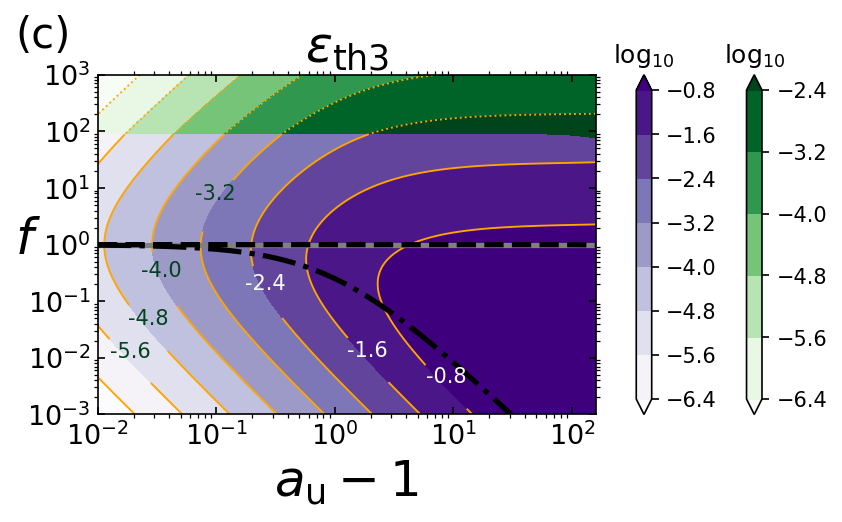} \\
     \vspace{-0.3cm}
     \hspace{-0.017cm}\includegraphics[width=0.46\textwidth]{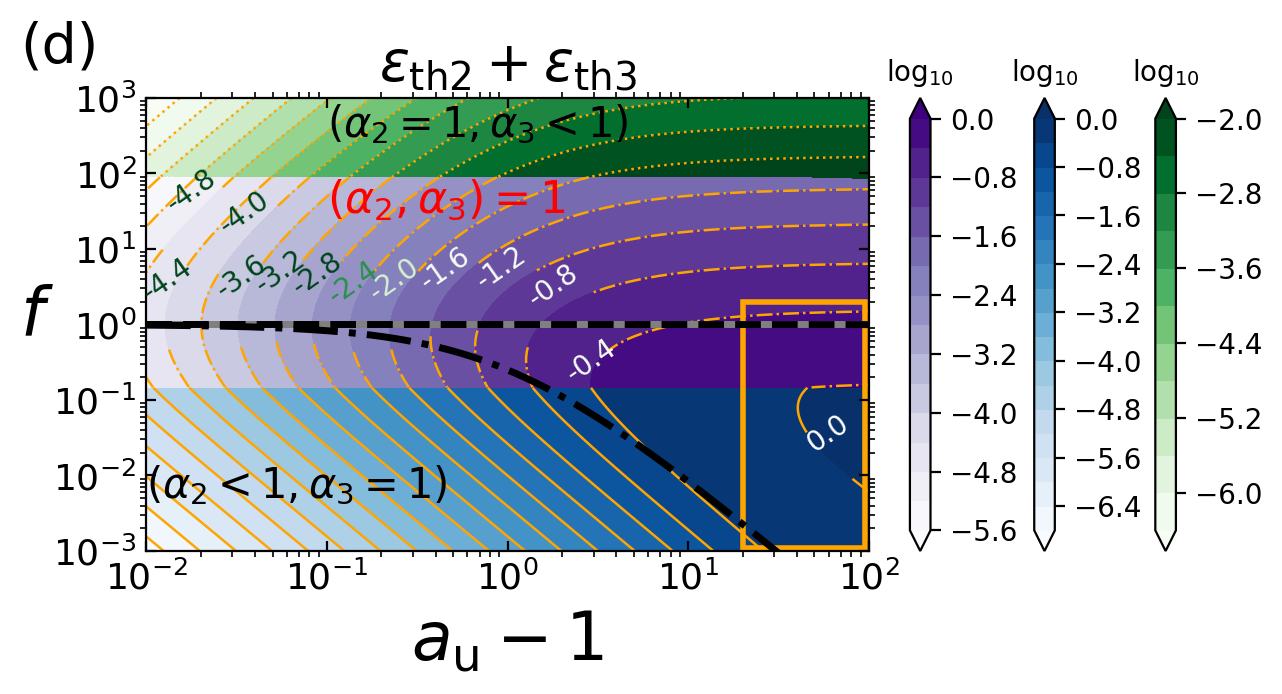} \\
     \end{tabular} \vspace{-0.2cm}
    \caption{The figure corresponds to collision of \textbf{Newtonian} shells of equal initial radial widths ($\chi=1$) for a fixed proper speed of $u_1 = 10^{-3}$. The orange rectangle at the bottom right corner in panel (d) is zoomed in Fig. \ref{newt_part}. 
    }
    \label{newt_critical}
\end{figure}

\begin{figure}
    \centering
    \includegraphics[width=\columnwidth]{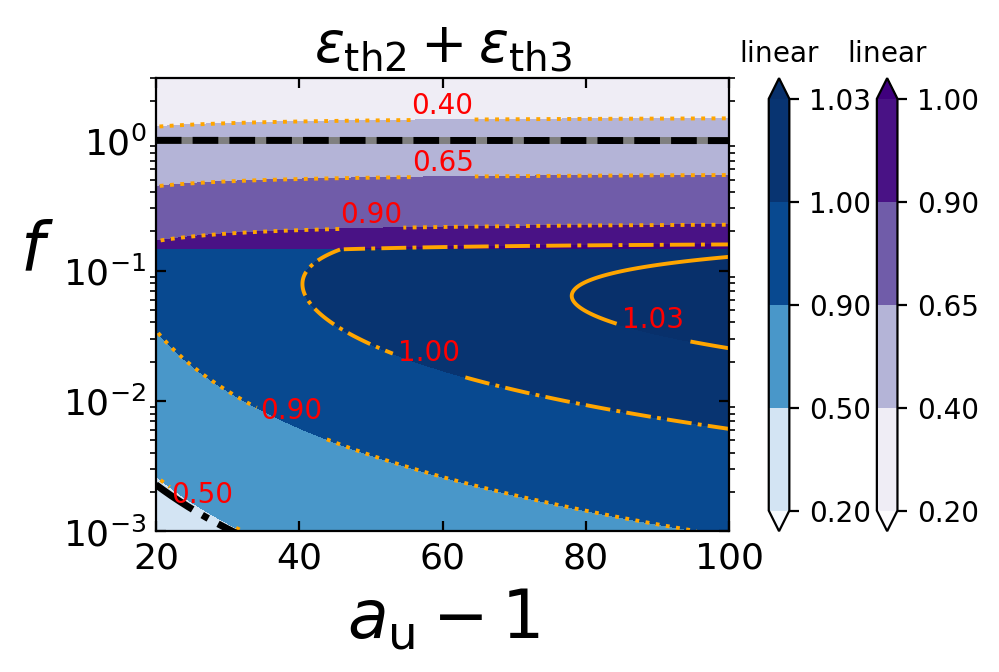}\vspace{-0.7cm}
    \caption{A part of panel (d) from Fig. \ref{newt_critical} where the combined efficiency of both shocks is equal to unity. Note the contour plot is linear in scale.} 
    \label{newt_part}
\end{figure}

 Fig.~\ref{newt_critical} shows the parameter space of $(\alpha_2, \alpha_3)$ (Panel (a)), as well as $\epsilon_\mathrm{th2}$,  $\epsilon_\mathrm{th3}$ and $\epsilon_\mathrm{th,tot}=\epsilon_\mathrm{th2}+\epsilon_\mathrm{th3}$ (panels (b), (c) and (d)), for collision of  two Newtonian shells ($u_1<u_4\ll1$) of equal initial radial width ($\chi=1$). The 5 critical lines divide the $f$\,--\,$a_u$ parameter space of proper density contrast $f$ and proper speed contrast $a_u$ into six cases. It can be seen that the equal mass collision corresponds to the $f=1$ case and lies on top of the L3 line defined by $t_{\rm RS}=t_{\rm FS}$ (i.e. dividing between cases III and IV). The equal energy  collision corresponds to case III at low $a_\mathrm{u}$ values, transitions to case II at moderate $a_\mathrm{u}$ values and finally at intermediate to high $a_\mathrm{u}$ values it enters the case I regime. Fig.~\ref{newt_part} shows a zoomed in version of the parameter space presented in panel (d) of Fig.~\ref{newt_critical}, where the total thermal efficiency of both shocks equals and marginally exceeds unity (see \S \ref{upp_lim}).

\subsection{Comparison of dissipated energy with plastic collision case}

Out of convenience and simplicity, the collision of two shells is often approximated as a
plastic collision of two infinitely thin shells \citep[e.g.][]{Kobayashi+97,Daigne-Mochkovitch98,Guetta+01,Kobayashi-Sari01,Tanihata+03,Barraud+05,Granot+06,Suzuki-Kawai06,Krimm+07,Jamil+10}. In this case, the merged shell's Lorentz factor is
\begin{equation}
\Gamma =    \frac{\Gamma_{1} M_{1,0} + \Gamma_{4} M_{4,0} }{\sqrt{M^2_{1,0} + M^2_{4,0} + 2 \Gamma_{41} M_{1,0} M_{4,0}}}\ ,
\end{equation}
where $\Gamma_{41} = \Gamma_1 \Gamma_4 (1 - \beta_1 \beta_4)$ and the total initial and final kinetic energies are
\begin{eqnarray}\nonumber
E_{\rm k,0}&=&(\Gamma_{1}-1)M_{1,0}c^2+(\Gamma_{4}-1)M_{4,0}c^2\ ,\quad\quad
\\
E_{\rm k,f}&=&(\Gamma-1)(M_{1,0}+M_{4,0}) c^2\ ,
\end{eqnarray}
the internal energy produced by the collision
\begin{equation}
E_{\rm int} = E_{\rm k,0}-E_{\rm k,f}=\left[\Gamma_{1}M_{1,0}+\Gamma_{4}M_{4,0}-\Gamma(M_{1,0}+M_{4,0})\right]c^2\ ,
\end{equation}
 is dissipated, and the thermal efficiency is given by
\begin{equation}
 \epsilon_\mathrm{th,ball} = \frac{E_\mathrm{int}}{E_{\rm k,0}} =
 1-\frac{E_{\rm k,f}}{E_{\rm k,0}}= 1 - \frac{(\Gamma-1)\left( 1 + \frac{M_{4,0}}{M_{1,0}} \right)}{(\Gamma_1 - 1) + (\Gamma_4 -1) \frac{M_{4,0}}{M_{1,0}} }\ , 
\end{equation}

For ultra-relativistic shells ($u_4>u_1\gg1$), the thermal efficiency is given by
\begin{equation}
\epsilon_\mathrm{th,Rel,plastic} = \begin{cases} &\ 1-\frac{a_u+1}{\sqrt{2(a_u^2+1)}}   \hspace{1cm} \text{for $E_\mathrm{k,4,0} = E_\mathrm{k,1,0}$}\ , 
\\
\\
&\  \frac{(\sqrt{a_u}-1)^2}{a_u+1}  \hspace{1.7cm} \text{for $M_{4,0} = M_{1,0}$}\ , 
\end{cases} \label{plastic_effi}
\end{equation}
which for high proper speed contrast ($a_\mathrm{u}\gg1$) approaches 100\% for equal masses, but only $1-1/\sqrt{2}\approx29.3$\% for equal energies.

\begin{figure}
    \centering
    \begin{tabular}{c}
    \includegraphics[scale=0.51]{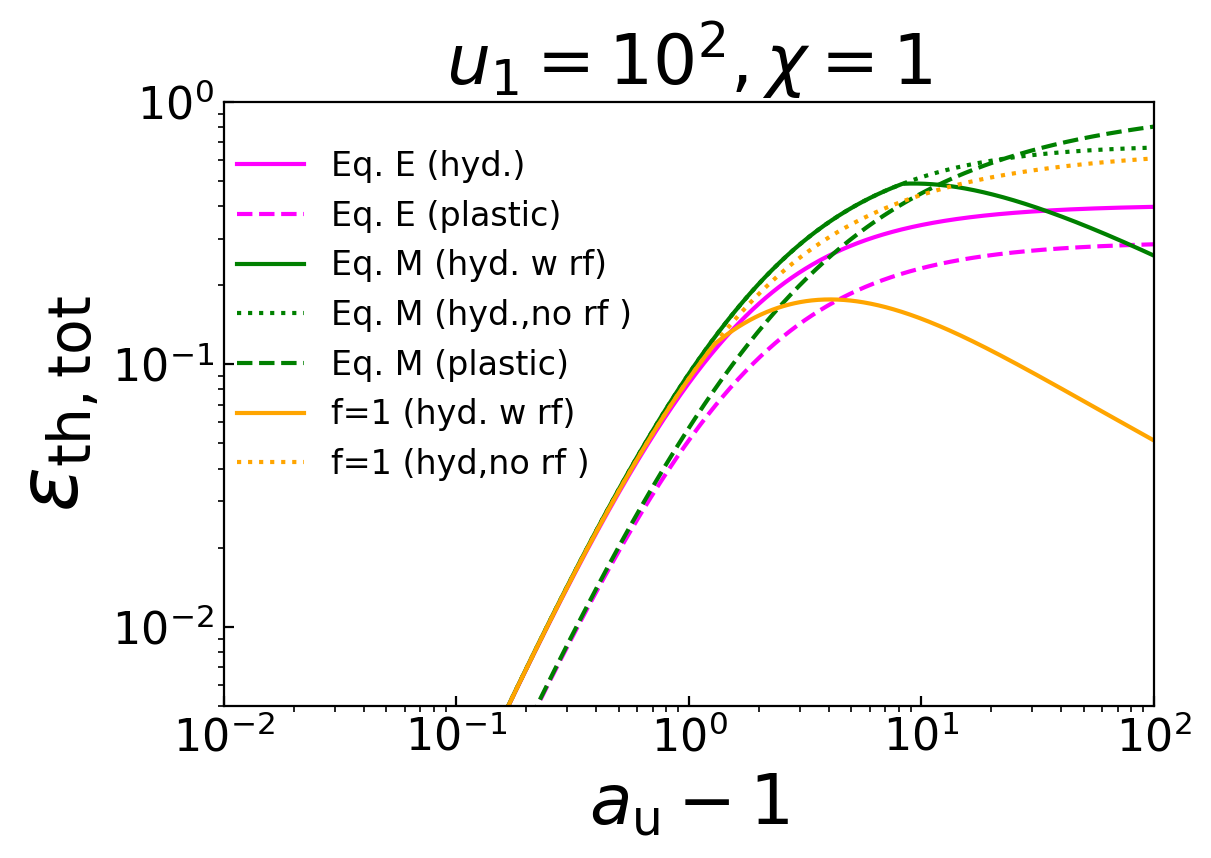} \\
    \includegraphics[scale=0.50]{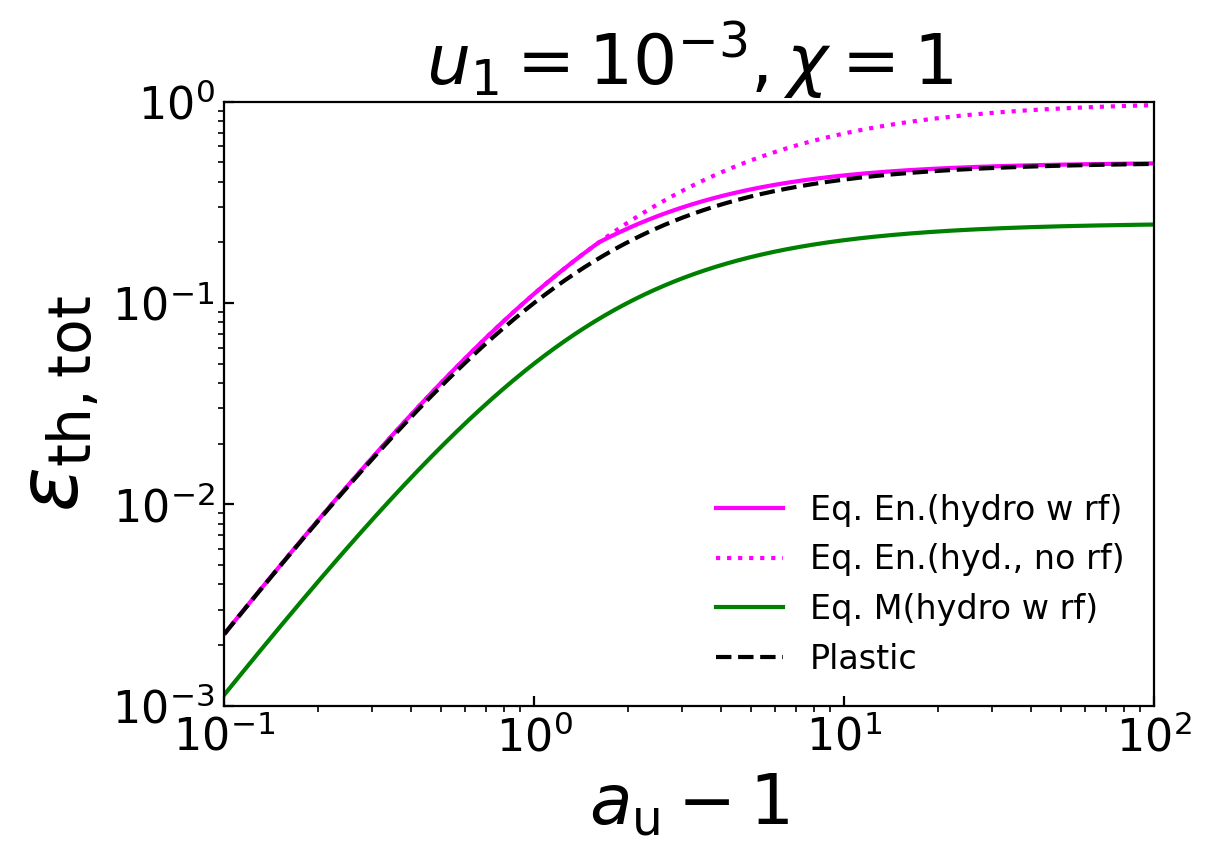}
    \end{tabular}\vspace{-0.3cm}
    \caption{Comparison of the overall thermal efficiency between the ballistic approach and the hydrodynamic approach for $\chi = 1$. The magenta, green and orange lines represent  collision between equal energy, equal mass and equal proper density shells. \textit{\textbf{Top:}} The collision of two ultra-relativistic shells with $u_{1}=10^{2}$.  The dashed magenta and green lines represent the thermal efficiency from the plastic collision approach (equation~(\ref{plastic_effi})). The solid lines represent the overall hydrodynamic efficiency (after accounting for rf wave propagation) . The dotted  lines represent the overall hydrodynamic efficiency if rf wave propagation are ignored.    \textbf{\textit{Bottom:}} The collision of two Newtonian shells with $u_\mathrm{1} = 10^{-3}$. The black dashed line represents the thermal efficiency (which is the same for equal energy and equal mass Newtonian shells) from the plastic collision approach (equation~(\ref{plastic_newt_effi})) (see text for details). }
    \label{comp_effi}
\end{figure}

Fig.~\ref{comp_effi} compares the thermal efficiency estimated from plastic collision and that estimated from shock hydrodynamics. The top panel shows that for collisions of equal energy and equal mass ultra-relativistic shells at low to moderate values of proper speed contrast $a_\mathrm{u}$, the hydrodynamic efficiency is $\sim 1.3$ times higher than the plastic collision efficiency. At higher $a_\mathrm{u}$, the trend continues for equal energy shells and the hydrodynamic efficiency saturates at $\sim 40 \%$. For equal mass shells, however, the overall hydrodynamic thermal efficiency reaches a maximum of $\sim 50\%$ and then starts decreasing monotonically at around $a_\mathrm{u} \sim 10$. This is because at higher $a_\mathrm{u} \geq 10$, the shell S4 which carries most of the initial available energy is only partially shocked due to (-)rf wave catching up with RS. The dotted green line shows the trend if rf propagation were not taken into account. Thus, we have a stark contrast for equal mass collision between the plastic approach which predicts $\sim 100 \%$ thermal efficiency at large values of $a_\mathrm{u}$  and the hydrodynamic approach which limits it at around $\sim 50 \%$. Due to partial shocking of S4, for collision of equal proper density shells, the overall hydrodynamic efficiency is capped at $\sim 10 \%$ at very moderate $a_\mathrm{u} \sim 3$. This is because for $f=1$, almost the entire initial kinetic energy is in S4.

For plastic collision of two equal energy and equal mass shells moving with Newtonian velocities , the thermal efficiency is given by
\begin{equation}
    \epsilon_\mathrm{th,ball,newt} = \frac{E_\mathrm{int}}{E_\mathrm{k,1,0} + E_\mathrm{k,4,0}} = \frac{(a_\mathrm{u} - 1)^2}{2(a^2_\mathrm{u} + 1)} \leq 0.5\ , \label{plastic_newt_effi}
\end{equation} 
which shows that for both equal mass and equal energy plastic collision, the thermal efficiency cannot exceed 50\%. 

The bottom panel of Fig.~\ref{comp_effi} represents collisions of Newtonian shells. For equal energy shells, there is partial shocking of shell S1 for $a_\mathrm{u} \geq 2$ and the overall thermal efficiency is capped at $\sim 50 \%$ at high $a_\mathrm{u}$. The plastic approximation closely follows the overall thermal efficiency of equal energy shells and is $\sim 1.2$ times higher than the overall hydrodynamic efficiency for equal mass shells.

\subsection{The upper limit on thermal efficiency}\label{upp_lim} 

The purpose of this subsection is to investigate whether the combined thermal efficiency of both shocks 
can significantly exceed unity for \textit{planar shocks}. The best-case scenario for this to happen is for $f=1$, for which the strength of both shock fronts are equal, and for ultra-relativistic shells ($u_4>u_1\gg1$) of high proper speed contrast ($a_u\gg1$), they are both relativistic. However, despite this we saw in \S\;\ref{ultra_high} that the combined thermal efficiency is still negligible, because of two factors. Firstly, the FS crossing timescale is much shorter than that of the RS, $t_{\rm FS}\ll t_{\rm RS}$. The energy of region R2, which is mostly internal, comes primarily from $pdV$ work by region R3 across the CD. In time $t_\mathrm{RS}$, about one-third of $E_\mathrm{k,4,0}$ could be transferred from S4 to S1. But since $t_{\rm FS}\ll t_{\rm RS}$ for $\chi = 1$,  a negligible fraction of this transfer actually takes place, leading to a negligible FS thermal efficiency. 
Secondly, the radial width of region R2 reduces drastically due to shock compression, allowing the backward propagating rf wave to very quickly catch up with the RS. As a consequence, much of the material in the shell S4 remains unshocked, leading to a very low RS thermal efficiency.

If we allow for the condition $\chi \geq \chi_{c3}$ (see \S \ref{sec:rarefactionmotivation}), the thermal efficiency by the RS can attain the maximum value $\epsilon_\mathrm{th,3} = \epsilon_\mathrm{th, max} = \frac{2}{3}$. Thus, next, we need to find the cases for which $\epsilon_\mathrm{th,2}$ can be maximized. Since for $(f=1,a_\mathrm{u} \gg 1)$ we have
\begin{equation}
    E_\mathrm{k,4,0} \approx \frac{a^2_\mathrm{u}}{\chi} E_\mathrm{k,1,0}\ ,
\end{equation}
which shows that the total initially available energy is entirely in shell S4 $E_\mathrm{k,0} = E_\mathrm{k,1,0} + E_\mathrm{k,4,0} \approx E_\mathrm{k,4,0}$. For $ \chi_\mathrm{c3} \leq \chi \leq \chi_\mathrm{c1} $ , the weighting factor $ (\alpha_{2},\alpha_{3}) = 1$ and we have
\begin{equation}
    \epsilon_\mathrm{th,tot} = \epsilon_\mathrm{th,3} + \epsilon_\mathrm{th,2} = \frac{2}{3} \left[ 1 + \frac{\chi}{a_\mathrm{u}}\right]  \hspace{1cm} \text{for $\chi_\mathrm{c3} \leq \chi \leq \chi_\mathrm{c1}$}\ ,\label{max_effi}
\end{equation}
and for $\chi > \chi_\mathrm{c1}$ we have the limiting value for the total thermal efficiency as 
\begin{equation}
    \epsilon_\mathrm{th,tot} = \frac{2}{3} \left[ 1 + \alpha_{2} \frac{\chi}{a_\mathrm{u}}\right] \; =  \frac{2}{3} \left[ 1 +  \frac{\chi_\mathrm{c1}}{a_\mathrm{u}} \right] \hspace{1cm} \text{For $\chi >$ $\chi_{c1}$}\ ,
\end{equation}
where we have used the definition of $\alpha_2 = \chi_{c1}/\chi$. The initial ratios of the radial widths, $\chi = (\chi_\mathrm{c3},\chi_\mathrm{c2},\chi_\mathrm{c1})$, can be estimated by equating $t_\mathrm{FS}$ to $(t_\mathrm{RS}, t_\mathrm{RS} + t_\mathrm{3rf+}, t_\mathrm{RS}+ t_\mathrm{3rf+} + t_\mathrm{2rf+})$ where 
\begin{equation}
t_\mathrm{RS} \approx \frac{a_\mathrm{u}}{2 \chi} t_\mathrm{FS}\ ,\quad \ \ 
t_\mathrm{3rf+} \approx 1.37 t_\mathrm{RS}\ ,\quad\ \   t_\mathrm{2rf+} \approx 0.71 t_\mathrm{RS}\ ,
\end{equation}
which gives $(\chi_\mathrm{c3},\chi_\mathrm{c2},\chi_\mathrm{c1}) \approx (0.50,0.90,1.25) a_\mathrm{u}$.
These values when substituted in equation~(\ref{max_effi}) give
\begin{equation}
    \epsilon_\mathrm{th,tot} \approx 
        \begin{cases}  1.00  \hspace{3cm} \text{For $\chi = \chi_\mathrm{c3}$ } \\ 
         1.26  \hspace{3cm} \text{For $\chi = \chi_\mathrm{c2}$ } 
         \\
        1.51  \hspace{3cm} \text{For $\chi = \chi_\mathrm{c1}$ } \\ 
       \end{cases} 
\end{equation}
and $ \epsilon_\mathrm{tot,th}< 1.51 $ for $\chi > \chi_\mathrm{c1}$

To summarize, the combined thermal efficiency of both shock fronts can exceed unity for $f=1$ only if the forward shock front persists longer than the RS front. The longer time allows a greater amount of the $pdV$ work to be transferred from shell S4 to S1. However, the combined thermal efficiency saturates at a maximum value of 1.5. All our estimates are based on assuming a planar geometry. The limitation of our approach is discussed in the next section.  

\section{Limitations of our Analysis}\label{sphere_geom}

The following assumptions have been made in the course of our analysis. Firstly, we have used the \textbf{ planar geometry} approximation. Under this approximation, all physical quantities remain homogeneous and unchanged in regions (R1, R2, R3, R4). The planar approximation breaks when the radius reaches about twice its value at $t_{\rm o}$, i.e. at $R\gtrsim2R_{\rm o}$. 
Beyond this, spherical geometrical effects need to be taken into account. In spherical geometry, the proper speed of the shocked fluid in regions R2 and R3, remains continuous across the CD but develops a radial profile in proper speed with a positive gradient in the radially outward direction.  As an illustrative example, we consider the collision of equal energy ultra-relativistic shells of equal initial radial width. Since both shells are ultra-relativistic, the assumption of equal initial radial width is similar to assuming equal ejection timescale $t_\mathrm{on}$ for both shells. The collision radius $R_\mathrm{o}$ is given by 
\begin{equation}
    R_\mathrm{o} = \frac{\beta_1 \beta_4  c t_\mathrm{off}}{\beta_4 - \beta_1} \approx \frac{2 a_\mathrm{u}^2}{a_\mathrm{u}^2-1} \Gamma^2_1 c t_\mathrm{off}  \hspace{1.2cm} \text{for $\Gamma_4>\Gamma_1\gg 1$\ ,} 
\end{equation}
such that the radius doubles in a lab-frame time $t_\mathrm{2R} \approx R_\mathrm{o}/c \approx 2 (1-a_u^{-2})^{-1} \Gamma^2_1 t_\mathrm{off}$ such that for $a_\mathrm{u} \gg 1$ we have $R_\mathrm{o}/c \sim 2 \Gamma^2_1 t_\mathrm{off}$. From equations~(\ref{cross_RS})-(\ref{cross_FS}) for collision of equal energy shells at $a_\mathrm{u} \gg 1$, we have $ t_\mathrm{FS} = \frac{5}{3} t_\mathrm{RS} = \frac{5}{3} (2 \Gamma^2_1 t_\mathrm{on})$. Requiring $t_\mathrm{2R} = t_\mathrm{FS}$ gives  $t_\mathrm{off} \sim \frac{5}{3} t_\mathrm{on}$, which if satisfied means the planar assumption is approximately valid till the time FS takes to reach the edge of shell S1. Secondly, we have assumed that there is \textbf{no spread in the proper speed}
of the shells S1 and S4. For ultra-relativistic shells, if there is a spread in the Lorentz factor of the shells, their radial width $\Delta$ increases compared to its initial value $\Delta_\mathrm{o}$ as the shells move away from the central engine such that $\Delta\sim\Delta_\mathrm{o}+R/\Gamma^2$ for a spread $\Delta\Gamma\sim\Gamma$, and the shell increases its radial width significantly at a radius $R_\Delta\sim\Delta_\mathrm{o}\Gamma^2$. For a small proper speed spread, $\Delta\Gamma/\Gamma\ll1$, we have $\Delta\sim\Delta_\mathrm{o}+R\Delta\Gamma/\Gamma^3$ and $R_\Delta\sim\Delta_\mathrm{o}\Gamma^3/\Delta\Gamma$. Besides, one could also consider a realistic situation where the source power and asymptotic LF smoothly varies with ejection time, leading to spontaneous formation of shocks whose strength varies with radius. This will be explored in a follow-up work. Thirdly, we have assumed \textbf{no radiative losses} in our analysis. We have assumed that total energy post-collision is the summation of kinetic and internal energy only. Lastly, we have assumed \textbf{cold shells}. \citealt{Peer+17} pointed out that if the shells were to be hot, then no shocks would be generated if the proper speed contrast does not exceed a critical value. We note, however, that in spherical geometry the shells cool adiabatically on the radius doubling time, so they are expected to greatly cool before reaching $R_{\rm o}$, and also significantly cool  between subsequent collisions.  

In the next section, we explore a few representative astrophysical scenarios where our analysis can be applied to understand some generic features.

\section{Application to few representative cases}\label{appl_cases}

In the following subsections, we explore the internal shocks parameter space for several astrophysical scenarios. In each subsection, we briefly introduce the model associated with the astrophysical transient and then make some general remarks.

\subsection{GRB prompt emission internal shocks model}\label{app_1}

One of the leading models for producing the extremely bright, short-lived and highly time-variable prompt $\gamma$-ray emission in GRBs features internal shocks. The latter may naturally arise from time-variability in the central source's activity that leads to variations in the asymptotic Lorentz factor (that is reached at large distances from the source) of the ultra-relativistic outflow that it launches. Faster pasts of the outflow catch up with slower parts and collide with them, each collision creating a pair of shocks: FS and RS.

The typical inferred parameter values in such prompt GRB internal shocks models are: $10^2\lesssim u_1\lesssim10^{2.5}$, $10^{-1}\lesssim a_{u} - 1\lesssim10$, $10^{-0.5}\lesssim\chi\lesssim10^{0.5}$. 
While the prompt GRB emission is highly variable, consisting of multiple sharp spikes, when averaging over these spikes there is no clear temporal trend, e.g. the fluences in the first and second halves of the prompt GRB emission episode appear to be similar. This suggests an approximately constant power of the outflow emanating from the central engine during its activity period. The time between pulses in the prompt GRB lightcurve is typically comparable to the pulse widths, suggesting that $t_{\rm off}\sim t_{\rm on}$ (see \citealt{Nakar2002}). This suggests that shells are ejected with roughly similar kinetic energy at very short intervals.   

For the collision of equal energy and equal mass shells moving at ultra-relativistic speeds, the RS is relativistic and dominates the thermal efficiency. At very large proper speed contrast $a_\mathrm{u}$, for collision of equal energy shells, the overall efficiency of $\sim$ 40 $\%$ while RS (ultra-relativistic strength) dissipates internal energy $\sim$ 5 times more efficiently than the FS (mildly sub-relativistic strength).  For equal mass collision, the overall efficiency reaches a maximum of $\sim 50 \%$ and actually decreases at very high proper speed contrast due to partial shocking of the trailing shell S4. For equal mass collision, the RS is ultra-relativistic and FS is mildly-relativistic.
The inferred prompt $\gamma$-ray efficiencies in GRBs, of order $\sim 15\%$ \citep{Beniamini2015}, are consistent with these values, considering that there is a further efficiency reduction between dissipated energy and observed $\gamma$-rays. Recently,  \cite{2024MNRAS.528L..45R} has shown that the variability in the lightcurves and the spectrum of GRBs can be explained when contributions from both shocked regions are taken into account.

\subsection{FRB blastwave model}\label{app_2}
One class of fast radio burst (FRB) models involves synchrotron maser emission from relativistic outflows. There are different variants of this model. We discuss below two of these, which involve different types of shocks.

\subsubsection{Model 1 of fast radio bursts}

Model 1: (internal collisions between magnetar giant flare outflows) $10^{1.5}\lesssim u_1 \lesssim10^{2.5}$, $a_{u} - 1 \sim 1$, equal energy, refer to \S\;\ref{app_1} 

This model involves the collision of two ultra-relativistic shells at moderate proper speed contrast. Here at moderate values of $a_\mathrm{u} \sim 2$, the RS is still stronger than the FS while the overall efficiency is $\sim 10 \%$. We note that this efficiency reduction comes in addition to the already tiny estimated efficiency in this model resulting from: (i) the efficiency of converting shock heated plasma to maser radiation, (ii) the efficiency loss due to the requirement that the optical depth for induced Compton close to the peak of the observed spectrum should not be too large, (iii) the efficiency loss due to the requirement that the bursts could reproduce the high observed level of temporal and spectral variability and (iv) the efficiency suppression in magnetar models due to the fact that escaping outflow should be moving along open field lines \citep{2019MNRAS.485.4091M,BK2020,BK2023}.

\subsubsection{Model 2 of Fast radio bursts}

\begin{table}
    \centering
    \caption{Parameters for model 2 of the FRB blast wave model}
    \begin{tabular}{c|c|c} \hline 
       Quantity  & Description & Typical values \\ \hline 
        $u_\mathrm{1,w}$  & proper speed of wind shell S1 &  0.5 \\
        $a_\mathrm{u,ej}$    & ratio of proper speed of ejecta $u_\mathrm{4,ej}$ to   $u_\mathrm{1,w}$ & 100 \\
        $t_\mathrm{on1}$ & Wind shell S1 ejection timescale & $\sim 10^{5}$ s \\ 
        $t_\mathrm{on4}$ & Ejecta shell S4 ejection timescale & $\sim 10^{-4} - 10^{-3}$ s \\ 
        $\Dot{M}_{1}$ & Mass injection rate for wind shell 1 & $10^{19} - 10^{21}$ g/s \\ 
        $E_\mathrm{k,4,0}$ & The initial kinetic energy of ejecta S4 & $10^{43} - 10^{45}$ erg \\  
        $\delta$ & Ratio of $r$ to $r_\mathrm{s}$ & $10^{-3}$ \\ \hline 
    \end{tabular}
    \label{FRB2}
\end{table}

This model proposed by \citet{2019MNRAS.485.4091M} requires the collision of an ultra-relativistic shell S4 with a mildly relativistic shell S1. It has the following set-up. The central engine injects a mildly relativistic wind of proper speed $u_\mathrm{1,w}$  with a mass loss rate of $\Dot{M}_{1}$ for time $t_\mathrm{on1}$. The material injected is uniformly spread  up to a radius $r_\mathrm{s} = v_\mathrm{w} t_\mathrm{on1}$. Shortly afterward, the central engine injects an ultra-relativistic shell
over a timescale $t_\mathrm{on4}$ with Lorentz factor $\Gamma_\mathrm{4,ej}$ and kinetic energy $E_\mathrm{k,4,GF}$. The collision takes place at a distance $r \ll r_\mathrm{s}$ from the  central engine such that we have
\begin{equation}
    \delta = \frac{r}{r_\mathrm{s}} \ll 1\ ,  
\end{equation}
where  $r_\mathrm{s} = \beta_\mathrm{1,w} c t_\mathrm{on,1} = v_\mathrm{1,w} t_\mathrm{on,1}$. The typical values of the parameters for this model are summarized in Table \ref{FRB2}. 
The lab frame density of the wind (shell S1) for $\delta \ll 1$ is given by
\begin{equation}
    n_{1} =  \frac{3 \Dot{M} t_\mathrm{on,1}}{4 \pi m_\mathrm{p} r^3_\mathrm{s}} = \frac{3 \Dot{M} }{4 \pi m_\mathrm{p} c^3 \beta^3_\mathrm{1,w} t^2_\mathrm{on,1} }\ . 
\end{equation}
The proper number density of the wind shell S1 is given by
\begin{equation}
    n'_\mathrm{1,w} = \frac{n_{1}}{\Gamma_\mathrm{1,w}} = \frac{3 \Dot{M} }{4 \pi m_\mathrm{p} c^3 \; u_\mathrm{1,w} \beta^2_\mathrm{1,w} t^2_\mathrm{on1} }\ . \label{wind_density}
\end{equation}
The proper number density of the ejecta shell S4 is given by
\begin{equation}
\begin{split}
    n'_\mathrm{4,ej} &\ = \frac{E_\mathrm{k,4,0}}{t_\mathrm{on,4}} \frac{1}{4 \pi \delta^2 r^2_\mathrm{s} m_\mathrm{p} c^3} \frac{1}{\Gamma^2_\mathrm{4,ej}} \\
    &\ = \frac{1}{\delta^2} \left( \frac{E_\mathrm{k,4,0}}{t_\mathrm{on4}} \right) \frac{1}{4 \pi m_\mathrm{p} c^5  \beta^2_\mathrm{1,w} t^2_\mathrm{on1}} \frac{1}{\Gamma^2_\mathrm{4,ej}}\ .
\end{split} \label{ejecta_density}
\end{equation}


Using equations (\ref{wind_density})-(\ref{ejecta_density}) the proper density contrast $f$ can be expressed as  
\begin{equation}
\begin{split}
      &\ f  = \frac{n'_\mathrm{4,ej}}{n'_\mathrm{1,w}}  = \frac{\eta}{\delta^2}  \left( \frac{E_\mathrm{k,4,0}}{E_\mathrm{k,1,0}} \right) \left[ \frac{u_\mathrm{1,w} (\Gamma_\mathrm{1,w} - 1)}{3 \Gamma^2_\mathrm{4,ej}}\right] \\
      &\ \approx 3 \times 10^4 \; \eta_{8} \; \delta^2_{-3} \; E_\mathrm{ej,43} \; E^{-1}_\mathrm{w,46} \; u^3_\mathrm{w,-0.3} \Gamma^{-2}_\mathrm{ej,4}\ ,  
\end{split}  
\end{equation} 
where $\eta_{8} = \eta/ 10^{8} = (t_\mathrm{on1}/10^5\,\mathrm{s}) (t_\mathrm{on4}/10^{-3}\,\mathrm{s})$, $\delta_{-3}=\delta/10^{-3}$, $E_\mathrm{ej,43}=E_\mathrm{k,4,0}/10^{43} \,\mathrm{erg}$, $E_\mathrm{w,46} = E_\mathrm{k,1,0}/10^{46} \mathrm{erg} = (\Dot{M}/ 10^{21}\,\mathrm{g/s} ) (t_\mathrm{on1}/10^{5}\,\mathrm{s})c^2$, $u_\mathrm{w,-0.3}=u_\mathrm{1,w}/0.5$ and $\Gamma_\mathrm{ej,2} = \Gamma_\mathrm{4,ej}/100$.  

This case corresponds to the external shock scenario wherein the forward shock is relativistic. Since the FS shock strength is ultra-relativistic almost all the initially available energy in shell S4 is reprocessed into the thermal energy of shell S1. However, the radiated energy can be much lower due  an efficiency of converting only a fraction of the internal energy into energy of non-thermal electrons. This is in addition to the efficiency factors alluded to in the previous subsection.

\subsection{Deceleration of ejecta from SLSN by collision with a pre-ejected massive shell}\label{app_3}

Superluminous Supernova(SLSNe) are the brightest among core-collapse supernova. In a matter of few months, the radiated energy is close to $\sim 10^{50} -10^{51}$ ergs, comparable to the kinetic energy of associated with standard supernova explosion $\sim 10^{51}$ ergs. This in turn requires that the kinetic energy of explosion be turned into radiation very early on and very efficiently. To achieve the same, interaction powered models (see \citealt{2018SSRv..214...59M}) have been proposed involving  the collision of two shells moving at Newtonian velocities. In this model, a massive progenitor star suffers two episodic instability events spaced a few years apart. In the first event it ejects a massive shell  $M_{4,0} \sim\textrm{few}\times M_\odot$ 
at speeds of $\sim10^{3.5}\;$km/s. A few years later, in a second episodic event another less massive but faster shell is ejected. The second shell has comparable kinetic energy to the first shell. Typical values of the parameters of this model are summarized in Table \ref{SLSN}.   
The proper density contrast $f$ is given by
\begin{equation}
    f \approx \chi \frac{M_{4,0} }{M_{1,0} } \sim 0.2 \; \chi \; \left( \frac{M_{4,0}}{5 M_{\odot}}\right) \; \left( \frac{M_{1,0}}{25 M_{\odot}}\right)^{-1}\ .  
\end{equation}
In this case, while RS tends to be stronger, the thermal efficiencies associated with both shock fronts are comparable $\sim 25 \%$. 

\begin{table}
    \centering
    \caption{Parameter space for interaction of SLSN ejecta with a pre-ejected massive shell (The parameters are quoted for SN 2006 gy)}
    \begin{tabular}{c|c|c} \hline 
    Quantity & Description & Typical values \\ \hline
    $M_{1,0}$ & Mass of pre-ejected shell 1 & $24.5 M_{\odot}$ \\ 
     $M_{4,0}$ & Mass of SLSN shell 4 & $5.1 M_{\odot}$ \\
    $E_\mathrm{k,1,0}$ & kinetic energy of  shell 1 & $1.4 \times 10^{50}$ erg \\
     $E_\mathrm{k,4,0}$ & kinetic energy of  shell 4 & $6 \times 10^{50}$ erg \\ \hline 
    \end{tabular}
    \label{SLSN}
\end{table}

\subsection{Deceleration of magnetar Giant flare by bow-shock shell}\label{app_4}

\begin{figure}
    \centering
    \includegraphics[scale=0.52]{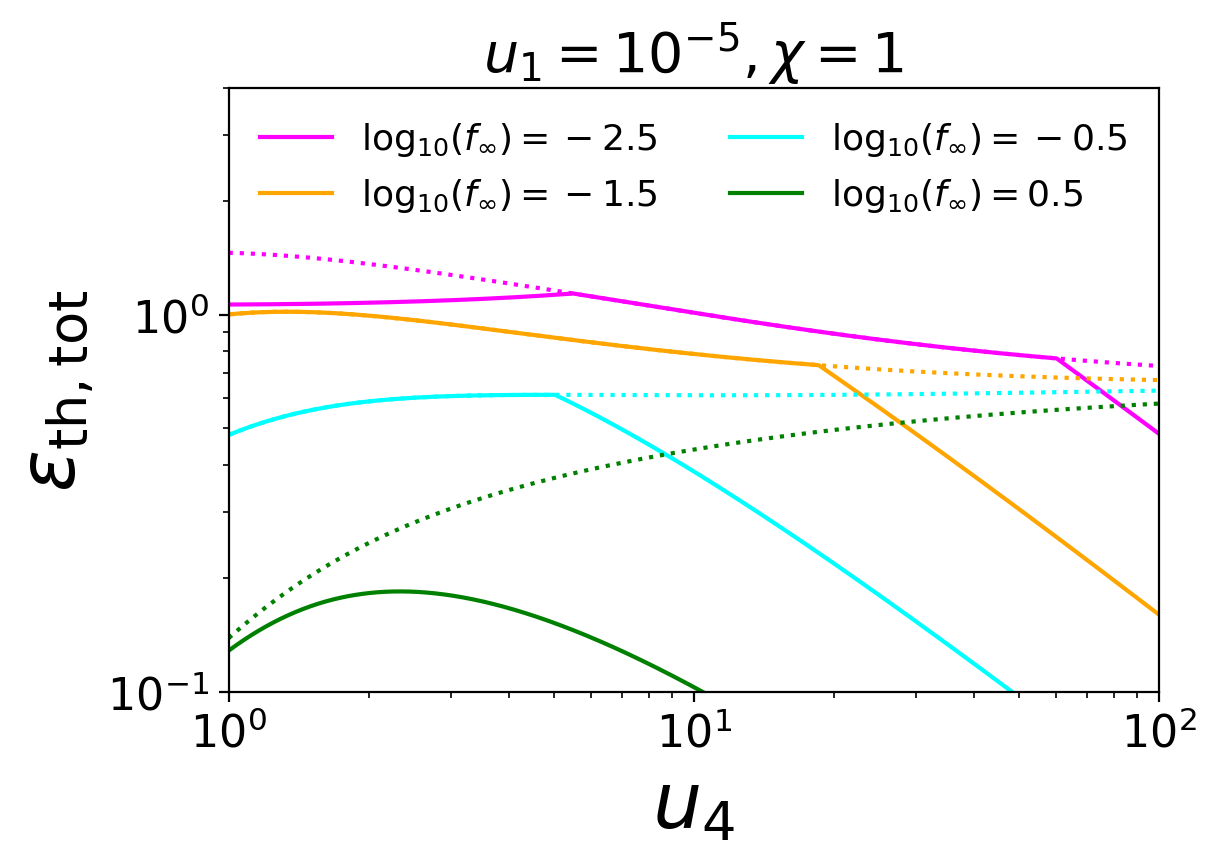} \vspace{-0.2cm}
    \caption{Hydrodynamic thermal efficiency for collision of  magnetar giant flare shell S4 of isotropic equivalent kinetic energy $E_\mathrm{GF,46}$ with a newtonian bow-shock shell S1 with proper number density $4 n_\mathrm{o}$. Radial widths of both shells are taken to be equal $\chi = 1$. The dotted lines show efficiencies without consideration of rf waves. The magenta, orange, cyan and green lines correspond to $\log_{10}(f_{\infty}) = -2.5, -1.5, -0.5, 0.5$ (see text for more details).    }
    \label{therm_bow}
\end{figure}

\begin{table}
    \centering
    \caption{Parameter space for interaction of magnetar giant flare with bow-shock shell}
    \begin{tabular}{c|c|c} \hline 
      Quantity & Description & Typical values \\ \hline  
       $v_\mathrm{NS}$ &  velocity of neutron star & 300 km/s \\ 
       $L_\mathrm{sd}$ & Spin-down luminosity of the neutron star & $10^{34.5}$ erg/s \\
       $E_\mathrm{k,4,GF}$ & Outflow isotropic equivalent kinetic energy  & $10^{44}\!-\!10^{46}\,$erg \\
       $t_\mathrm{on,4}$ & The time taken for shell 4 to be ejected & $10^{-0.5}$ s\\
       $n$ & The typical particle number density in ISM & 1 cm$^{-3}$ \\ 
       $n_\mathrm{1,bS}$  & Number density in bow-shock shell & 4$n$ \\ \hline  
    \end{tabular}
    \label{bowShock}
\end{table}

This model involves the collision of a mildly-relativistic \citep[$u_4\sim1$;][]{Gaensler+05,Gelfand+05,Granot+06} up to an ultra-relativistic shell \citep[$u_4\sim100$;][]{Sculptor-GF-21} S4, with a stationary shell S1. The setup of the model is as follows: In  pulsars  most of the spin-down power is carried by an ultra-relativistic MHD wind ($L_\mathrm{w} \approx L_\mathrm{sd}$). The pulsar itself has a  systemic velocity $v_{\rm NS}\sim10^{2.5}\;$km/s relative to the ISM \citep{2005MNRAS.360..974H, 2019ApJ...877...78S,2022ApJ...932..117L}. The pulsar wind interacts with the ISM leading to the formation of a bow shock shell. The lab frame is identified with the bow shock shell. In the lab frame, the steady state radius of the bow shock shell is determined by the  balance of the ram pressure due to pulsar wind and the ram pressure due to ISM. During a flaring event, the  magnetar gives rise to a giant flare, ejecting an outflow on timescales of $\sim10^{-0.5}\;$s of (isotropic equivalent) kinetic energy $E_\mathrm{k,4,GF}$, which can then collide with the bow shock shell. The typical parameters for this model are summarized in Table \ref{bowShock}.      

The radius of the (head of the) bow shock shell can be obtained by equating the ram pressure due to ISM ($\rho v^2_\mathrm{NS}$) and the pulsar wind $(L_\mathrm{sd}/4\pi R^2_\mathrm{bs}c)$ as 
\begin{equation}
    \begin{split}
        R_\mathrm{bs} = \frac{1}{v_\mathrm{NS}} \sqrt{ \frac{L_\mathrm{sd}}{4 \pi m_\mathrm{p} c n }  } \approx 7.08 \times 10^{15} \; n^{-1/2}_\mathrm{o} \;  v^{-1}_\mathrm{NS,2.5} \;  L^{1/2}_\mathrm{sd,34.5}\,\text{cm}\ ,
    \end{split}
\end{equation}
where $n_\mathrm{o} = n/(1\,\mathrm{cm}^{-3})$,  $v_\mathrm{NS,2.5} = v_\mathrm{NS}/(10^{2.5}\,  \text{km/s})$  and $L_\mathrm{sd,34.5}= L_\mathrm{sd}/(10^{34.5}\,\text{erg/s})$.

The initial radial width of the giant flare shell S4 is given by $\Delta_{4,0} = \beta_\mathrm{4} c t_\mathrm{on4}$. If this shell has a Lorentz factor spread $\Delta\Gamma_4\sim\Gamma_{4}$, then by the time it (S4) reaches the bow shock shell (S1), its radial width has expanded to 
\begin{equation}
 \Delta_\mathrm{4}  = \Delta_\mathrm{4,0} + \Psi \frac{R_\mathrm{bs}}{\Gamma^2_{4}} \approx \Psi \frac{R_\mathrm{bs}}{\Gamma^2_4}\ ,
\end{equation}
 where $\Psi$ is a factor of order unity. 

The lab frame density of Giant flare shell 4 can be estimated as
\begin{equation}
    n_\mathrm{4,GF} = \frac{E_\mathrm{k,4.GF}}{m_\mathrm{p} c^2 (\Gamma_4 - 1)} \frac{1}{V_{4}} = \frac{E_\mathrm{k,4,GF}}{m_\mathrm{p} c^2 (\Gamma_4 - 1)} \frac{1}{4 \pi R^2_\mathrm{bs}\Delta_{4}}\ .   
\end{equation}
The comoving density of Giant flare shell 4 can be estimated as
\begin{equation}
    n'_\mathrm{4,GF} = \frac{n_\mathrm{4,GF}}{\Gamma_4} =  \frac{E_\mathrm{k,4.GF}}{m_\mathrm{p} c^2 \Gamma_4 (\Gamma_4 - 1)} \frac{1}{4 \pi R^2_\mathrm{bs} \Delta_4 }\ . \label{n4GF}
\end{equation}  
For the comoving particle density in the bow shock we can use
\begin{equation}
    n'_\mathrm{1,bs} = n_\mathrm{1,bs} = 4 n\ , \label{n1BS}
\end{equation}
since $u_\mathrm{1,bs} = 0$ and the shock compression ensures that the particle density in the bow shocked region is four times the external density (for a Newtonian strong shock). Using equations~(\ref{n4GF})-(\ref{n1BS}) the proper density contrast can be estimated as
\begin{equation}\label{f_bS}
 f = \frac{n'_\mathrm{4,GF}}{n'_\mathrm{1,bs}}  =  \frac{1}{12\Psi}  \left( \frac{\Gamma_4}{ \Gamma_4 - 1} \right)  \frac{E_\mathrm{k,4,GF}}{\frac{4}{3} \pi R^3_\mathrm{bs} n m_\mathrm{p} c^2 } \equiv\frac{\Gamma_4 f_\infty}{\Gamma_4-1}\ ,   
\end{equation}
which for $\Gamma_4 \gg 1$ approaches
\begin{equation}
    \begin{split}
        f_{\infty} &\ =  \frac{1}{2 \Psi} \sqrt{\frac{\pi m_\mathrm{p} n }{c}} v^3_\mathrm{NS} L^{-3/2}_\mathrm{sd} E_\mathrm{k,4,GF} \hspace{1cm} \text{for $\Gamma_4 \gg 1$} \\
       &\  \approx 0.37 \; \Psi^{-1} \; n^{1/2}_\mathrm{o} \; v^3_\mathrm{NS,2.5} \; L^{-3/2}_\mathrm{sd,34.5} \; E_\mathrm{GF,46}\ , 
    \end{split} \label{f_infty}
\end{equation}
where  $E_\mathrm{GF,46} = E_\mathrm{k,4,GF}/(10^{46}\, \text{erg})$ is the isotropic equivalent energy of the shell ejected during the giant flare, and generally.

Equation~(\ref{f_bS}) shows that for $\Gamma_4 \gg 1$, the proper density contrast is roughly equal to the ratio of the kinetic energy in the giant flare to the rest mass energy of the ISM mass within a sphere of radius $R_\mathrm{bs}$, which is roughly the isotropic equivalent mass of the bow shock shell, $M_1$. Thus $f\propto M_1^{-1}\propto n^{-1}R_{\rm bs}^{-3}\propto n^{1/2}v_{\rm NS}^3L_{\rm sd}^{-3/2}$. Equation~(\ref{f_infty}) shows the asymptotic value $f_{\infty}$ of proper density contrast $f$ at large $\Gamma_4 \gg 1$. 

Figure~\ref{therm_bow} shows 
the hydrodynamic thermal efficiency of the collision for $\log_{10}(f_{\infty}) = -2.5, -1.5, -0.5, 0.5$. It demonstrates that $f_{\infty}\ll1$ is required for high thermal efficiency ($\epsilon_{\rm th,tot}\gtrsim0.5$) with a relativistic outflow ($u_4\gg1$).  For $f_{\infty} \ll 1$, the thermal efficiency becomes limited by partial shocking of S1 at lower $u_{4}$ and partial shocking of S4 at higher $u_4$. As $f$ approaches unity, the thermal efficiency decreases drastically as the rf wave catches up with RS very close to the CD. It must be noted that in order to get the observed radiation the thermal efficiency must be multiplied by additional efficiency factors related to conversion of internal energy to observed radiation. 

The elaborate observation of the 2004 giant flare from the Galactic magnetar SGR\,1806$-$20 imply $u_4\sim1$ and $f\sim100$ \citep{Gaensler+05,Gelfand+05,Granot+06}, implying a low thermal efficiency ($\epsilon_{\rm th,tot}\sim10^{-2}$), which is nonetheless consistent with the observations of this event. As observations imply $L_\mathrm{sd,34.5}\approx1$ over the relevant timescale before the giant flare \citep{Woods+07}, the required $f\sim100$ suggests a fairly high systemic velocity for this source, $v_{\rm NS}\sim(1-1.5)\times10^3\;$km/s, which is again consistent with observations.

On the other hand, the observation of GeV photons associated with a magnetar giant flare in the Sculptor galaxy imply $u_4\sim100$ and a high thermal efficiency \citep{Sculptor-GF-21}, which in turn require $f_{\infty}\lesssim10^{-2}$. As an illustration, for a given ($E_\mathrm{GF,46},n_\mathrm{o}$) to get $f_{\infty} = (10^{-2.5},10^{-1.5},10^{-0.5},10^{0.5})$ at a fixed $v_\mathrm{NS} = 10^{2.5}$ km/s, one requires $L_\mathrm{sd} = (10^{35.8},10^{35.1},10^{34.5},10^{33.8})$ ergs/s. Conversely, at a fixed $L_\mathrm{sd} = 10^{34.5}$ ergs/s, the required neutron star velocities would be $v_\mathrm{NS} = (10^{1.8},10^{2.1},10^{2.5},10^{2.8})$ km/s.

\section{Conclusions}
\label{sec:conclusions}

The objective of the present work was to provide a comprehensive self-consistent framework for characterizing the dynamics of shock propagation for collision between two cold shells. We find the reverse shock to be a leading candidate for internal energy dissipation for a generic parameter space for astrophysical transients. We find that the overall thermal efficiency at higher proper speed contrast is majorly affected by the rarefaction waves catching up the shock fronts and halting further dissipation of internal energy. This is not captured by the plastic collision approach which instead predicts unrealistically very high values of thermal efficiency at these limits. The analytical parameter space presented here will be useful for calibrating more computationally expensive hydrodynamical simulations. 

\section*{Acknowledgements}

This research was funded in part by the ISF-NSFC joint research program under grant no. 3296/19 (S.M.R., J.G.). PB was supported by a grant (no. 2020747) from the United States-Israel Binational Science Foundation (BSF), Jerusalem, Israel and by a grant (no. 1649/23) from the Israel Science foundation. The authors would like to thank the referee for useful comments and suggestions.

\section*{Data Availability}

No new data were generated during the analysis of this project.



\bibliographystyle{mnras}
\bibliography{References} 

\appendix
\onecolumn 


\section{Deriving the Blandford-Mckee conditions for collision between two cold shells} \label{App_A}

The \cite{BM76} (hereafter BM76) shock conditions  can be summarized as 
\begin{eqnarray}
  &\  \frac{e'_2}{n'_2} = \Gamma_{21} \; \frac{w'_1}{n'_1} \\
  &\ \frac{n'_2}{n'_1} = \frac{\hat{\gamma}_2 \; \Gamma_{21} + 1}{\hat{\gamma}_2 - 1} \\ 
  &\ \frac{e'_3}{n'_3} = \Gamma_{34} \; \frac{w'_4}{n'_4} \\
  &\ \frac{n'_3}{n'_4} = \frac{\hat{\gamma}_3 \; \Gamma_{34} + 1}{\hat{\gamma}_3 - 1}
\end{eqnarray}
where
\begin{eqnarray*}
 & w'_1 = e'_1 + p_1 =  \left( \rho'_1  c^2 + e'_\mathrm{int,1}\right) + p_1 = \left( n'_1 m_\mathrm{p} c^2 + e'_\mathrm{int,1}\right) + p_1 \\
& w'_4 = e'_4 + p_4 = \left( \rho'_4 c^2 + e'_\mathrm{int,4} \right) + p_4  = \left( n'_4 m_\mathrm{p} c^2 + e'_\mathrm{int,4} \right) + p_4 \\
& p_2 = \left( \hat{\gamma}_2 - 1 \right) e'_\mathrm{int,1} \\ 
& p_3 = \left( \hat{\gamma}_3 - 1 \right) e'_\mathrm{int,4} \\ 
& e'_2 = n'_2 m_\mathrm{p}c^2 + e'_\mathrm{int,2} \\
& e'_3 = n'_3 m_\mathrm{p}c^2 + e'_\mathrm{int,3} 
\end{eqnarray*}
such that the relative LFs are defined as
\begin{eqnarray}
   & \Gamma_{21} =  \Gamma_2 \Gamma_1 (1 - \beta_2 \beta_1) \\ 
   & \Gamma_{34} =  \Gamma_3 \Gamma_4 (1 - \beta_3 \beta_4) 
\end{eqnarray}
and where following \citet{2003ApJ...591.1075K} we assume the adiabatic constants $(\hat{\gamma}_2, \hat{\gamma}_3)$ to be  
\begin{eqnarray}
    & \hat{\gamma}_2 = \frac{4 \Gamma_{21} + 1}{3 \Gamma_{21}} \\
& \hat{\gamma}_3 = \frac{4 \Gamma_{34} + 1}{3 \Gamma_{34}}
\end{eqnarray}

Since shells 1 and 4 are cold we have, 
\begin{eqnarray}
   &\ p_1 = 0, \;  e'_\mathrm{int,1} = 0  \\
   &\ p_4 = 0, \; e'_\mathrm{int,4} = 0  
\end{eqnarray}

Using equations (A9)-(A10), the equations (A1)-(A4) gives the shock conditions for two cold shell collision as
\begin{eqnarray}
     & \frac{e'_\mathrm{int,2}}{n'_{2}} = \left(\Gamma_{21} - 1 \right) m_\mathrm{p} c^2 \\ 
     & \frac{n'_2}{n'_1} = 4 \Gamma_{21} \\
     & \frac{e'_\mathrm{int,3}}{n_{3}} = \left(\Gamma_{34} - 1 \right) m_\mathrm{p} c^2   \\ 
     & \frac{n'_3}{n'_4} = 4 \Gamma_{34} 
\end{eqnarray}

We assume that the pressures and velocities on either side of the contact discontinuity to be equal which can be expressed as
\begin{eqnarray}
     & \Gamma_2 = \Gamma_{3} \\ 
     & p_{2} = p_{3}  
\end{eqnarray}

Using equations (A15)-(A16) in equations (A11)-(A14) we obtain the following equation,
\begin{equation}
    (\Gamma^2_{21} - 1) = f (\Gamma^2_{34} - 1)
\end{equation}

\section{Solving for the proper speed of the shocked fluid $\lowercase{u}_{21}$ in the rest frame of R1} \label{App_B}

The relative LF $\Gamma_{34}$ is given as
\begin{equation}
     \Gamma_{34} = \Gamma_{31} \Gamma_{41} (1 - \beta_{31} \beta_{41})  =  \Gamma_{41} \Gamma_{31}  - u_{41} u_{31} = \Gamma_{41} \Gamma_{21}  - u_{41} u_{21}
\end{equation}
where we have used $u_{21} = u_{31}$ since the velocities are equal across the contact discontinuity.

Plugging equation (B1) in equation (A17) gives us
\begin{equation}
u^2_{21} = f \left[ (u^2_{41} + \Gamma^2_{41} ) u^2_{21} + u^2_{41} \right] - 2 f u_{21} \Gamma_{21} \Gamma_{41} u_{41}
\end{equation}
which can be solved for $u_{21}$ as 
\begin{equation}
\begin{split}
     u_{21} = u_{31} &\ = u_{41} \sqrt{\frac{  2 \sqrt{f^3(1 + u^2_{41} )} - f(1+f)  }{ 2 f(2 u^2_{41} + 1 ) - (1 + f^2)}}  =  u_{41} \sqrt{\frac{  2 f^{3/2} \Gamma_{41} - f(1+f)  }{ 2 f( u^2_{41} + \Gamma^2_{41} ) - (1 + f^2)}}  \label{soln_com}
\end{split}
\end{equation}

\section{Estimating shell crossing timescales}\label{App_C}

The velocity for the forward shock front in the lab frame can be estimated by equating the rate at which particles are being added to region R2 and are being lost from region R1. Let $\mathcal{A}$ be the planar area (which remains constant for our planar geometry). If $\Dot{M}_2$ is the rate at which particles are added to region R2 and $\Dot{M}_1$ is the rate at which particles are being lost from region R1, then we have   
\begin{align*}
&\   \Dot{M}_2 = - \Dot{M}_1 
\Rightarrow  \rho_\mathrm{1} \; \mathcal{A} (\beta_\mathrm{FS} - \beta_1)c  = \rho_\mathrm{2} \; \mathcal{A}(\beta_\mathrm{FS} - \beta_2)c   
\Rightarrow   \Gamma_1 n'_1 \; (\beta_\mathrm{FS} - \beta_1)  = \Gamma_2 n'_2 \; (\beta_\mathrm{FS} - \beta_2) 
\end{align*}
which gives
\begin{equation}
    \beta_\mathrm{FS} = \frac{\left( \frac{ n_1}{ n_2} \right) \beta_1 - \beta_2 }{\left( \frac{ n_1}{ n_2} \right) - 1} = \frac{\left( \frac{\Gamma_1 n'_1}{\Gamma_2 n'_2} \right) \beta_1 - \beta_2 }{\left( \frac{\Gamma_1 n'_1}{\Gamma_2 n'_2} \right) - 1}, 
\end{equation}

The equation (C1) can be simplified using the BM76 shock condition $n'_2/n'_1 = 4 \Gamma_{21}$ and noting $(\Gamma_2,\beta_2) = (\Gamma, \beta)$ to give
\begin{equation}
    \beta_\mathrm{FS} = \frac{ \frac{1}{4 \Gamma_{21}}\left( \frac{\Gamma_1 }{\Gamma } \right) \beta_1 - \beta_2 }{\frac{1}{4 \Gamma_{21}} \; \left( \frac{\Gamma_1 }{\Gamma } \right) - 1} = \frac{ \frac{1}{4 \Gamma_{21}}\left( \frac{u_1 }{\Gamma } \right)  - \beta_2 }{\frac{1}{4 \Gamma_{21}} \; \left( \frac{\Gamma_1 }{\Gamma } \right) - 1} ,
\end{equation}

Using equations (C1)-(C2) the time it takes for the forward shock front to cross the radial width $\Delta_{1,0}$ is given by
\begin{equation}
    \begin{split}
         t_\mathrm{FS}  = \frac{\Delta_{1,0}}{c(\beta_\mathrm{FS} - \beta_1)}  = \frac{\Delta_{1,0} }{c(\beta_2 - \beta_1)} \left[ 1 -  \left( \frac{n_1}{n_2} \right) \right]   
          = \frac{\Delta_{1,0} }{c(\beta_2 - \beta_1)} \left[ 1 - \left( \frac{\Gamma_1}{\Gamma_2} \right) \left( \frac{n'_1}{n'_2} \right) \right]  =   \frac{\Delta_{1,0} }{c(\beta - \beta_1)} \left[ 1 - \left( \frac{\Gamma_1}{\Gamma} \right) \left( \frac{1}{4 \Gamma_{21}} \right) \right] 
    \end{split}
\end{equation}

Similarly, the velocity for the reverse shock front in the lab frame can be estimated by equating the rate at which particles are being added to region R3 and are being lost from region R4. If $\Dot{M}_3$ is the rate at which particles are added to region R3 and $\Dot{M}_4$ is the rate at which particles are being lost from region R4  , then we have
\begin{align*}
    &\  \Dot{M}_3 = - \Dot{M}_4 
    \Rightarrow  \rho_\mathrm{4} \mathcal{A} (\beta_4 - \beta_\mathrm{RS}) c = \rho_\mathrm{3} \mathcal{A} (\beta_3 - \beta_\mathrm{RS})c
    \Rightarrow  \Gamma_{4} n'_\mathrm{4} (\beta_4 - \beta_\mathrm{RS}) = \Gamma_{3} n'_\mathrm{3} (\beta_3 - \beta_\mathrm{RS})
\end{align*}
which gives
\begin{equation}
    \beta_\mathrm{RS} = \frac{\beta_4 - \beta_3 \left(\frac{n_3}{n_4} \right) }{1 - \left(\frac{n_3}{ n_4} \right) } = \frac{\beta_4 - \beta_3 \left(\frac{\Gamma_3 n'_3}{\Gamma_4 n'_4} \right) }{1 - \left(\frac{\Gamma_3 n'_3}{\Gamma_4 n'_4} \right) }
\end{equation}

The equation (C4) can be simplified using the BM76 shock condition $n'_3/n'_4 = 4 \Gamma_{34}$ and noting $(\Gamma_3,\beta_3) = (\Gamma, \beta)$ to give
\begin{equation}
    \beta_\mathrm{RS} = \frac{\beta_4 - 4 \Gamma_{34}  \left(\frac{\Gamma }{\Gamma_4 }  \right) \beta }{1 - 4 \Gamma_{34} \left(\frac{\Gamma }{\Gamma_4 } \right) } =  \frac{\beta_4 - 4 \Gamma_{34}  \left(\frac{u }{\Gamma_4 }  \right) }{1 - 4 \Gamma_{34} \left(\frac{\Gamma }{\Gamma_4 } \right) }
\end{equation}

Using equations (C4)-(C5) the time it takes for the reverse shock front to cross the radial width $\Delta_{4,0}$ is given by
\begin{equation}
    \begin{split}
         t_\mathrm{RS}  = \frac{\Delta_{4,0}}{c(\beta_\mathrm{4} - \beta_\mathrm{RS} )}  = \frac{\Delta_{4,0} }{c(\beta_4 - \beta_3)} \left[ 1 -  \left( \frac{n_4}{n_3} \right) \right] 
          = \frac{\Delta_{4,0} }{c(\beta_4 - \beta_3)} \left[ 1 - \left( \frac{\Gamma_4}{\Gamma_3} \right) \left( \frac{n'_4}{n'_3} \right) \right] 
         =   \frac{\Delta_{4,0} }{c(\beta_4 - \beta)} \left[ 1 - \left( \frac{\Gamma_4}{\Gamma} \right) \left( \frac{1}{4 \Gamma_{34}} \right) \right] 
    \end{split}
\end{equation} 

\section{Estimating the maximum kinetic energy and the internal energy associated with the shocked regions}\label{App_D}

The energy-momentum tensor is given by
\begin{equation}
    T^{\mu \nu} = w u^{\mu} u^{\nu} + p \eta^{\mu \nu}
\end{equation}
such that
\begin{equation}
    e = T^{00} = w' \Gamma^2 - p  
\end{equation}
where the enthalpy $w$ and the pressure $p$ in the comoving frame is given as
\begin{eqnarray}
    &\ w' = \rho' c^2 + e'_\mathrm{int} + p \\
    &\ p = (\hat{\gamma} - 1) e'_\mathrm{int} 
\end{eqnarray}
where $\hat{\gamma}$ is the adiabatic index of the gas.

Using equation (D3) and equation (D4) in equation (D2) we have
\begin{equation}
    e = \Gamma^2 \rho' c^2 + \left[ \Gamma^2 + (\hat{\gamma} - 1) u^2  \right] e'_\mathrm{int} = \Gamma^2 \rho' c^2 + \Gamma^2 \left[ 1 + (\hat{\gamma} - 1) \beta^2  \right] e'_\mathrm{int} 
\end{equation}
where we have used the identity $u = \Gamma \beta$.

The energy density in the lab frame $e$ can be written as
\begin{equation}
    e = e_\mathrm{rest} + e_\mathrm{kin} + e_\mathrm{int}
\end{equation}

Equation (D5) can be re-arranged in the form
\begin{equation}
    e = \Gamma \rho' c^2 + \Gamma (\Gamma - 1) \rho' c^2 +  \Gamma^2 \left[ 1 + (\hat{\gamma} - 1) \beta^2  \right] e_\mathrm{int} 
\end{equation}

Comparing equation (D6) and equation (D7) we identify the following mapping 
\begin{eqnarray}
    &\ e_\mathrm{rest} = \Gamma \rho' c^2 = \rho c^2 \\ 
    &\ e_\mathrm{kin} = \Gamma (\Gamma - 1) \rho' c^2 = (\Gamma - 1) \rho c^2 \\
    &\ e_\mathrm{int} = \Gamma^2 \left[ 1 + (\hat{\gamma} - 1) \beta^2  \right] e'_\mathrm{int} 
\end{eqnarray}
where $e_\mathrm{rest}$ is the rest mass energy density associated with the outflow  in the lab frame, $e_\mathrm{kin}$ is the kinetic energy associated with the bulk motion of the outflow in the lab frame and $e_\mathrm{int}$ is the internal/thermal energy associated with the outflow in the lab frame. 

Let us recall that $\Dot{M}_2$ is the rate at which particles are added to region R2. Since the internal energy per unit particle in region R2 is $e_\mathrm{int,2}/\rho_\mathrm{2} c^2$, the rate of increase of the internal energy is the product of the two. The $\Dot{M}_2$  can be calculated by noting that in time $t_\mathrm{FS}$, the mass swept by the forward shockfront would be $M_{1,0}$. Using equation (D10) the quantity $e_\mathrm{int,2}/\rho_\mathrm{2} c^2$ can be expressed in terms of the comoving fluid quantities. The internal energy dissipation rate in region R2 is given by
\begin{equation}
\begin{split}
        \Dot{E}_\mathrm{int,2} &\  = \Dot{M}_2 c^2 \left( \frac{e_\mathrm{int,2}}{\rho_2 c^2}\right)  = \left( \frac{M_{1,0} c^2}{t_\mathrm{FS}} \right)  \frac{\Gamma^2_2 [1 + \beta^2_2 (\hat{\gamma}_2 - 1)] e'_\mathrm{int,2}}{\Gamma_2 \rho'_2 c^2}  = \left( \frac{M_{1,0} c^2}{t_\mathrm{FS}} \right)  \frac{\Gamma^2 [1 + \beta^2(\hat{\gamma}_2 - 1)] e'_\mathrm{int,2}}{\Gamma \rho'_2 c^2} \\
        &\ = \left( \frac{M_{1,0} c^2}{t_\mathrm{FS}} \right) \Gamma [1 + \beta^2(\hat{\gamma}_2 - 1)] (\Gamma_{21} - 1) \\ 
        &\ = \left( \frac{M_{1,0} c^2}{t_\mathrm{FS}} \right) \Gamma \left[1 + \beta^2 \left(\frac{\Gamma_{21} + 1}{3 \Gamma_{21}}\right) \right] (\Gamma_{21} - 1)  
\end{split}
\end{equation}
where in the last line we have expressed adiabatic constant $\hat{\gamma}_{2}$ from equation (A7). 

From equation (D11) we can multiply by $t_{\rm FS}$ to estimate the maximum internal energy in region R2
\begin{equation}
    E_\mathrm{int,2,max} = M_{1,0} c^2 \Gamma \left[1 + \beta^2 \left(\frac{\Gamma_{21} + 1}{3 \Gamma_{21}}\right) \right] (\Gamma_{21} - 1) 
\end{equation}

Similarly, the  internal energy dissipation rate in region R3 is given by
\begin{equation}
\begin{split}
    \Dot{E}_\mathrm{int,3} &\ =  \left( \frac{M_{4,0} c^2}{t_\mathrm{RS}} \right) \Gamma [1 + \beta^2(\hat{\gamma}_3 - 1)] (\Gamma_{34} - 1)   = \left( \frac{M_{4,0} c^2}{t_\mathrm{RS}} \right) \Gamma \left[1 + \beta^2 \left( \frac{\Gamma_{34} + 1}{3 \Gamma_{34}}\right) \right] (\Gamma_{34} - 1) 
\end{split}
\end{equation}

 From equation (D13) the maximum internal in region R3 is given by
\begin{equation}
   E_\mathrm{int,3,max} =  M_{4,0} c^2 \Gamma \left[1 + \beta^2 \left( \frac{\Gamma_{34} + 1}{3 \Gamma_{34}}\right) \right] (\Gamma_{34} - 1) 
\end{equation}

The rate of increase of the kinetic energy in regions R2 and R3 is given by
\begin{eqnarray}
    &\ \Dot{E}_\mathrm{k,2} =   (\Gamma -1) \Dot{M}_2 c^2   = (\Gamma -1) \left( \frac{M_{1,0} c^2}{t_\mathrm{FS}} \right)   \\ 
    &\ \Dot{E}_\mathrm{k,3} =  (\Gamma -1) \Dot{M}_3 c^2   = (\Gamma - 1) \left( \frac{M_{4,0} c^2}{t_\mathrm{RS}} \right) 
\end{eqnarray}

\section{Change in radial widths of regions R2 and R3 due to shock passage}\label{App_E}

The conservation of rest mass requires that the mass $M_\mathrm{3,f}$ in region R3 after one complete sweep by reverse shock must be equal to the total mass in the trailing shell $M_{4,0}$. If $\Delta_\mathrm{3f}$ is the radial width of region R3 after one complete sweep by the reverse shock, then we have
\begin{eqnarray}
    &\ M_\mathrm{3,f} = \Delta_\mathrm{3f} \; \rho_3 \mathcal{A} =  \Delta_\mathrm{3f} \; \Gamma_3 \; \rho'_3 \mathcal{A}  \\ 
    &\ M_\mathrm{4,0} = \Delta_\mathrm{4,0} \; \rho_4 \mathcal{A} =  \Delta_\mathrm{4,0} \; \Gamma_4 \; \rho'_4 \mathcal{A} 
\end{eqnarray} 

Equating equations (E1) and (E2) gives
\begin{equation}
    \Delta_\mathrm{3f} = \left( \frac{\rho'_4}{\rho'_3} \right) \; \left( \frac{\Gamma_4}{\Gamma_3} \right)  \Delta_{4,0} =  \frac{1}{4 \Gamma_{34}} \left( \frac{\Gamma_4}{\Gamma}\right) \; \Delta_{4,0}
\end{equation}

Using similar arguments, we can estimate $\Delta_\mathrm{2f}$, radial width of region R2 after one complete sweep of the trailing shell by the reverse shock as
\begin{equation}
    \Delta_\mathrm{2f} = \left( \frac{\rho'_1}{\rho'_2} \right) \left( \frac{\Gamma_1}{\Gamma_2} \right) \Delta_\mathrm{1,0} = \frac{1}{4 \Gamma_{21}} \left( \frac{\Gamma_1}{\Gamma} \right) \; \Delta_{1,0}   
\end{equation}

\section{Estimating the transfer of $\lowercase{p} \lowercase{d} V$ work across the contact discontinuity}\label{App_F}

The rate of pdV transfer of work from shell S4 to shell S1 can be estimated as
\begin{equation}
\begin{split}
    \Dot{W}_\mathrm{pdV} &\ = p_{3} \mathcal{A} \times \beta c  = \left( \hat{\gamma}_3 - 1 \right) e'_\mathrm{int,3} \mathcal{A} \times \beta c  = \left( \frac{\Gamma_{34} + 1}{3 \Gamma_{34}} \right)  e'_\mathrm{int,3}\mathcal{A} \times \beta c 
\end{split}
\end{equation}

The internal energy density $e'_\mathrm{int,3}$ can be expressed as
\begin{equation}
    \begin{split}
        e'_\mathrm{int,3} &\ = (\Gamma_{34} - 1) \rho'_3 c^2   = \left( \frac{\rho'_3}{\rho'_4} \right) (\Gamma_{34} - 1)  \rho'_4 c^2 = 4 \Gamma_{34} (\Gamma_{34} - 1) \left( \frac{M_{4} c^2}{\Gamma_4 V_{4,0}} \right)  = \frac{4 \Gamma_{34} (\Gamma_{34} - 1)}{\Gamma_4 (\Gamma_4 - 1)} \frac{E_\mathrm{k,4,0}}{V_{4,0}}  = \frac{4 \Gamma_{34} (\Gamma_{34} - 1)}{\Gamma_4 (\Gamma_4 - 1)} \frac{E_\mathrm{k,4,0}}{\mathcal{A} \Delta_{4,0}}
    \end{split}
\end{equation}
where in the last line we have used the relation $V_{4,0} = \mathcal{A} \Delta_{4,0}$. 

Using equation (F2) in equation (F1) we have
\begin{equation}
    \Dot{W}_\mathrm{pdV} = \frac{4}{3} \frac{(\Gamma^2_{34} - 1)}{\Gamma_4 (\Gamma_4 - 1)} \frac{\beta c}{\Delta_{4,0}} E_\mathrm{k,4,0} 
\end{equation}

Next we can evaluate the pdV work done in time $t_\mathrm{RS}$ as 
\begin{equation}
\begin{split}
W_\mathrm{pdV,RS} &\ =  \Dot{W}_\mathrm{pdV} \times t_\mathrm{RS} =  \Dot{W}_\mathrm{pdV} \times \frac{\Delta_{4,0}}{(\beta_4 - \beta)} \left[ 1 - \frac{1}{4 \Gamma_{34}} \left( \frac{\Gamma_4}{\Gamma}\right) \right]  = \frac{4}{3} \frac{(\Gamma^2_{34} - 1)}{\Gamma_4 (\Gamma_4 - 1)} \frac{\beta}{(\beta_4 - \beta)} \left[ 1 - \frac{1}{4 \Gamma_{34}} \left( \frac{\Gamma_4}{\Gamma}\right) \right] E_\mathrm{k,4,0}
\end{split}
\end{equation}

\section{Useful results for collision of ultra-relativistic shells}\label{App_G}

For ultra-relativistic shells, we have (see \citealt{1995ApJ...455L.143S}) 
\begin{equation}
    \Gamma_{34} \approx \frac{1}{2} \frac{\Gamma_4 }{\Gamma}
\end{equation}

Using equation (G1) in equation (D14) gives,
\begin{equation}
    E_\mathrm{int,3,max} \approx \left( \frac{\Gamma}{\Gamma_4}\right) \left( \frac{4}{3} \right) \left( \frac{\Gamma_4}{2 \Gamma} \right) E_\mathrm{k,4,0} = \frac{2}{3} E_\mathrm{k,4,0}
\end{equation}

Using equation (G1) in equation (C6) the time taken for complete sweep of shell 4 by the relativistic reverse shock front is given by
\begin{equation}
    \begin{split}
        t_\mathrm{RS}  \approx  \frac{\Delta_{4,0} (1 + \beta)}{c(1 - \beta^2)} \left[ 1 - \frac{1}{4 \left( \frac{\Gamma_4}{2 \Gamma} \right) } \left( \frac{\Gamma_4}{\Gamma} \right) \right]  =  \frac{\Gamma^2 \Delta_{4,0}}{c}   
    \end{split}
\end{equation}

Using equation (G1) in equation (F4) the rate of pdV work done for relativistic reverse shock can be estimated as
\begin{equation}
   \begin{split}
       \Dot{W}_\mathrm{pdV} &\ \approx \frac{4}{3} \frac{\Gamma^2_{34}}{\Gamma^2_4} \frac{c}{\Delta_{4,0}} E_\mathrm{k,4,0}  =   \frac{4}{3} \left( \frac{\Gamma^2_4}{4 \Gamma^2 }\right) \left( \frac{1}{\Gamma^2_4} \right)  \frac{c}{\Delta_{4,0}} E_\mathrm{k,4,0}  = \frac{1}{3} E_\mathrm{k,4,0} \left( \frac{c}{\Gamma^2 \Delta_{4,0}} \right) 
   \end{split}
\end{equation}

Multiplying equation (G3) and equation (G4) the pdV work transferred during one complete sweep by the relativistic reverse shock is given by
\begin{equation}
    W_\mathrm{pdV,RS} \approx \frac{1}{3} E_\mathrm{k,4,0} \sim 0.33 E_\mathrm{k,4,0} 
\end{equation}

Using equation (E4) the final width of shell 4 after one complete sweep by the relativistic reverse shock is given by
\begin{equation}
    \Delta_\mathrm{3f} = \frac{1}{4 \Gamma_{34}} \left( \frac{\Gamma_4}{\Gamma} \right) \Delta_{4,0} \approx \frac{1}{4 \left( \frac{\Gamma_4}{2 \Gamma} \right) } \left( \frac{\Gamma_4}{\Gamma} \right) \Delta_{4,0} = \frac{\Delta_{4,0}}{2}
\end{equation}

Next we can evaluate the time taken by the $+$ rarefaction wave to reach CD starting from the edge of shell S4 given by 
\begin{equation}
    t_\mathrm{3rf+} = \frac{\Delta_\mathrm{3f}}{c (\beta_\mathrm{3rf+} - \beta)} \approx \frac{\Delta_{4,0}}{2 c (\beta_\mathrm{3rf+} - \beta)} 
\end{equation}

The quantity in the denominator of equation (G7) can be estimated as
\begin{equation}
    \beta_\mathrm{3rf+} - \beta = \frac{\beta + \beta'_\mathrm{s3}}{1 + \beta \beta'_\mathrm{s3}} - \beta = \frac{\beta'_\mathrm{s3} (1 - \beta^2)}{1 + \beta \beta'_\mathrm{s3}} = \frac{1}{\Gamma^2} \frac{\beta'_\mathrm{s3}}{1 + \beta \beta'_\mathrm{s3}} \approx \frac{1}{\Gamma^2} \frac{\sqrt{3}}{1 + \sqrt{3}}
\end{equation}
where in the last step we have used the sound speed in the comoving frame to be $\beta'_{s3} \rightarrow \frac{1}{\sqrt{3}}$ which is true for relativistic reverse shock. 

Using equation (G8) in equation (G7) we obtain
\begin{equation}
    t_\mathrm{3rf+} \approx  \frac{\Gamma^2 \Delta_{4,0}}{c} \left( \frac{1 + \sqrt{3}}{2 \sqrt{3}} \right) = t_\mathrm{RS} \left( \frac{1 + \sqrt{3}}{2 \sqrt{3}} \right) 
\end{equation}

The pdV work done from the launch of $+$ rf till the head of rf wave reaches CD  is given by 
\begin{equation}
    W_\mathrm{pdV,3rf+} = \Dot{W}_\mathrm{pdV} \times t_\mathrm{3rf+} \approx \frac{1 + \sqrt{3}}{6 \sqrt{3}} E_\mathrm{k,4,0} 
\end{equation}

Using equation (G5) and equation (G10) we can put a limit on the pdV workdone from the onset of collision till the $+$ rf wave reaches CD as 
\begin{eqnarray}
\begin{split}
     W_\mathrm{pdV,+}  = W_\mathrm{pdV,RS} + W_\mathrm{pdV,3rf+}  \approx \frac{1}{3} \left[ 1 + \frac{1 + \sqrt{3}}{2 \sqrt{3}}\right] E_\mathrm{k,4,0}  \sim 0.59 \; E_\mathrm{k,4,0} 
\end{split}
\end{eqnarray}

\section{Estimating the five critical width ratio $\chi_\mathrm{cX}$ for $X=(1,2,3,4,5)$ }\label{App_H}

In order to estimate the speed of the $\pm$ rf wave in the lab frame, we use the TM equation of state (\citealt{1971ApJ...165..147M}) to get the expression for the sound speed $\beta'_\mathrm{sj}$ in the comoving frame of the region $j=(2,3)$(see Eq.14 of \cite{2006ApJS..166..410R} and Eq. 17  of \cite{2007MNRAS.378.1118M}) is given by
\begin{equation}
    \beta'^{2}_\mathrm{sj}  = \frac{\Theta_\mathrm{j}}{3 h_\mathrm{j}} \frac{5 h_\mathrm{j} - 8 \Theta_\mathrm{j}}{h_\mathrm{j} - \Theta_\mathrm{j}} = \frac{5 \Theta \sqrt{\Theta^2 + 4/9} + 3 \Theta^2}{ 12 \Theta \sqrt{\Theta^2 + 4/9} + 12 \Theta^2 + 2}
\end{equation}
where
\begin{equation}
    h_\mathrm{j} = \frac{5}{2} \Theta_\mathrm{j} + \sqrt{\frac{9}{4} \Theta^2_\mathrm{j} + 1}
\end{equation}
where the quantity $\Theta_\mathrm{j}$ is defined as
\begin{equation}
\Theta_\mathrm{j} = \frac{p_\mathrm{j}}{\rho'_\mathrm{j} c^2} = \left( \hat{\gamma}_\mathrm{j} - 1\right)  (\Gamma_\mathrm{ij} - 1) = \frac{\Gamma^2_\mathrm{ij} - 1}{3 \Gamma_\mathrm{ij}}  
\end{equation}

\begin{table*}
    \centering
     \caption{Expressions for the various timescales used to evaluate the five critical widths}
    \begin{tabular}{c|c|c} \hline 
      Symbol   & Definition & Expression \\ \hline \\ 
       $\beta_\mathrm{2rf\pm}$ & The speed of the $\pm$ rf wave in region R2 & $\frac{\beta \pm \beta'_{s2}}{1 \pm \beta \beta'_\mathrm{s2}}$ \\ \\  
        $\beta_\mathrm{3rf\pm}$ & The speed of the $\pm$ rf wave in region R3 & $\frac{\beta \pm \beta'_{s3}}{1 \pm \beta \beta'_\mathrm{s3}}$ \\ \\  \hline 
       $t_\mathrm{3rf+}$  & Time taken for the $(+)$ rf wave in region R3 to reach CD &  $\frac{\Delta_\mathrm{3f}}{c(\beta_\mathrm{3rf+} - \beta)}$ \\  \\ 
       $t_\mathrm{2rf+}$  & Time taken by the $(+)$ rf wave in region R2 to catch up with the FS & $\frac{(\beta_\mathrm{FS} - \beta)(t_\mathrm{RS} + t_\mathrm{3rf+})}{(\beta_\mathrm{2rf+} - \beta_\mathrm{FS})}$ \\ \\
       $t_\mathrm{2rf-}$ & Time taken for the $(-)$ rf wave in region R2 to reach CD & $\frac{\Delta_\mathrm{2f}}{c(\beta - \beta_\mathrm{2rf-})}$ \\ \\ 
       $t_\mathrm{3rf-}$ & Time taken by the $(-)$ rf wave in region R3 to catch up with RS & $\frac{(\beta - \beta_\mathrm{RS})(t_\mathrm{FS} + t_\mathrm{2rf-})} {(\beta_\mathrm{RS} - \beta_\mathrm{3rf-})}$  \\ \\ \hline      
    \end{tabular}
   
    \label{tab:my_label}
\end{table*}

The procedure to get the critical widths corresponding to the five lines defined in \S 3 is as follows.

We begin by estimating $t_\mathrm{RS}, t_\mathrm{3rf+},t_\mathrm{2rf+}$ considering a hypothetical scenario in which the width of shell 1 were infinite. Then, we ask if the radial width of shell 1 were to be finite, for what ratio $\chi$ would we obtain the same values for the quantities? As an illustration, consider the L1 corresponding to $t_\mathrm{FS} = t_\mathrm{RS} + t_\mathrm{3rf+ } + t_\mathrm{2rf+}$. Assuming the infinite radial width of shell S1, we first estimate 
\begin{equation}
     t_\mathrm{RS} + t_\mathrm{3rf+ } + t_\mathrm{2rf+} = \left[ 1 + \frac{(\beta_\mathrm{FS} - \beta)}{(\beta_\mathrm{2rf+} - \beta_\mathrm{FS})}\right] (t_\mathrm{RS} + t_\mathrm{3rf+})
\end{equation}
Next assuming that shell S1 has finite radial width $\Delta_{1,0} = \chi_{c1} \Delta_{4,0}$ we have
\begin{equation}
    t_\mathrm{FS} = \frac{\chi_{c1} \; \Delta_{4,0}}{c(\beta_\mathrm{FS} - \beta_1 )}
\end{equation}

Equating equation (H4) and equation (H5) the critical width $\chi_\mathrm{c1}$ corresponding to L1 is given by 
\begin{equation}
    \begin{split}
           \chi_\mathrm{c1} &\ =  (\beta_\mathrm{FS} - \beta_1) \left[ 1 + \frac{(\beta_\mathrm{FS} - \beta)}{(\beta_\mathrm{2rf+} - \beta_\mathrm{FS})}\right] \left[ \frac{1}{(\beta_4 - \beta_\mathrm{RS})}  + \frac{1}{(\beta_\mathrm{3rf+} - \beta )} \left( \frac{\Delta_\mathrm{3f}}{\Delta_{4,0}}\right)\right] \\
            &\ =  (\beta_\mathrm{FS} - \beta_1) \left[ 1 + \frac{(\beta_\mathrm{FS} - \beta)}{(\beta_\mathrm{2rf+} - \beta_\mathrm{FS})}\right] \left[ \frac{1}{(\beta_4 - \beta_\mathrm{RS})}  + \frac{1}{4 \Gamma_{34}} \left( \frac{\Gamma_4}{\Gamma}\right) \frac{1}{(\beta_\mathrm{3rf+} - \beta )}   \right] 
    \end{split}
\end{equation}
where the second line is simplified using equation (E4).

If $\chi > \chi_{c1}$, shell S1 is partially shocked. The fraction of the mass in shell S1 that is shocked is given by
\begin{equation}
    \alpha_{2} = \frac{\chi_{c1}}{\chi} \hspace{2cm} \text{For $\chi > \chi_\mathrm{c1}$} 
\end{equation}

Similarly, the critical width ratio for line L2 is given by
\begin{equation}
     \chi_\mathrm{c2}  =  (\beta_\mathrm{FS} - \beta_1) \left[ \frac{1}{(\beta_4 - \beta_\mathrm{RS})}  + \frac{1}{(\beta_\mathrm{3rf+} - \beta )}  \frac{1}{4 \Gamma_{34}} \left( \frac{\Gamma_4}{\Gamma}\right) \right] 
\end{equation}

 The critical width ratio for line L3 is given by
 \begin{equation}
     \chi_\mathrm{c3} = \frac{(\beta_\mathrm{FS} - \beta_1)}{(\beta_4 - \beta_\mathrm{RS})}
 \end{equation}

The critical width ratio for L4 is given by
\begin{equation}
    \chi^{-1}_\mathrm{c4} = (\beta_4 - \beta_\mathrm{RS}) \left[ \frac{1}{(\beta_\mathrm{FS} - \beta_1)} + \frac{1}{4 \Gamma_{21}} \left( \frac{\Gamma_1}{\Gamma} \right) \left( \frac{1}{(\beta - \beta_\mathrm{2rf-})}\right) \right] 
\end{equation}

The critical width ratio for L5 is given by
\begin{equation}
\begin{split}
    &\ \chi^{-1}_\mathrm{c5}  = (\beta_4 - \beta_\mathrm{RS}) \left[ 1 + \left( \frac{\beta - \beta_\mathrm{RS}}{\beta_\mathrm{RS} - \beta_\mathrm{3rf-}} \right) \right] \left[ \frac{1}{(\beta_\mathrm{FS} - \beta_1)} + \frac{1}{4 \Gamma_{21}} \left( \frac{\Gamma_1}{\Gamma} \right) \left( \frac{1}{(\beta - \beta_\mathrm{2rf-})}\right) \right]
\end{split}     
\end{equation}

For $\chi < \chi_\mathrm{c5}$ shell S4 is partially shocked. The fraction $\alpha_3$ of the mass shocked in shell S4 is given by
\begin{equation}
    \alpha_3 = \frac{\chi}{\chi_\mathrm{c5}} \hspace{2cm} \text{For $\chi < \chi_\mathrm{c5}$}
\end{equation}

\section{Shock hydrodynamics in CD frame: An illustrative example}

The goal of the present appendix is to outline the similarities $\&$ differences between our approach and that of \citealt{2004ApJ...611.1021K} (hereafter KMY04). Table \ref{CD_defn} summarizes the physical quantities in the CD frame (indicated by tilde). Comoving quantities are primed. The objective is to analyse shock (both rs and rf) propagation in the CD rest frame for a collision of two equal proper density shells ($f=1$) of equal lab frame radial widths ($\Delta_{1,0} = \Delta_{4,0} = \Delta_{0}$). The $f=1$ case is chosen because of two reasons. Firstly, KMY04 assume the adiabatic constant to be equal in both the shocked regions. In our assumed equation of state, this is true only if the shock strengths are equal which is the case for $f=1$. In order to be \textbf{consistent} with their assumption, we choose the $f=1$ scenario to have equal adiabatic indices in both regions by construction. Secondly, $f=1$ is a \textbf{simple scenario} because  the shock strengths ($\Gamma_{21} = \Gamma_{34}$), the speed of shock fronts $(\widetilde{\beta}_\mathrm{FS} = \widetilde{\beta}_\mathrm{RS})$  and the sound speeds ($\widetilde{\beta}_\mathrm{s3} = \widetilde{\beta}_\mathrm{s2} = \widetilde{\beta}_\mathrm{s}$) in the two shocked regions R3 and R2 are all equal. In particular, we are interested in the ultra-relativistic case for which $\Gamma_{4}>\Gamma_1\gg1$. Like the rest of our analysis, we will assume a planar geometry where the planar area is $\mathcal{A}$. The quantities $(\widetilde{\beta}_1,\widetilde{\beta}_4,\widetilde{\beta}_\mathrm{RS},\widetilde{\beta}_\mathrm{FS})$ are the absolute speeds measured
in the CD frame. As for the velocity directions, in the CD frame, both the front edge of shell S1 and the rear edge of shell S4 move towards the CD (the former to the left and the latter to the right). While both the shock fronts move away from the CD (The RS to the left and the FS to the right). In the rest of the analysis, it must be noted that that the vector difference of the velocities is the summation of the absolute speeds. Table \ref{Expr_compare} compares our notation with that by KMY04.

In what follows, we will first derive the general expression in the CD frame and then write the solutions for the particular scenario of $f=1$ (indicated at the end of the equation).

\begin{table}
    \centering
    \caption{Symbols and definitions of quantities in the CD frame (comoving frame associated with shocked fluid). The (-) rf wave refers to a backward propagating rf wave. }
    \begin{tabular}{c|c} \hline 
       Symbol & Definition  \\ \hline 
        $\widetilde{\Delta}_\mathrm{1}$  & The radial width of shell S1 in the CD frame     \\ 
        $\widetilde{\Delta}_\mathrm{4}$  & The radial width of shell S4 in the CD frame \\
        $\widetilde{\beta}_\mathrm{FS}$  & The speed of the FS in the CD frame \\
        $\widetilde{\beta}_\mathrm{RS}$ & The speed of the RS in the CD frame \\
        $\widetilde{\rho}_\mathrm{1}$ &  density of region R1 as seen from the CD frame \\
        $\widetilde{\rho}_\mathrm{4}$ & density of region R4 as seen from the CD frame \\
        $\widetilde{t}_\mathrm{FS}$  &   The time for the FS to propagate from the CD to the edge of S1 (CD frame)\\
        $\widetilde{t}_\mathrm{RS}$  &  The time for the RS to propagate from the CD to the edge of S4 (CD frame) \\
        $\widetilde{\Delta}_\mathrm{2f}$ & The radial width of shell S1 just after the FS reaches its front edge (CD frame)\\  
        $\widetilde{\beta}_\mathrm{s}$   & The co-moving sound speed in the CD frame (equal in R2 and R3 for $f=1$)\\
        $\widetilde{t}_\mathrm{2rf-}$ & The time taken for the head of (-) rf wave to propagate from the front edge of S1 to the CD (CD frame)\\
        $\widetilde{t}_\mathrm{3rf-}$ & The time taken for the head of (-) rf wave to propagate from the CD to the RS (CD frame) \\
        $\widetilde{t}_\mathrm{FR-RS}$ & $ \widetilde{t}_\mathrm{FS} + \widetilde{t}_\mathrm{2rf-} + \widetilde{t}_\mathrm{3rf-}$ \\ \hline 
    \end{tabular}
    \label{CD_defn}
\end{table}

\begin{table}
\centering 
\caption{Comparing the notation in our work and that by KMY04.   }
    \begin{tabular}{c|c} \hline 
      KMY04  & Our work  \\ \hline 
       $\Gamma_\mathrm{r}$ & $\Gamma_{4}$ \\
       $\Gamma_\mathrm{s}$  & $\Gamma_{1}$ \\ 
       $\rho_\mathrm{r}$   & $\rho'_{4}$  \\ 
       $\rho_\mathrm{s}$    & $\rho'_{1}$ \\ 
        $t'_\mathrm{FS}$ &   $\widetilde{t}_\mathrm{FS}$ \\     
        $t'_\mathrm{FR-CD}$  & $\widetilde{t}_\mathrm{FS} + \widetilde{t}_\mathrm{2rf-}$ \\
        $t'_\mathrm{FR-RS}$ & $ \widetilde{t}_\mathrm{FR-RS}(= \widetilde{t}_\mathrm{FS} + \widetilde{t}_\mathrm{2rf-} + \widetilde{t}_\mathrm{3rf-})$ \\ \hline  
    \end{tabular}
    \label{Expr_compare}
\end{table}

The radial width of shells S1 and S4 in the CD frame is given by
\begin{equation}
    \widetilde{\Delta}_\mathrm{1} = \frac{\Delta_{1,0} \Gamma_1}{\Gamma_{21}} 
    = \frac{\Delta_{0} \Gamma_1}{\widetilde{\Gamma}_{1}}\ , \quad \quad \quad \quad  \widetilde{\Delta}_\mathrm{4} = \frac{\Delta_{4,0} \Gamma_4}{\Gamma_{34}} 
    = \frac{\Delta_{0} \Gamma_4}{\widetilde{\Gamma}_{4}}=\frac{\widetilde{\Gamma}_{1}}{\widetilde{\Gamma}_{4}}\frac{\Gamma_4}{\Gamma_1} \widetilde{\Delta}_\mathrm{1}\ , \label{CD_width}
\end{equation}
which shows that while lab frame radial widths are equal, in the CD frame they are not equal. Note that $\widetilde{\Gamma}_1 = \Gamma_{21}$ and $\widetilde{\Gamma}_{4}= \Gamma_{34}$, while for $f=1$ all four are equal.

Next, we estimate the speed $\widetilde{\beta}_\mathrm{FS}$ of the FS in the CD frame by estimating the rate at which mass in region R2 changes. This gives 
\begin{equation}
    \frac{dM_{2}}{dt'} = \rho'_\mathrm{2}\, \mathcal{A} \; \widetilde{\beta}_\mathrm{FS}\,c = \tilde{\rho_\mathrm{1}}\, \mathcal{A} \; (\widetilde{\beta}_{1} + \widetilde{\beta}_\mathrm{FS})\, c \quad\Longrightarrow\quad \rho'_{2} \; \widetilde{\beta}_\mathrm{FS} = \rho'_{1} \widetilde{\Gamma}_\mathrm{1} (\widetilde{\beta}_{1} + \widetilde{\beta}_\mathrm{FS})  \quad \Longrightarrow\quad 4 \widetilde{\Gamma}_1 \rho'_{1} \; \widetilde{\beta}_\mathrm{FS} = \rho'_{1} \widetilde{\Gamma}_\mathrm{1} (\widetilde{\beta}_{1} + \widetilde{\beta}_\mathrm{FS})  \quad \Longrightarrow\quad \widetilde{\beta}_\mathrm{FS} = \frac{1}{3} \widetilde{\beta}_{1}
    \quad   \label{FS_CD} 
\end{equation}
where $\rho'_{1}$ and $\rho'_{2}$ are the proper density in regions R1 and R2 respectively. In the second equality we use the transformation $\widetilde{\rho}_\mathrm{1} = \Gamma_{21} \rho'_{1} = \widetilde{\Gamma}_{1} \rho'_{1} $. In the third equality, we have used the BM76 condition for our adopted equation of state, $\rho'_{2} = 4 \Gamma_{21} \rho'_{1} = 4 \widetilde{\Gamma}_{1} \rho'_{1}$. Similarly we have,
\begin{equation}
 \widetilde{\beta}_\mathrm{RS} = \frac{1}{3} \widetilde{\beta}_{4} \label{RS_CD} 
\end{equation}
From symmetry,  $f=1$ implies that $\widetilde{\beta}_\mathrm{RS} = \widetilde{\beta}_\mathrm{FS}$ and $\widetilde{\beta}_4=\widetilde{\beta}_1$,
\begin{equation}
    \widetilde{\beta}_\mathrm{RS} = \widetilde{\beta}_\mathrm{FS} =  \frac{1}{3} \widetilde{\beta}_{4} = \frac{1}{3} \widetilde{\beta}_{1}  \hspace{0.8cm} \text{for $f=1$\ .}
\end{equation}

In the CD frame, the time it takes for the FS to reach the front edge of S1 and the RS to reach the rear edge of S4 are
\begin{equation}
    \widetilde{t}_\mathrm{FS} = \frac{\widetilde{\Delta}_\mathrm{1}}{(\widetilde{\beta}_{1} + \widetilde{\beta}_\mathrm{FS})c} = \frac{3}{4} \frac{\widetilde{\Delta}_{1}}{\widetilde{\beta}_{1} c}\  = \frac{3}{4} \frac{\Delta_{0} \Gamma_{1}}{\widetilde{u}_{1} c}\ ,   \quad \quad \quad \quad \widetilde{t}_\mathrm{RS} =  \frac{\widetilde{\Delta}_\mathrm{4}}{(\widetilde{\beta}_{4} + \widetilde{\beta}_\mathrm{RS})c} = \frac{3}{4} \frac{\Delta_{0} \Gamma_{4}}{\widetilde{u}_{4} c}\ , \label{Crossing_CD}
\end{equation}
where from Eq.~(\ref{FS_CD})-(\ref{RS_CD}), 
we have   $\widetilde{\beta}_{1} + \widetilde{\beta}_\mathrm{FS} = \frac{4}{3} \widetilde{\beta}_{1}$ and $\widetilde{\beta}_{4} + \widetilde{\beta}_\mathrm{RS} = \frac{4}{3} \widetilde{\beta}_{4}$.  

In order to investigate the propagation of rf wave, one must estimate the ratio of the crossing times. Using eqn. (\ref{Crossing_CD}), the ratio of the crossing times can be written as 
\begin{equation}
    \frac{\widetilde{t}_\mathrm{RS}}{\widetilde{t}_\mathrm{FS}} = \frac{\widetilde{\Delta}_{4}}{\widetilde{\Delta}_{1}} \frac{\widetilde{\beta}_1}{\widetilde{\beta}_4} = \frac{\Gamma_{4}}{\Gamma_{1}} \frac{\widetilde{u}_1}{\widetilde{u}_4}\ ,
\end{equation}
which shows that for $f=1$ for which $\widetilde{\beta}_1 = \widetilde{\beta}_4$ ($\widetilde{u}_1 = \widetilde{u}_4$),  the ratio of the crossing times is
\begin{equation}
    \frac{\widetilde{t}_\mathrm{RS}}{\widetilde{t}_\mathrm{FS}} = \frac{\widetilde{\Delta}_4}{\widetilde{\Delta}_1}=\frac{\Gamma_4}{\Gamma_{1}} \approx a_\mathrm{u} \hspace{0.8cm} \text{for $f=1$}\ .
\end{equation}
Thus, for $f=1$ at a high proper speed contrast $a_\mathrm{u} \gg 1$, and for an observer in the CD frame, the FS reaches the front edge of S1 much before RS can reach the rear edge of S4. The crossing times become equal in the limit of low proper speed contrast $(a_\mathrm{u} -1) \ll 1$. In the range of moderate to high proper speed contrast, a backward propagating (-) rf wave is launched just after the FS reaches the edge of shell S1. 

The final (CD frame) radial width of S1 just after FS reaches its front edge, $\widetilde{\Delta}_\mathrm{2f}$, can be obtained using conservation of mass in shell S1
\begin{equation}
    \rho'_{2} \; \mathcal{A} \widetilde{\Delta}_\mathrm{2f} = \widetilde{\rho}_\mathrm{1}  \; \mathcal{A} \widetilde{\Delta}_\mathrm{1}  \Rightarrow   \widetilde{\Delta}_\mathrm{2f} = \frac{\Delta_{0} \Gamma_1}{4 \widetilde{\Gamma}_{1}}\ . 
\end{equation}
Just after the FS reaches the front edge of S1, a backward propagating (-)rf wave is launched towards the CD. The head of the rf wave travels at the comoving sound speed $\widetilde{\beta}_\mathrm{s2}$. 
The time it takes for the head of rf wave to reach the CD is estimated as 
\begin{equation}\label{t2_com} 
    \widetilde{t}_\mathrm{2rf-} = \frac{\widetilde{\Delta}_\mathrm{2f}}{\widetilde{\beta}_\mathrm{s2} c} = \frac{1}{\widetilde{\beta}_\mathrm{s2} c} \left( \frac{\Delta_{0} \Gamma_1}{4 \widetilde{\Gamma}_{1}} \right)\ .  
\end{equation}
The time it takes for the (-)rf to catch up with the RS, starting from the CD can be estimated from
\begin{equation}
    \widetilde{\beta}_\mathrm{RS} c (\widetilde{t}_\mathrm{3rf-} + \widetilde{t}_\mathrm{2rf-} + \widetilde{t}_\mathrm{FS}) = \widetilde{\beta}_\mathrm{s3} c \widetilde{t}_\mathrm{3rf-} \Rightarrow \widetilde{t}_\mathrm{3rf-} = \frac{\widetilde{\beta}_\mathrm{RS}}{\widetilde{\beta}_\mathrm{s3} - \widetilde{\beta}_\mathrm{RS}} (\widetilde{t}_\mathrm{FS} + \widetilde{t}_\mathrm{2rf-})
    = \frac{\widetilde{\beta}_{4}}{3\widetilde{\beta}_\mathrm{s3} - \widetilde{\beta}_\mathrm{4}} \left[ \frac{3}{4} \frac{\Delta_{0} \Gamma_{1}}{\widetilde{u}_{1} c} +  \frac{1}{\widetilde{\beta}_\mathrm{s2} c} \left( \frac{\Delta_{0} \Gamma_1}{4 \widetilde{\Gamma}_{1}} \right)   \right] \ , 
\end{equation}
where we used $\widetilde{\beta}_\mathrm{RS} = \frac{1}{3} \widetilde{\beta}_\mathrm{4}$ in the last equality. 

Using $\widetilde{\beta}_{1} = \widetilde{\beta}_{4}$ and $\widetilde{\beta}_\mathrm{s2}= \widetilde{\beta}_\mathrm{s3} = \widetilde{\beta}_\mathrm{s}$ for $f=1$ we can further simplify the expression above as 
\begin{equation}\label{t3_com} 
    \widetilde{t}_\mathrm{3rf-} = \left( \frac{\Delta_{0} \Gamma_1}{4c}\right) \frac{\widetilde{\beta}_{1}}{3 \widetilde{\beta}_\mathrm{s} - \widetilde{\beta}_{1}} \left[ \frac{3}{\widetilde{\beta}_{1} \widetilde{\Gamma}_{1}} + \frac{1}{\widetilde{\beta}_\mathrm{s} \widetilde{\Gamma}_{1}} \right] =  \frac{1}{\widetilde{\Gamma}_{1}} \left( \frac{\Delta_{0} \Gamma_1}{4c}\right) \frac{1}{\widetilde{\beta}_\mathrm{s}} \; \left[ \frac{3 \widetilde{\beta}_\mathrm{s} + \widetilde{\beta}_{1}}{3 \widetilde{\beta}_\mathrm{s} - \widetilde{\beta}_{1}} \right] \hspace{1.4cm} \text{for $f=1$}\ .  
\end{equation}

Next, we want to express  $(\widetilde{u}_{1},\widetilde{\beta}_{1}, \widetilde{\Gamma}_{1})$ as a function of $\Gamma_{41}$. For $f = 1$, from Eq. \ref{soln_com} we have
\begin{equation}\label{u_com}
\widetilde{u}_{1} = u_{21} = u_{34}  = \sqrt{\frac{\Gamma_{41} - 1}{2}} \hspace{1.8cm} \text{for $f=1$}\ ,  
\end{equation}
which leads to 
\begin{align}
  \label{gamma_com}  &\ \widetilde{\Gamma}_{1} = \sqrt{\frac{1+ \Gamma_{41}}{2}}  \hspace{3.3cm} \text{for $f=1$}\ , \\ 
    &\ \widetilde{\beta}_{1} = \frac{\widetilde{u}_{1}}{\widetilde{\Gamma}_{1}} = \sqrt{\frac{\Gamma_{41} - 1}{\Gamma_{41} + 1}} = \frac{u_{41}}{ \Gamma_{41} + 1} \hspace{1.1cm} \text{for $f=1$}\ . \label{beta_com}
\end{align}

Using eqns. (\ref{beta_com})-(\ref{gamma_com}) in eqns. (\ref{Crossing_CD}),(\ref{t2_com}) and (\ref{t3_com}) we can summarize the various timescales for $f=1$ as
\begin{align}
    &\ \widetilde{t}_\mathrm{FS} =  \frac{3}{2 \sqrt{2}} \left( \frac{\Delta_{0} \Gamma_1}{c}\right) \frac{1}{\sqrt{\Gamma_{41} - 1}}\ = \frac{3}{4 u_{34}} \left( \frac{\Delta_{0} \Gamma_1}{c}\right)  , \\ 
    &\ \widetilde{t}_\mathrm{RS} = \frac{3}{2 \sqrt{2}} \left( \frac{\Delta_{0} \Gamma_4}{c}\right) \frac{1}{\sqrt{\Gamma_{41} - 1}} = \frac{3}{4 u_{34}} \left( \frac{\Delta_{0} \Gamma_4}{c}\right) \ ,  \\ 
    &\ \widetilde{t}_\mathrm{2rf-} = \frac{1}{2\sqrt{2}} \frac{1}{\widetilde{\beta}_\mathrm{s}}\; \left( \frac{\Delta_{0} \Gamma_1}{c}\right) \frac{1}{\sqrt{\Gamma_{41} + 1}} = \frac{1}{4 \Gamma_{34}} \frac{1}{\widetilde{\beta}_\mathrm{s}}\; \left( \frac{\Delta_{0} \Gamma_1}{c}\right) \ , \\
    &\ \widetilde{t}_\mathrm{3rf-} = \frac{1}{2 \sqrt{2}} \frac{1}{\widetilde{\beta}_\mathrm{s}} \left( \frac{\Delta_{0} \Gamma_1}{c} \right) \frac{1}{\sqrt{\Gamma_{41} + 1}} \left[ \frac{3 \widetilde{\beta}_\mathrm{s} ( \Gamma_{41}+1)+ u_{41}} {3 \widetilde{\beta}_\mathrm{s} (\Gamma_{41}+1)- u_{41}} \right] = \frac{1}{4\Gamma_{34}} \left( \frac{\Delta_{0} \Gamma_1}{c}\right) \frac{1}{\widetilde{\beta}_\mathrm{s}} \; \left[ \frac{3 \widetilde{\beta}_\mathrm{s} + \beta_{34}}{3 \widetilde{\beta}_\mathrm{s} - \beta_{34}} \right] \ ,
\end{align}
where $\widetilde{\beta}_\mathrm{s} \rightarrow \frac{1}{\sqrt{3}}$ and $\beta_{34} \rightarrow 1$ for the ultra-relativistic shock limit ($\widetilde{\Gamma}_1=\Gamma_{12},\,\widetilde{\Gamma}_4=\Gamma_{34}\gg1$).  For ultra-relativistic collisions  ($\Gamma_4>\Gamma_1\gg1$) we can make use of $\Gamma_{3} = \Gamma \approx \sqrt{a_\mathrm{u}}\,\Gamma_1$:
\begin{equation}\label{eq:G34_au}
   \Gamma_{34} - 1 \approx \frac{1}{2} \left( \frac{\Gamma_4}{\Gamma_3} + \frac{\Gamma_3}{\Gamma_4} \right) - 1 = \frac{(\sqrt{a}_\mathrm{u} - 1)^2}{2 \sqrt{a}_\mathrm{u}} \hspace{2cm}  \text{For $f=1$}\ ,  
\end{equation}
which gives $\Gamma_{34} - 1 \approx 0.423$ for $\Gamma_4/\Gamma_1 \approx a_\mathrm{u} = 6$. 

It must be emphasized that while the \textit{times} may not be simultaneous in the lab frame and the CD frame, quantities like the shocked fraction $\alpha_3$ of shell S4 must remain invariant. The shocked fraction $\alpha_3$ can be represented as 
\begin{equation}\label{eq:alpha3}
    \alpha_{3} = \frac{\widetilde{t}_\mathrm{RR-FR}}{\widetilde{t}_\mathrm{RS}} = \left( \frac{\Gamma_1}{\Gamma_4}\right) \left( \frac{3 \widetilde{\beta}_\mathrm{s} + \widetilde{\beta}_1}{3 \widetilde{\beta}_\mathrm{s} - \widetilde{\beta}_1}  \right)  \hspace{2cm}  \text{For $f=1$ and $\widetilde{t}_\mathrm{RR-FR} < \widetilde{t}_\mathrm{RS}$}\ ,
\end{equation}
where $\widetilde{t}_\mathrm{RR-FR} = \widetilde{t}_\mathrm{FS} + \widetilde{t}_\mathrm{2rf-} + \widetilde{t}_\mathrm{3rf-}$. For $f=1$ , the condition $\widetilde{t}_\mathrm{RR-FR} = \widetilde{t}_\mathrm{RS}$ (corresponding to the line L5 referred to in \S 3 of the main text) is satisfied (in the ultra-relativistic limit $\Gamma_4>\Gamma_1 \gg 1$) for a shock strength of $\Gamma_{34} - 1 \approx 0.136$ (or equivalently a proper speed contrast of $a_\mathrm{u} = 2.81$). The asymptotic behaviour at the ultra-relativistic shock limit is
\begin{equation}
  \alpha_{3}  \approx \frac{1}{a_\mathrm{u}}\left( \frac{\sqrt{3} +1}{\sqrt{3} - 1} \right) = \frac{2 + \sqrt{3}}{a_\mathrm{u}} \sim \frac{3.73}{a_\mathrm{u}} \hspace{2cm} \text{For $a_\mathrm{u} \gg 1$}\ ,
\end{equation}
i.e. the same as in the lab frame (see Eq.~(53) in the main text), as it should be. Besides, for the illustrative scenario from KMY04 of $a_\mathrm{u} = 6$, we verified that the shocked fraction is $\alpha_{3}\approx0.514$ in both the lab and CD frames.

The adiabatic index $\hat{\gamma}$ for KMY04 and our work (based on \citealt{2003ApJ...591.1075K}) can be represented as a function of the proper speed contrast $a_\mathrm{u}$ (in the ultra-relativistic regime $\Gamma_4 > \Gamma_1 \gg 1$) as 
\begin{align}
  \label{adia_kmy04} &\  \text{KMY04:} \; \quad\quad \hat{\gamma} = 
    \begin{cases}
         \frac{4}{3}  \hspace{2cm}\text{For $\Gamma_{34}>2$ ( or $a_\mathrm{u} >$  13.92)}  \\ 
        \frac{5}{3}  \hspace{2cm}\text{For $\Gamma_{34}<2$ (or $a_\mathrm{u} < 13.92$ )}  
    \end{cases}    \\ 
    \label{adia_our}&\ \text{Our work (based on \citealt{2003ApJ...591.1075K}):} \quad \quad \hat{\gamma} = \frac{4 \Gamma_{34} + 1}{3 \Gamma_{34}} \approx  \frac{4}{3} + \frac{2}{3} \left(\frac{\sqrt{a_\mathrm{u}}}{1+a_\mathrm{u}} \right)    \hspace{2cm} \text{For $f=1$}  
\end{align}
The form as represented in Eq.~(\ref{adia_our}) corresponds to the Matthews Equation of state (\citealt{1971ApJ...165..147M}) for a cold upstream medium.

\begin{figure*}
  \begin{tabular}{c|c}
   \includegraphics[scale=0.55]{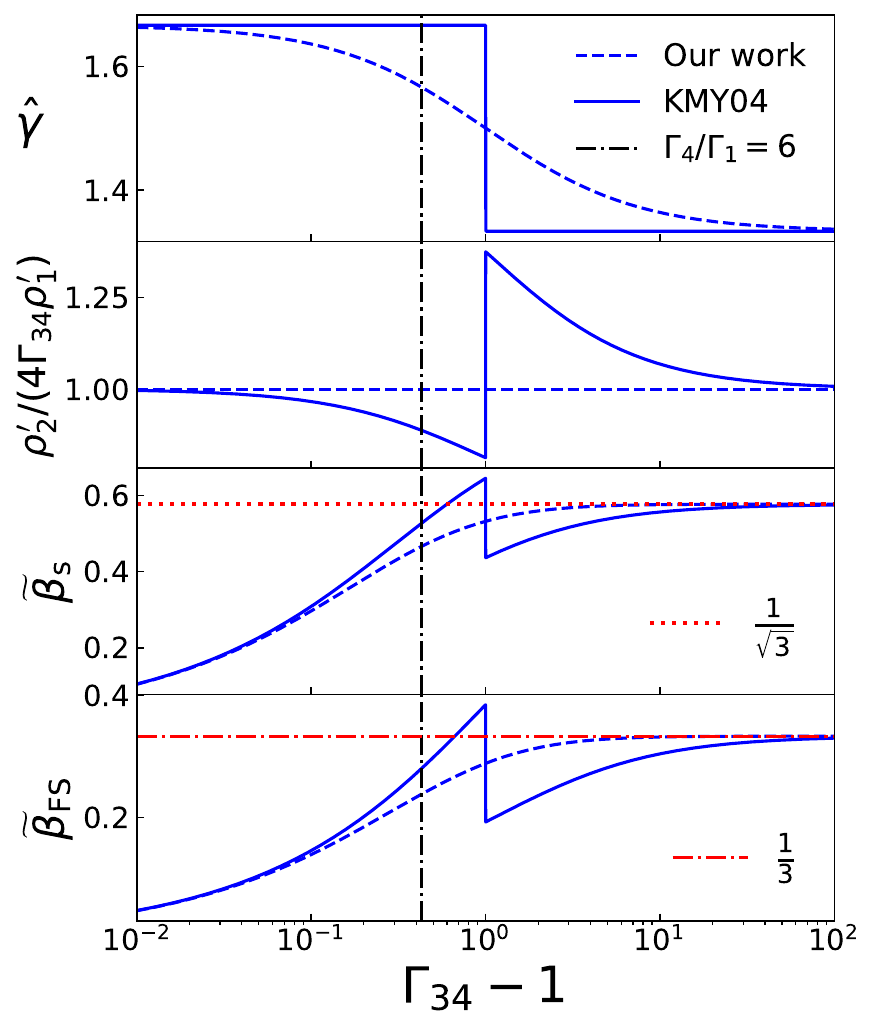} &\includegraphics[scale=0.55]{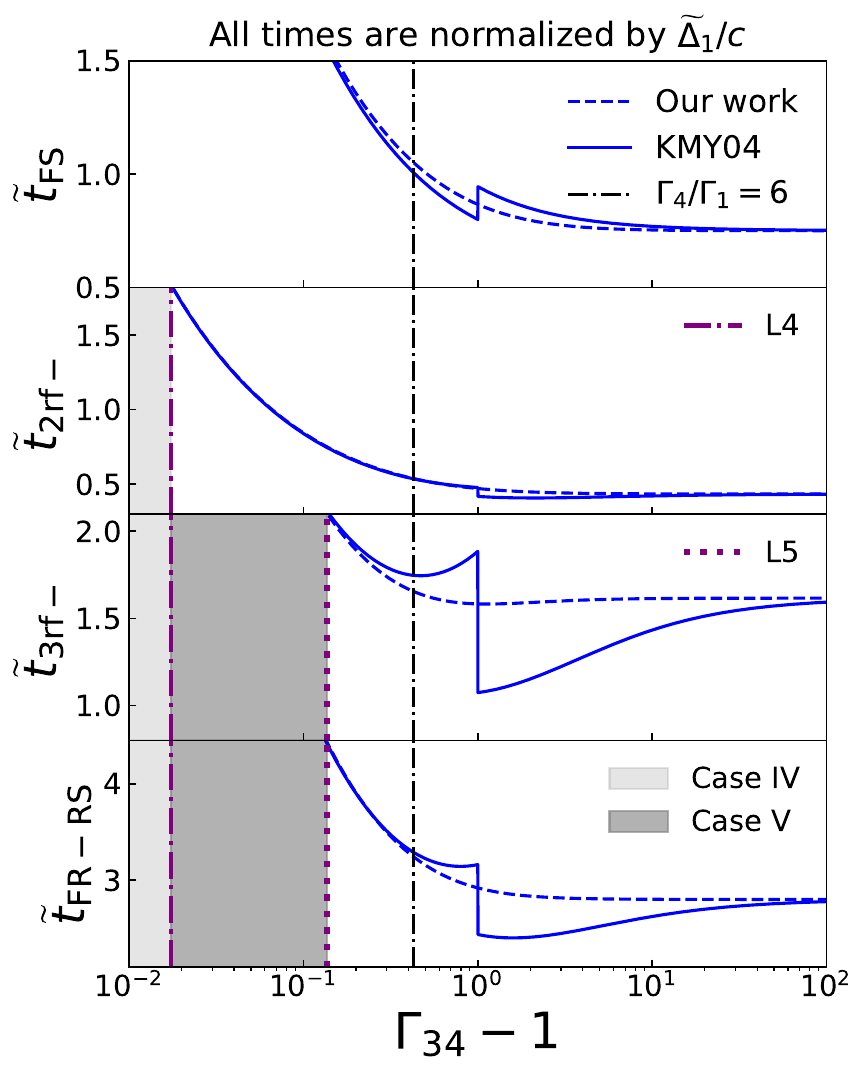}    \\ 
  \end{tabular}
\caption{Comparison between our work and that of KMY04 for a collision of two shells with equal lab frame widths and $f=1$ in the ultra-relativistic regime $(\Gamma_4 > \Gamma_1 \gg 1)$. We note that for $f=1$ we have $\Gamma_{34} = \Gamma_{21}$. While the representation shows the physical quantities as a function of shock strength $\Gamma_{34} - 1$, it corresponds to the proper density contrast according to Eq.~(\ref{eq:G34_au}).  In all panels the solid and dashed blue lines correspond to KMY04 and our work, respectively. The vertical dashed-dotted line corresponds to the shock strength $\Gamma_{34} - 1 \approx 0.429$ (or $a_\mathrm{u} = 6$). Table~\ref{diff@6} shows the difference of various quantities related to this shock strength. \textbf{Left:} From top to bottom panels shows the adiabatic index, the compression ratio, the CD frame sound speed and the CD frame forward shock speed, as a function of the shock strength $\Gamma_{34} - 1$. The horizontal red dotted and dashed-dotted lines correspond to the asymptotic values of the CD frame sound speed and the forward shock speed, respectively, at very high shock strengths ($\Gamma_{34}-1\gg1$).  \textbf{Right:} From top to bottom panels show the CD frame time as a function of the shock strength $\Gamma_{34} -1$ for forward shock crossing, the time for the head of the rf wave to reach CD, the time for rf wave to catch up with RS and the sum of all the times. All times are normalized to the light crossing time of shell S1 in the CD frame $\widetilde{\Delta}_1/c$. The vertical purple dashed-dotted and dotted lines correspond to L4 and L5 demarcating cases IV and V (indicated in Table 6 of the main text). The shell S4 is partially shocked in the unshaded region where the shocked fraction $\alpha_3$ is given by Eq.~(\ref{eq:alpha3}). The grey horizontal line in Fig. 8 in the main text provides an alternative representation for $f=1$ scenario in the lab frame.}
\label{Compare_works}
\end{figure*}

Fig. \ref{Compare_works} shows the difference between various physical quantities between our work and that of KMY04. Fig. 11 of KMY04 uses $a_\mathrm{u}  \approx \Gamma_4/\Gamma_1 = 6$. This corresponds to the vertical dashed-dotted black line in all the panels of Fig.~\ref{Compare_works}. The difference in the values of the various quantities in our work and that by KMY04 is summarized in table \ref{diff@6} corresponding to $a_\mathrm{u} = 6$. The rest of the discussion that follows corresponds to this specific value of proper speed contrast. It can be seen that in the case of KMY04, the compression ratio is smaller compared to our result which means the radial width of the shocked shell S1 is higher for KMY04. However, both the sound speed and the forward shock speed are higher compared to our case. As a result, these two competing effects compensate each other and the timescales are very similar (with less than 3 $\%$ difference). In fact, most of the deviation occurs near $\Gamma_{34} - 1 = 2$ where the differences can be at the level $\sim 14 \%$.    

\begin{table}
    \centering
\caption{Absolute difference and  fractional difference (between our work RGB24 and that by KMY04) of the values of the various quantities for a proper speed contrast of $a_\mathrm{u} = 6$, as shown along the vertical dot dashed black line shown in Fig. \ref{Compare_works}. For the fractional difference, the value is calculated relative to that found in our analysis. The negative sign in the difference and the fractional difference indicates that the values in KMY04 for that quantity is higher than in our approach and vice versa. All times are normalized to the light crossing time of shell S1 in the CD frame $\widetilde{\Delta}_1/c$.   }
    \begin{tabular}{c|c|c} \hline 
        Quantity & Absolute difference & Fractional difference \\
           $X$     & $X_\mathrm{RGB24} - X_\mathrm{KMY04}$    &   $(X_\mathrm{RGB24} - X_\mathrm{KMY04})/X_\mathrm{RGB24}$           \\  \hline 
          $\hat{\gamma}$  &   -0.100   & -0.064 \\
          $\rho_2'/(4 \Gamma_{34}\rho_1')$  & 0.112 & 0.112  \\
          $\widetilde{\beta}_\mathrm{s}$    &  -0.061 & -0.130\\
          $\widetilde{\beta}_\mathrm{FS}$   &  -0.042 & -0.177\\
          $\widetilde{t}_\mathrm{FS}$       &  0.044  & 0.042 \\
          $\widetilde{t}_\mathrm{2rf-}$     &  0.002  &  0.003\\
          $\widetilde{t}_\mathrm{3rf-}$     &  -0.092 &  -0.056 \\
          $\widetilde{t}_\mathrm{FR-RS}$    &  -0.046 &  -0.014\\ \hline 
    \end{tabular}
    \label{diff@6}
\end{table}

At this point we contrast the motivation of our work with that by KMY04. Table \ref{Expr_compare} compares the notation used in this work with KMY04. KMY04 carried a numerical simulation of shock propagation using a relativistic Riemann solver in the CD frame in the moderate $(a_\mathrm{u} = 3)$ to high proper speed contrast $(a_\mathrm{u} = 6)$ in the ultra-relativistic $(\Gamma_1 \gg 1)$ limit. Their simulation was carried for very long times, as a result they had scenarios where two rarefaction waves (a forward propagating and a backward propagating RF wave) can meet and interact among themselves. The simulation was carried out on these long times so as to investigate the evolution of the different proper density profiles with comoving time. In our approach, we limit our analysis to the time when a particular shock front either crosses its shell or is caught by a rarefaction wave (whichever occurs earlier). This is because our main objective is to estimate the energy dissipated by the shock fronts.  

To summarize, we find that 
\begin{itemize}
    \item The principal difference between our work and KMY04 is the method of estimation of thermal efficiency. We estimate thermal efficiency in the lab frame where we explicitly take into account the $p dV$ work done across CD from region R3 to R2. In the KMY04 CD rest frame approach, the CD is at rest by construction and there is no notion of $p dV$ work. There are two main limitations with a CD frame analysis: (i) if one needs the thermal energy dissipated over a spatial region in the lab frame at a given lab frame time, the spatial integration involves quantities in CD frame evaluated at different CD frame times (see discussion after eq. 26 of KMY04). The issue has to do with ``simultaneity" in relativity. The times in CD frame are not simultaneous with times in the lab frame. (ii) when Lorentz transforming quantities from the CD rest frame to the lab frame, KMY04 miss out on the $p dV$ work done and systematically underestimate the thermal efficiency.\\ \indent 
    Since we calculate the thermal efficiency in the lab frame, our expressions can be used more readily to infer radiation properties in the observer frame for a variety of astrophysical scenarios (albeit after being multiplied by additional efficiency factor(s)).
   \item The numerical analysis of KMY04 was restricted to moderately high proper speed contrasts $a_\mathrm{u} \sim 3 -6$ and ultra-relativistic colliding shells. Our analytical expressions are more general -- applicable for arbitrary proper speed contrast $a_\mathrm{u}$ and proper speeds of the colliding shells.    
   \item In our approach, we use the \cite{2003ApJ...591.1075K} representation of the Mathews equation of state (\citealt{1971ApJ...165..147M}) for a cold upstream. KMY04 use an \textit{ad hoc} equation of state (EoS) where the adiabatic constant is taken to be $5/3$ below $\Gamma_{34} < 2$ and $4/3$ for $\Gamma_{34} > 2$. Thus, their EoS has a discontinuous change at $\Gamma_{34} = 2$ (between the two asymptotic values). The authors further assume the adiabatic constant to be equal in both shocked regions. This is not true in our formalism as in general the shock strengths are different in two regions and so does the adiabatic constant which depends on the shock strength. In fact, one of the primary motivation of choosing to analyse the $f=1$ collision is to compare a scenario consistent with their assumption of equal adiabatic indices in both shocked regions.   
   \item We find the crossing times are similar. However, this similarity of crossing times comes with few caveats which we discuss further. Most of the differences (in our work) in the sound speed and the shock speed arises near $\Gamma_{34}=2$ (the point of discontinuity in the KMY04 equation of state). In KMY04, for $\Gamma_{34}<2$, the compression ratio is lower (indicating a wider radial width of the shocked region) while both the sound speed and the shock speeds are higher compared to our case (the trend reverses for $\Gamma_{34}>2$). These two opposing effects gives similar crossing timescales. The maximum difference in the crossing timescales is of the order of $\sim 14 \%$.    
\end{itemize}

\bsp	
\label{lastpage}
\end{document}